\documentclass[11pt,twoside]{article}

\usepackage{fullpage}

\usepackage{epsf}
\usepackage{fancyhdr}
\usepackage{arydshln}
\usepackage{graphics}
\usepackage{graphicx}
\usepackage{psfrag}
\usepackage{microtype}
\usepackage{algorithmic}
\usepackage{tikz}
\usetikzlibrary{arrows.meta}
\usepackage{subfig}
\usepackage{multirow}
\usepackage{wrapfig}

\usepackage[linesnumbered,ruled]{algorithm2e}% http://ctan.org/pkg/algorithm2e
\DontPrintSemicolon
\usepackage{color}

\usepackage{amsthm}
\usepackage{amsfonts}
\usepackage{amsmath}
\usepackage{amssymb,bbm}

\usepackage{natbib}
\usepackage{url}
\usepackage[colorlinks,linkcolor=magenta,citecolor=blue, pagebackref=true,backref=true]{hyperref}
\renewcommand*{\backrefalt}[4]{%
    \ifcase #1 \footnotesize{(Not cited.)}%
    \or        \footnotesize{(Cited on page~#2.)}%
    \else      \footnotesize{(Cited on pages~#2.)}%
    \fi}

\long\def\comment#1{}
\usepackage{nicefrac}
\usepackage[normalem]{ulem}
\usepackage{chngpage}

\usepackage{enumitem}
\usepackage{booktabs}
\usepackage{caption}

\usepackage{mathtools}

\usepackage{fullpage}
\newtheorem{theorem}{Theorem}[section]
\newtheorem{corollary}[theorem]{Corollary}
\newtheorem{lemma}[theorem]{Lemma}
\newtheorem{proposition}[theorem]{Proposition}

\newtheorem{definition}{Definition}[section]
\newtheorem{example}{Example}[section]

\newtheorem{remark}[theorem]{Remark}

\newcommand{\st}{\textnormal{s.t.}}

\newcommand{\diag}{\textnormal{diag}}

\newcommand{\vct}{\textnormal{vec}\,}

\newcommand{\x}{\mathbf x}
\newcommand{\y}{\mathbf y}

\newcommand{\argmin}{\mathop{\rm argmin}}

\newcommand{\LCal}{\mathcal{L}}

\newcommand{\OCal}{\mathcal{O}}
\newcommand{\PCal}{\mathcal{P}}

\newcommand{\br}{\mathbb{R}}

\newcommand{\ba}{\begin{array}}
\newcommand{\ea}{\end{array}}

\newcommand{\one}{\textbf{1}}
\newcommand{\zero}{\textbf{0}}
\newcommand{\bigO}{O}
\newcommand{\bigOtil}{\widetilde{O}}
\newcommand{\mydefn}{:=}

\newcommand{\stkout}[1]{\ifmmode\text{\sout{\ensuremath{#1}}}\else\sout{#1}\fi}

\definecolor{DSgray}{cmyk}{0,1,0,0}

\begin{document}

%%%%%%% TITLE PAGE %%%%%%%%%%%%%%%%%%%%%%%%%%%%%%%%%%%%%%%%%%%%%%%%%%%

\begin{center}
{\bf{\LARGE{Fixed-Support Wasserstein Barycenters:  \\ [.2cm] Computational Hardness and Fast Algorithm}}}

\vspace*{.2in}
{\large{ \begin{tabular}{ccccc}
Tianyi Lin$^\diamond$ & Nhat Ho$^\star$ & Xi Chen$^\ddagger$ &  Marco Cuturi$^{\triangleleft, \triangleright}$ & Michael I. Jordan$^{\diamond, \dagger}$ \\
\end{tabular}
}}

\vspace*{.2in}

\begin{tabular}{c}
Department of Electrical Engineering and Computer Sciences$^\diamond$ \\
Department of Statistics$^\dagger$ \\ 
University of California, UC Berkeley \\
Department of Statistics and Data Sciences, University of Texas, Austin$^\star$ \\ 
Stern School of Business, New York University$^\ddagger$ \\
CREST - ENSAE$^\triangleleft$, Google Brain$^\triangleright$
\end{tabular}
\vspace*{.2in}

\today

\vspace*{.2in}
\begin{abstract}
We study the fixed-support Wasserstein barycenter problem (FS-WBP), which consists in computing the Wasserstein barycenter of $m$ discrete probability measures supported on a finite metric space of size $n$. We show first that the constraint matrix arising from the standard linear programming (LP) representation of the FS-WBP is \textit{not totally unimodular} when $m \geq 3$ and $n \geq 3$. This result resolves an open question pertaining to the relationship between the FS-WBP and the minimum-cost flow (MCF) problem since it proves that the FS-WBP in the standard LP form is not an MCF problem when $m \geq 3$ and $n \geq 3$. We also develop a provably fast \textit{deterministic} variant of the celebrated iterative Bregman projection (IBP) algorithm, named \textsc{FastIBP}, with a complexity bound of $\bigOtil(mn^{7/3}\varepsilon^{-4/3})$, where $\varepsilon \in (0, 1)$ is the desired tolerance. This complexity bound is better than the best known complexity bound of $\bigOtil(mn^2\varepsilon^{-2})$ for the IBP algorithm in terms of $\varepsilon$, and that of $\bigOtil(mn^{5/2}\varepsilon^{-1})$ from accelerated alternating minimization algorithm or accelerated primal-dual adaptive gradient algorithm in terms of $n$. Finally, we conduct extensive experiments with both synthetic data and real images and demonstrate the favorable performance of the \textsc{FastIBP} algorithm in practice.
\end{abstract}
\end{center}

\section{Introduction}
Over the past decade, the Wasserstein barycenter problem~\citep{Agueh-2011-Barycenters} (WBP) has served as a foundation for theoretical analysis in a wide range of fields, including economics~\citep{Carlier-2010-Matching, Chiappori-2010-Hedonic} and physics~\citep{Buttazzo-2012-Optimal, Cotar-2013-Density, Trouve-2005-Local} to statistics~\citep{Munch-2015-Probabilistic, Ho-2017-Multilevel, Srivastava-2018-Scalable}, image and shape analysis~\citep{Rabin-2011-Wasserstein, Bonneel-2015-Sliced, Bonneel-2016-Barycoord} and machine learning~\citep{Cuturi-2014-Fast}. The WBP problem is related to the optimal transport (OT) problem, in that both are based on the Wasserstein distance, but the WBP is significantly harder.  It requires the minimization of the sum of Wasserstein distances, and typically considers $m > 2$ probability measures. Its closest relative is the multimarginal optimal transport problem~\citep{Gangbo-1998-Optimal}, which also compares $m$ measures; see~\citet{Villani-2008-Optimal} for a comprehensive treatment of OT theory and~\citet{Peyre-2019-Computational} for an introduction of OT applications and algorithms.

An ongoing focus of work in both the WBP and the OT problem is the design of fast algorithms for computing the relevant distances and optima and the delineation of lower bounds that capture the computational hardness of these problems~\citep{Peyre-2019-Computational}. For the OT problem,~\citet{Cuturi-2013-Sinkhorn} introduced the Sinkhorn algorithm which has triggered significant progress~\citep{Cuturi-2016-Smoothed, Genevay-2016-Stochastic, Altschuler-2017-Near, Dvurechensky-2018-Computational, Blanchet-2018-Towards, Lin-2019-Efficient, Lahn-2019-Graph, Quanrud-2019-Approximating, Jambulapati-2019-direct, Lin-2019-Acceleration}. Variants of the Sinkhorn and Greenkhorn algorithms~\citep{Altschuler-2017-Near, Dvurechensky-2018-Computational, Lin-2019-Efficient} continue to serve as the baseline approaches in practice. As for the theoretical complexity, the best bound is $\bigOtil(n^2\varepsilon^{-1})$~\citep{Blanchet-2018-Towards, Quanrud-2019-Approximating, Lahn-2019-Graph, Jambulapati-2019-direct}. Moreover,~\citet{Lin-2019-Complexity} provided a complexity bound for the multimarginal OT problem. 

There has been significant effort devoted to the development of fast algorithms in the case of $m > 2$ discrete probability measures~\citep{Rabin-2011-Wasserstein, Cuturi-2014-Fast, Carlier-2015-Numerical, Bonneel-2015-Sliced, Benamou-2015-Iterative, Anderes-2016-Discrete, Staib-2017-Parallel, Ye-2017-Fast, Borgwardt-2020-Improved, Puccetti-2020-Computation, Claici-2018-Stochastic, Uribe-2018-Distributed, Dvurechenskii-2018-Decentralize, Yang-2018-ADMM, Le-2019-Scalable, Kroshnin-2019-Complexity, Guminov-2019-Accelerated, Ge-2019-Interior, Borgwardt-2019-Computational}. This work has provided the foundation for progress on the WBP. An important step forward was the proposal of ~\citet{Cuturi-2014-Fast} to smooth the WBP using an entropic regularization, leading to a simple gradient-descent scheme that was later improved and generalized under the name of the iterative Bregman projection (IBP) algorithm~\citep{Benamou-2015-Iterative, Kroshnin-2019-Complexity}. Further progress includes the semi-dual gradient descent~\citep{Cuturi-2016-Smoothed, Cuturi-2018-Semidual}, accelerated primal-dual gradient descent (APDAGD)~\citep{Dvurechenskii-2018-Decentralize, Kroshnin-2019-Complexity}, accelerated IBP~\citep{Guminov-2019-Accelerated}, stochastic gradient descent~\citep{Claici-2018-Stochastic}, distributed and parallel gradient descent~\citep{Staib-2017-Parallel, Uribe-2018-Distributed}, alternating direction method of multipliers (ADMM)~\citep{Ye-2017-Fast, Yang-2018-ADMM} and interior-point algorithm~\citep{Ge-2019-Interior}. Very recently,~\citet{Kroshnin-2019-Complexity} and~\citet{Guminov-2019-Accelerated} have proposed a novel primal-dual framework that made it possible to derive complexity bounds for various algorithms, including IBP, accelerated IBP and APDAGD. 

Concerning the computational hardness of the WBP with free support,~\citet{Anderes-2016-Discrete} proved that the barycenter of $m$ empirical measures is also an empirical measure with support whose cardinality is at most the size of the union of the support of the $m$ measures, minus $m-1$. When $m=2$ and the measures are bound and the support is fixed, the computation of the barycenter amounts to solving a network flow problem on a directed graph.~\citet{Borgwardt-2019-Computational} proved that finding a barycenter of sparse support is NP hard even in the simple setting when $m=3$. However, their analysis cannot be extended to the fixed-support WBP, where the supports of the constituent $m$ probability measures are prespecified.

\paragraph{Contribution.} In this paper, we revisit the fixed-support Wasserstein barycenter problem (FS-WBP) between $m$ discrete probability measures supported on a prespecified set of $n$ points. Our contributions can be summarized as follows:
\begin{enumerate}
\item We prove that the FS-WBP in the standard LP form is not a minimum-cost flow (MCF) problem in general. In particular, we show that the constraint matrix arising from the standard LP representation of the FS-WBP is totally unimodular when $m \geq 3$ and $n = 2$ but not totally unimodular when $m \geq 3$ and $n \geq 3$. Our results shed light on \textit{the necessity of problem reformulation}---e.g., entropic regularization~\citep{Cuturi-2014-Fast, Benamou-2015-Iterative} and block reduction~\citep{Ge-2019-Interior}.
\item We propose a fast \textit{deterministic} variant of the iterative Bregman projection (IBP) algorithm, named \textsc{FastIBP}, and provide a theoretical guarantee for the algorithm. Letting $\varepsilon \in (0, 1)$ denote the target tolerance, the complexity bound of the algorithm is $\bigOtil(mn^{7/3}\varepsilon^{-4/3})$, which improves the complexity bound of $\bigOtil(mn^2\varepsilon^{-2})$ of the IBP algorithm~\citep{Benamou-2015-Iterative} in terms of $\varepsilon$ and the complexity bound of $\bigOtil(mn^{5/2}\varepsilon^{-1})$ from the accelerated IBP and APDAGD algorithms in terms of $n$~\citep{Kroshnin-2019-Complexity, Guminov-2019-Accelerated}. We conduct experiments on synthetic and real datasets and demonstrate that the \textsc{FastIBP} algorithm achieves the favorable performance in practice.
\end{enumerate}
\paragraph{Organization.} The remainder of the paper is organized as follows. In Section~\ref{sec:problem}, we provide the basic setup for the entropic-regularized FS-WBP and the dual problem. In Section~\ref{sec:hardness}, we present our computational hardness results for the FS-WBP in the standard LP form. In Sections~\ref{sec:fast-IBP}, we propose and analyze the \textsc{FastIBP} algorithm. We present experimental results on synthetic and real data in Section~\ref{sec:experiments} and conclude in Section~\ref{sec:discussions}.
\paragraph{Notation.} We let $[n]$ be the set $\{1, 2, \ldots, n\}$ and $\br^n_+$ be the set of all vectors in $\br^n$ with nonnegative components. $\one_n$ and $\zero_n$ are the $n$-vectors of ones and zeros. $\Delta^n$ stands for the probability simplex:  $\Delta^n = \{u \in \br^n_+: \one_n^\top u = 1\}$. For a smooth function $f$, we denote $\nabla f$ and $\nabla_\lambda f$ as the full gradient and the gradient with respect to a variable $\lambda$. For $x \in \br^n$ and $1 \leq p \leq \infty$, we write $\|x\|_p$ for its $\ell_p$-norm. For $X = (X_{ij}) \in \br^{n \times n}$, the notations $\vct(X) \in \br^{n^2}$ and $\det(X)$ stand for the vector representation and the determinant. The notations $\|X\|_\infty = \max_{1 \leq i, j \leq n} |X_{ij}|$ and $\|X\|_1 = \sum_{1 \leq i, j \leq n} |X_{ij}|$. The notations $r(X) = X\one_n$ and $c(X) = X^\top\one_n$. Let $X, Y \in \br^{n \times n}$, the Frobenius and Kronecker inner product are denoted by $\langle X, Y\rangle$ and $X \otimes Y$. Given the dimension $n$ and $\varepsilon$, the notation $a = \bigO(b(n,\varepsilon))$ stands for the upper bound $a \leq C \cdot b(n, \varepsilon)$ where $C>0$ is independent of $n$ and $\varepsilon$, and $a = \bigOtil(b(n, \varepsilon))$ indicates the previous inequality where $C$ depends only the logarithmic factors of $n$ and $\varepsilon$. 
%%%%%%%%%%%%%%%%%%%%%%%%%%%%%%%%%%%%%%%%%%%%%%%%%%%%%%%%%%%%%%%%%%%%%

\section{Preliminaries and Technical Background}\label{sec:problem}
In this section, we introduce the basic setup of the fixed-support Wasserstein barycenter problem (FS-WBP), starting with the linear programming (LP) presentation and entropic-regularized formulation and including a specification of an approximate barycenter. 

\subsection{Linear programming formulation}
For $p \geq 1$, let $\mathcal{P}_p(\Omega)$ be the set of Borel probability measures on $\Omega$ with finite $p$-th moment. The Wasserstein distance of order $p\geq 1$ between $\mu, \nu \in \mathcal{P}_{p}(\Omega)$ is defined by~\citep{Villani-2008-Optimal}:
\begin{equation}
W_p(\mu, \nu) \mydefn \inf_{\pi \in \Pi(\mu, \nu)} \left(\int_{\Omega \times \Omega} d^{p}(\x, \y) \ \pi(d\x, d\y)\right)^{1/p}, 
\end{equation}
where $d(\cdot, \cdot)$ is a metric on $\Omega$ and $\Pi(\mu, \nu)$ is the set of couplings between $\mu$ and $\nu$. Given a weight vector $(\omega_1, \omega_2, \ldots, \omega_m) \in \Delta^m$ for $m \geq 2$, the \emph{Wasserstein barycenter}~\citep{Agueh-2011-Barycenters} of $m$ probability measures $\{\mu_k\}_{k=1}^m$ is a solution of the following functional minimization problem 
\begin{equation}\label{prob:barycenter_original}
\min_{\mu \in \mathcal{P}_{p}(\Omega)} \ \sum_{k=1}^m \omega_k W_p^p(\mu, \mu_k). 
\end{equation}
Because our goal is to provide computational schemes to approximately solve the WBP, we need to provide a definition of an $\varepsilon$-approximate solution to the WBP.
\begin{definition}\label{def:approx_barycenter_original}
The probability measure $\widehat{\mu} \in \mathcal{P}_{p}(\Omega)$ is called an \emph{$\varepsilon$-approximate barycenter} if $\sum_{k=1}^m \omega_k W_p^p(\widehat{\mu}, \mu_k) \leq \sum_{k=1}^m \omega_k W_p^p(\mu^\star, \mu_k) + \varepsilon$ where $\mu^\star$ is an optimal solution to problem~\eqref{prob:barycenter_original}.
\end{definition}
There are two main settings: (i) \emph{free-support Wasserstein barycenter}, namely, when we optimize both the weights and supports of the barycenter in Eq.~\eqref{prob:barycenter_original}; and (ii) \emph{fixed-support Wasserstein barycenter}, namely, when the supports of the barycenter are obtained from those from the probability measures $\{\mu_k\}_{k=1}^m$ and we optimize the weights of the barycenter in Eq.~\eqref{prob:barycenter_original}. 

\textit{The free-support WBP problem is notoriously difficult to solve.} It can either be solved using a solution to the multimarginal-OT (MOT) problem, as described in detail by~\citet{Agueh-2011-Barycenters}, or approximated using alternative optimization techniques. Assuming that each measure is supported on $n$ distinct points, the WBP problem can be solved \emph{exactly} by solving first a MOT, to then compute $(n-1)m+1$ barycenters of \emph{points} in $\Omega$ (these barycenters are exactly the support of the barycentric measure). Solving a MOT is, however, equivalent to solving an LP with $n^m$ variables and $(n-1)m+1$ constraints. The other route, alternative optimization, requires specifying an initial guess for the barycenter, a discrete measure supported on $k$ weighted points (where $k$ is predefined). One can then proceed by updating the locations of $\mu$ (or even add new ones) to decrease the objective function in Eq.~\eqref{prob:barycenter_original}, before changing their weights. In the Euclidean setting with $p=2$, the free-support WBP is closely related to the clustering problem, and equivalent to $k$-means when $m=1$~\citep{Cuturi-2014-Fast}. Whereas solving the free-support WBP using MOT results in a convex (yet intractable) problem, the alternating mimimization approach is not, in very much the same way that the $k$-means problem is not, and results in the minimization of a piece-wise quadratic function. \textit{On the other hand, the fixed-support WBP is comparatively easier to solve, and as such has played a role in real-world applications.} For instance, in imaging sciences, pixels and voxels are supported on a predefined, finite grid. In these applications, the barycenter and $\mu_k$ measures share the same support. 

In view of this, throughout the remainder of the paper, we let $(\mu_k)_{k=1}^m$ be discrete probability measures and take the support points $\{\x_i^k\}_{i \in [n]}$ to be fixed. Since $\{\mu_k\}_{k=1}^m$ have the fixed support, they are fully characterized by the weights $\{u^k\}_{k=1}^m$. Accordingly, the support of the barycenter $\{\widehat{\x}_i\}_{i \in [n]}$ is also fixed and can be prespecified by $\{\x_i^k\}_{i \in [n]}$. Given this setup, the FS-WBP between $\{\mu_k\}_{k=1}^m$ has the following standard LP representation~\citep{Cuturi-2014-Fast, Benamou-2015-Iterative, Peyre-2019-Computational}:
\begin{equation}\label{prob:barycenter}
\min\limits_{\{X_i\}_{i=1}^m \subseteq \br_{+}^{n \times n}} \ \sum_{k=1}^m \omega_k \langle C_k, X_k\rangle, \quad \st 
\begin{array}{l}
r(X_k) = u^k \textnormal{ for all } k \in [m], \\
c(X_{k+1}) = c(X_k) \textnormal{ for all } k \in [m-1],
\end{array}
\end{equation}
where $\{X_k\}_{k=1}^m$ and $\{C_k\}_{k=1}^m \subseteq \br_+^{n \times \ldots \times n}$ denote a set of \emph{transportation plans} and \emph{nonnegative cost matrices} and $(C_k)_{ij} = d^{p}(\x_i^k, \widehat{\x}_j)$ for all $k \in [m]$. The fixed-support Wasserstein barycenter $u \in \Delta_n$ is determined by the weight $\sum_{k=1}^m \omega_k c(X_k)$ and the support $(\widehat{\x}_1, \widehat{\x}_2, \ldots, \widehat{\x}_n)$. 

From Eq.~\eqref{prob:barycenter}, the FS-WBP is an LP with $2mn-n$ equality constraints and $mn^2$ variables. This has inspired work on solving the FS-WBP using classical optimization algorithms~\citep{Ge-2019-Interior, Yang-2018-ADMM}. Although progress has been made, the understanding of the structure of FS-WBP via this approach has remained limited. Particularly, while the OT problem~\citep{Villani-2008-Optimal} is equivalent to a minimum-cost flow (MCF) problem, it remains unknown whether the FS-WBP is a MCF problem even in the simplest setting when $m = 2$.
%%%%%%%%%%%%%%%%%%%%%%%%%%%%%%%%%%%%%%%%%%%%%%%%%%%%%%%%%%%%%%%%%%%%%

\subsection{Entropic regularized FS-WBP}
Using Cuturi's entropic approach to the OT problem~\citep{Cuturi-2013-Sinkhorn}, we define a regularized version of the FS-WBP in Eq.~\eqref{prob:barycenter}, where an entropic regularization term is added to the Wasserstein barycenter objective. The resulting formulation is as follows: 
\begin{equation}\label{prob:barycenter_regularized}
\min\limits_{\{X_i\}_{i=1}^m \subseteq \br_+^{n \times n}} \ \sum_{k=1}^m \omega_k (\langle C_k, X_k\rangle - \eta H(X_k)), \quad \st 
\begin{array}{l}
r(X_k) = u^k \textnormal{ for all } k \in [m], \\
c(X_{k+1}) = c(X_k) \textnormal{ for all } k \in [m-1],
\end{array}
\end{equation}
where $\eta > 0$ is the parameter and $H(X) \mydefn - \langle X, \log(X)-\one_n\one_n^\top\rangle$ denotes the entropic regularization term. We refer to Eq.~\eqref{prob:barycenter_regularized} as \emph{entropic regularized FS-WBP}. When $\eta$ is large, the optimal value of entropic regularized FS-WBP may yield a poor approximation of the cost of the FS-WBP. To guarantee a good approximation, we scale the parameter $\eta$ as a function of the desired accuracy.
\begin{definition}\label{def:approx_barycenter}
The probability vector $\widehat{u} \in \Delta^n$ is called an \emph{$\varepsilon$-approximate barycenter} if there exists a feasible solution $(\widehat{X}_1, \widehat{X}_2, \ldots, \widehat{X}_m) \in \br_+^{n \times n} \times \cdots \times \br_+^{n \times n}$ for the FS-WBP in Eq.~\eqref{prob:barycenter} such that $\widehat{u} = \sum_{k=1}^m \omega_k  c(\widehat{X}_k)$ for all $k \in [m]$ and $\sum_{k=1}^m \omega_k \langle C_k, \widehat{X}_k\rangle \leq \sum_{k=1}^m \omega_k \langle C_k, X_k^\star\rangle + \varepsilon$ where $(X_1^\star, X_2^\star, \ldots, X_m^\star)$ is an optimal solution of the FS-WBP in Eq.~\eqref{prob:barycenter}.
\end{definition}
With these definitions in mind, we develop efficient algorithms for approximately solving the FS-WBP where the dependence on $m$, $n$ and $\varepsilon$ is competitive to state-of-the-art algorithms~\citep{Kroshnin-2019-Complexity, Guminov-2019-Accelerated}.

\subsection{Dual entropic regularized FS-WBP}\label{sec:dual_entropic_barycenter}
Using the duality theory of convex optimization~\citep{Rockafellar-1970-Convex}, one dual form of entropic regularized FS-WBP has been derived before~\citep{Cuturi-2014-Fast, Kroshnin-2019-Complexity}. Differing from the usual 2-marginal or multimarginal OT~\citep{Cuturi-2018-Semidual, Lin-2019-Complexity}, the dual entropic regularized FS-WBP is a convex optimization problem with an affine constraint set. Formally, we have 
\begin{equation}\label{prob:barycenter_regularized_dual_old}
\min\limits_{\lambda, \tau \in \br^{mn}} \varphi_{\textnormal{old}}(\lambda, \tau) \mydefn \sum_{k=1}^m \omega_k \left(\sum_{1 \leq i, j \leq n} e^{\lambda_{ki} + \tau_{kj} - \eta^{-1}(C_k)_{ij}} - \lambda_k^\top u^k\right), \quad \st \ \sum_{k=1}^m \omega_k \tau_k = \zero_n.
\end{equation}
However, the objective function in Eq.~\eqref{prob:barycenter_regularized_dual_old} is not sufficiently smooth because of the sum of exponents. This makes the acceleration very challenging. In order to alleviate this issue, we turn to derive another smooth dual form of entropic-regularized FS-WBP. 

By introducing the dual variables $\{\alpha_1, \alpha_2, \ldots, \alpha_m, \beta_1, \beta_2, \ldots, \beta_{m-1}\} \subseteq \br^n$, we define the Lagrangian function of the entropic regularized FS-WBP in Eq.~\eqref{prob:barycenter_regularized} as follows:
\begin{eqnarray}\label{opt:Lagrangian}
& & \LCal(X_1, \ldots, X_m, \alpha_1, \ldots, \alpha_m, \beta_1, \ldots, \beta_{m-1})   \\
& = & \sum_{k=1}^m \omega_k(\langle C_k, X_k\rangle - \eta H(X_k)) - \sum_{k=1}^m \alpha_k^\top(r(X_k) - u^k) - \sum_{k=1}^{m-1}\beta_k^\top(c(X_{k+1}) - c(X_k)). \nonumber 
\end{eqnarray}
By the definition of $H(X)$, the nonnegative constraint $X \geq 0$ can be neglected. In order to derive the smooth dual objective function, we consider the following minimization problem:
\begin{equation*}
\min_{\{(X_1, \ldots, X_m): \|X_k\|_1=1, \forall k \in [m]\}} \LCal(X_1, \ldots, X_m, \alpha_1, \ldots, \alpha_m, \beta_1, \ldots, \beta_{m-1}). 
\end{equation*}
In the above problem, the objective function is strongly convex. Thus, the optimal solution is unique. For the simplicity, we let $\alpha = (\alpha_1, \alpha_2, \ldots, \alpha_m) \in \br^{mn}$ and $\beta = (\beta_1, \beta_2, \ldots, \beta_{m-1}) \in \br^{(m-1)n}$ and assume the convention $\beta_0 = \beta_m = \zero_n$. After the simple calculations, the optimal solution $\bar{X}(\alpha, \beta) = (\bar{X}_1(\alpha, \beta), \ldots, \bar{X}_m(\alpha, \beta))$ has the following form:
\begin{equation}\label{opt:barycenter_plan}
(\bar{X}_k(\alpha, \beta))_{ij} = \frac{e^{\eta^{-1}(\omega_k^{-1}(\alpha_{ki} + \beta_{k-1, j} - \beta_{kj}) - (C_k)_{ij})}}{\sum_{1 \leq i, j \leq n} e^{\eta^{-1}(\omega_k^{-1}(\alpha_{ki} + \beta_{k-1, j} - \beta_{kj}) - (C_k)_{ij})}} \quad \textnormal{for all } k \in [m],
\end{equation}
Plugging Eq.~\eqref{opt:barycenter_plan} into Eq.~\eqref{opt:Lagrangian} yields that the dual form is: 
\begin{equation*}
\max_{\alpha_1, \ldots, \alpha_m, \beta_1, \ldots, \beta_{m-1}} \ \left\{-\eta\sum_{k=1}^m \omega_k\log\left(\sum_{1 \leq i, j \leq n} e^{\eta^{-1}(\omega_k^{-1}(\alpha_{ki} + \beta_{k-1, j} - \beta_{kj}) - (C_k)_{ij})}\right) + \sum_{k=1}^m \alpha_k^\top u_k\right\}.
\end{equation*}
In order to streamline our subsequent presentation, we perform a change of variables $\lambda_k = (\eta\omega_k)^{-1}\alpha_k$ and $\tau_k = (\eta\omega_k)^{-1}(\beta_{k-1}-\beta_k)$ for all $k \in [m]$. Recall that $\beta_0 = \beta_m = \zero_n$, we have $\sum_{k=1}^m \omega_k \tau_k = \zero_n$. Putting these pieces together, we reformulate the problem as
\begin{equation*}
\min\limits_{\lambda, \tau \in \br^{mn}} \varphi(\lambda, \tau) \mydefn \sum_{k=1}^m \omega_k\log\left(\sum_{1 \leq i, j \leq n} e^{\lambda_{ki}+\tau_{kj} - \eta^{-1}(C_k)_{ij}}\right) - \sum_{k=1}^m \omega_k\lambda_k^\top u^k, \quad \st \ \sum_{k=1}^m \omega_k \tau_k = \zero_n.
\end{equation*}
To further simplify the above formulation, we use the notation $B_k(\lambda, \tau) \in \br^{n \times n}$ by $(B_k(\lambda_k, \tau_k))_{ij} = e^{\lambda_{ki}+\tau_{kj} - \eta^{-1}(C_k)_{ij})}$ for all $(i, j) \in [n] \times [n]$. To this end, we obtain the \emph{dual entropic-regularized FS-WBP problem} defined by
\begin{equation}\label{prob:barycenter_regularized_dual}
\min\limits_{\lambda, \tau \in \br^{mn}} \varphi(\lambda, \tau) \mydefn \sum_{k=1}^m \omega_k\left(\log(\|B_k(\lambda_k, \tau_k)\|_1) - \lambda_k^\top u^k\right), \quad \st \ \sum_{k=1}^m \omega_k \tau_k = \zero_n.
\end{equation}
\begin{remark}\label{remark:barycenter_regularized_dual}
The first part of the objective function $\varphi$ is in the form of the logarithm of sum of exponents while the second part is a linear function. This is different from the objective function used in previous dual entropic regularized OT problem in Eq.~\eqref{prob:barycenter_regularized_dual_old}. We also note that Eq.~\eqref{prob:barycenter_regularized_dual} is a special instance of a softmax minimization problem, and the objective function $\varphi$ is known to be smooth~\citep{Nesterov-2005-Smooth}. Finally, we point out that the same problem was derived in the concurrent work by~\citet{Guminov-2019-Accelerated} and used for analyzing the accelerated alternating minimization algorithm. 
\end{remark}
In the remainder of the paper, we also denote $(\lambda^\star, \tau^\star) \in \br^{mn} \times \br^{mn}$ as an optimal solution of the dual entropic regularized FS-WBP problem in Eq.~\eqref{prob:barycenter_regularized_dual}.

\subsection{Properties of dual entropic regularized FS-WBP}
In this section, we present several useful properties of the dual entropic regularized MOT in Eq.~\eqref{prob:barycenter_regularized_dual}. In particular, we show that there exists an optimal solution $(\lambda^\star, \tau^\star)$ such that it has an upper bound in terms of the $\ell_\infty$-norm.  
\begin{lemma}\label{Lemma:dual-bound-infinity}
For the dual entropic regularized FS-WBP, let $\bar{C} = \max_{1 \leq k \leq m} \|C_k\|_\infty$ and $\bar{u} = \min_{1 \leq k \leq m, 1 \leq j \leq n} u_{kj}$, there exists an optimal solution $(\lambda^\star, \tau^\star)$ such that the following $\ell_\infty$-norm bound holds true: 
\begin{equation*}
\|\lambda_k^\star\|_\infty \leq R_\lambda, \quad \|\tau_k^\star\|_\infty \leq R_\tau, \quad \textnormal{for all } k \in [m],  
\end{equation*} 
where $R_\lambda = 9\eta^{-1}\bar{C} + 2\log(n) - \log(\bar{u})$ and $R_\tau = 4\eta^{-1}\bar{C}$. 
\end{lemma}
\begin{proof}
First, we show that there exists $m$ pairs of optimal solutions $\{(\lambda^j, \tau^j)\}_{j \in [m]}$ such that each of $(\lambda^j, \tau^j)$ satisfies that  
\begin{equation}\label{claim-dual-bound-first}
\max\limits_{1 \leq i \leq n} (\tau_k^j)_i \geq 0 \geq \min\limits_{1 \leq i \leq n} (\tau_k^j)_i, \quad \text{for all} \ k \neq j. 
\end{equation}
Fixing $j \in [m]$, we let $(\widehat{\lambda}, \widehat{\tau})$ be an optimal solution of the dual entropic regularized FS-WBP in Eq.~\eqref{prob:barycenter_regularized_dual}. If $\widehat{\tau}$ satisfies Eq.~\eqref{claim-dual-bound-first}, the claim holds true for the fixed $j \in [m]$. Otherwise, we define $m-1$ shift terms given by
\begin{equation*}
\Delta\widehat{\tau}_k = \frac{\max_{1 \leq i \leq n} (\widehat{\tau}_k)_i + \min_{1 \leq i \leq n} (\widehat{\tau}_k)_i }{2} \in \br \quad \text{for all} \ k \neq j, 
\end{equation*}
and let $(\lambda^j, \tau^j)$ with 
\begin{equation*}
\begin{array}{lll}
\tau_k^j = \widehat{\tau}_k - \Delta\widehat{\tau}_k\one_n, & \lambda_k^j = \widehat{\lambda}_k + \Delta\widehat{\tau}_k\one_n, & \textnormal{for all} \ k \neq j, \\
\tau_j^j = \widehat{\tau}_j + (\sum_{k \neq j} (\frac{\omega_k}{\omega_j})\Delta\widehat{\tau}_k)\one_n, & \lambda_j^j = \widehat{\lambda}_j - (\sum_{k \neq j} (\frac{\omega_k}{\omega_j})\Delta\widehat{\tau}_k)\one_n. &
\end{array}
\end{equation*}
Using this construction, we have $(\lambda_k^j)_i + (\tau_k^j)_{i'} = (\widehat{\lambda}_k)_i + (\widehat{\tau}_k)_{i'}$ for all $i, i' \in [n]$ and all $k \in [m]$. This implies that $B_k(\widehat{\lambda}_k, \widehat{\tau}_k) = B_k(\lambda_k^{k'}, \tau_k^{k'})$ for all $k \in [m]$. Furthermore, we have
\begin{equation*}
\sum_{k=1}^m \omega_k\tau_k^j = \sum_{k=1}^m \omega_k\widehat{\tau}_k, \qquad \sum_{k=1}^m \omega_k (\lambda_k^j)^\top u^k = \sum_{k=1}^m \omega_k \widehat{\lambda}_k^\top u^k. 
\end{equation*}
Putting these pieces together yields $\varphi(\lambda^j, \tau^j) = \varphi(\widehat{\lambda}, \widehat{\tau})$. Moreover, by the definition of $(\lambda^j, \tau^j)$ and $m-1$ shift terms, $\tau^j$ satisfies Eq.~\eqref{claim-dual-bound-first}. Therefore, we conclude that $(\lambda^j, \tau^j)$ is an optimal solution that satisfies Eq.~\eqref{claim-dual-bound-first} for the fixed $j \in [m]$. Since $j \in [m]$ is chosen arbitarily, we can find the desired pairs of optimal solutions $\{(\lambda^j, \tau^j)\}_{j \in [m]}$ satisfying Eq.~\eqref{claim-dual-bound-first} by repeating the above argument $m$ times.   

Furthermore, each of $(\lambda^j, \tau^j)$ must satisfy the optimality condition for Eq.~\eqref{prob:barycenter_regularized_dual} for all $k \in [m]$. Fixing $j \in [m]$, there exists $z \in \br^n$ such that 
\begin{equation}\label{opt:KKT-tau}
\sum_{k=1}^m \omega_k\tau_k^j = \zero_n \quad \textnormal{and} \quad \frac{B_k(\lambda_k^j, \tau_k^j)^\top\one_n}{\|B_k(\lambda_k^j, \tau_k^j)\|_1} - z = \zero_n \quad \textnormal{for all } k \in [m]. 
\end{equation}
By the definition of $B_k(\cdot, \cdot)$, we have
\begin{equation*}
\tau_k^j = \log(z) + \log(\|B_k(\lambda_k^j, \tau_k^j)\|_1)\one_n - \log(e^{-\eta^{-1}C_k}\diag(e^{\lambda_k^j})\one_n) \textnormal{ for all } k \in [m].
\end{equation*}
This together with the first equality in Eq.~\eqref{opt:KKT-tau} yields that 
\begin{equation*}
\tau_k^j = \sum_{l=1}^m \omega_l\log(e^{-\eta^{-1}C_l}\diag(e^{\lambda_l^j})\one_n) - \log(e^{-\eta^{-1}C_k}\diag(e^{\lambda_k^j})\one_n) \textnormal{ for all } k \in [m].  
\end{equation*}
For each $i \in [n]$ and $l \in [m]$, by the nonnegativity of each entry of $C_l$, we have
\begin{equation*}
-\eta^{-1}\|C_l\|_\infty + \log(\one_n^\top e^{\lambda_l^j}) \leq [\log(e^{-\eta^{-1}C_l} \diag(e^{\lambda_l^j})\one_n)]_i \leq \log(\one_n^\top e^{\lambda_l^j}). 
\end{equation*}
Putting these pieces together yields 
\begin{equation}\label{claim-dual-bound-second}
\max_{1 \leq i \leq n} (\tau_k^j)_i - \min_{1 \leq i \leq n} (\tau_k^j)_i \leq \eta^{-1}\|C_k\|_\infty + \sum_{l=1}^m \omega_l\eta^{-1}\|C_l\|_\infty \textnormal{ for all } k \in [m]. 
\end{equation}
Combining Eq.~\eqref{claim-dual-bound-first} and Eq.~\eqref{claim-dual-bound-second} yields that 
\begin{equation}\label{claim-dual-bound-tau}
\|\tau_k^j\|_\infty \leq \eta^{-1}\|C_k\|_\infty + \sum_{l=1}^m \omega_l\eta^{-1}\|C_l\|_\infty \textnormal{ for all } k \neq j. 
\end{equation}
Since $\sum_{k=1}^m \omega_k \tau_k^j = \zero_n$, we have 
\begin{equation*}
\|\tau_j^j\|_\infty \leq \omega_j^{-1}\sum_{k \neq j} \omega_k\|\tau_k^j\|_\infty \leq (\eta\omega_j)^{-1}\sum_{k \neq j} \omega_k\|C_k\|_\infty + (\eta\omega_j)^{-1}(1 - \omega_j)\sum_{k=1}^m \omega_k\|C_k\|_\infty. 
\end{equation*}
Finally, we proceed to the key part and define the averaging iterate 
\begin{equation*}
\lambda^\star = \sum_{j=1}^m \omega_j \lambda^j, \qquad \tau^\star = \sum_{j=1}^m \omega_j \tau^j. 
\end{equation*}
Since $\varphi$ is convex and $(\omega_1, \omega_2, \ldots, \omega_m) \in \Delta^m$, we have $\varphi(\lambda^\star, \tau^\star) \leq \sum_{j=1}^m \omega_j\varphi(\lambda^j, \tau^j)$ and $\sum_{k=1}^m \omega_k \tau_k^\star = \zero_n$. Since $(\lambda^j, \tau^j)$ are optimal solutions for all $j \in [m]$, we conclude that $(\lambda^\star, \tau^\star)$ is an optimal solution. Without loss of generality, we assume that $(\lambda^\star, \tau^\star)$ satisfies that  
\begin{equation}\label{opt-shift-lambda}
\max\limits_{1 \leq i \leq n} (\lambda_k^\star)_i \geq 0 \geq \min\limits_{1 \leq i \leq n} (\lambda_k^\star)_i, \quad \text{for all} \ k \in [m]. 
\end{equation}
Indeed, if $\lambda^\star$ does not satisfy the above condition, we define $m$ shift terms given by
\begin{equation*}
\Delta\lambda_k^\star = \frac{\max_{1 \leq i \leq n} (\lambda_k^\star)_i + \min_{1 \leq i \leq n} (\lambda_k^\star)_i }{2} \in \br \quad \text{for all} \ k \in m, 
\end{equation*}
and let 
\begin{equation*}
\lambda_k^\star = \lambda_k^\star - \Delta\lambda_k^\star\one_n, \quad \textnormal{for all} \ k \in [m]. 
\end{equation*}
The above operation will not change the value of $\varphi(\lambda^\star, \tau^\star)$ such that $(\lambda^\star, \tau^\star)$ is a desired optimal solution which satisfies Eq.~\eqref{opt-shift-lambda}. 

The remaining step is to show that $\|\lambda_k^\star\|_\infty \leq R_\lambda$ and $\|\tau_k^\star\|_\infty \leq R_\tau$ for all $k \in [m]$. More specifically, we have
\begin{eqnarray*}
\|\tau_k^\star\|_\infty & \leq & \sum_{j=1}^m \omega_j\|\tau_k^j\|_\infty = \omega_k\|\tau_k^k\|_\infty + \sum_{j \neq k} \omega_j\|\tau_k^j\|_\infty \\
& \leq & \eta^{-1}\sum_{l \neq k} \omega_l\|C_l\|_\infty + \eta^{-1}(1 - \omega_k)\sum_{l=1}^m \omega_l\|C_l\|_\infty + \eta^{-1}(1 - \omega_k)(\|C_k\|_\infty + \sum_{l=1}^m \omega_l\|C_l\|_\infty) \\ 
& \leq & \eta^{-1}\|C_k\|_\infty + 3\eta^{-1}\sum_{l=1}^m \omega_l\|C_l\|_\infty \\ 
& \leq & 4\eta^{-1}\bar{C} = R_\tau.  
\end{eqnarray*}
Since $(\lambda^\star, \tau^\star)$ is an optimal solution, it satisfies the optimality condition for Eq.~\eqref{prob:barycenter_regularized_dual}. Formally, we have 
\begin{equation}\label{opt:KKT-lambda}
\frac{B_k(\lambda_k^\star, \tau_k^\star)\one_n}{\|B_k(\lambda_k^\star, \tau_k^\star)\|_1} - u^k = \zero_n \quad \textnormal{for all } k \in [m]. 
\end{equation}
By the definition of $B_k(\cdot, \cdot)$, we have
\begin{equation*}
\lambda_k^\star = \log(u^k) + \log(\|B_k(\lambda_k^\star, \tau_k^\star)\|_1)\one_n - \log(e^{-\eta^{-1}C_k}\diag(e^{\tau_k^\star})\one_n) \textnormal{ for all } k \in [m].
\end{equation*}
This implies that 
\begin{eqnarray*}
\max_{1 \leq i \leq n} (\lambda_k^\star)_i & \leq & \eta^{-1}\|C_k\|_\infty + \log(n) + \|\tau_k^\star\|_\infty + \log(\|B_k(\lambda_k^\star, \tau_k^\star)\|_1), \\
\min_{1 \leq i \leq n} (\lambda_k^\star)_i & \geq & \log(\bar{u}) - \log(n) - \|\tau_k^\star\|_\infty + \log(\|B_k(\lambda_k^\star, \tau_k^\star)\|_1). 
\end{eqnarray*}
Therefore, we have
\begin{equation*}
\max_{1 \leq i \leq n} (\lambda_k^\star)_i - \min_{1 \leq i \leq n} (\lambda_k^\star)_i \leq \eta^{-1}\|C_k\|_\infty + 2\log(n) - \log(\bar{u}) + 2\|\tau_k^\star\|_\infty. 
\end{equation*}
Together with Eq.~\eqref{opt-shift-lambda}, we conclude that 
\begin{equation*}
\|\lambda_k^\star\|_\infty \leq \eta^{-1}\|C_k\|_\infty + 2\log(n) - \log(\bar{u}) + 2\|\tau_k^\star\|_\infty. 
\end{equation*}
This completes the proof. 
\end{proof}
\begin{remark}
Lemma~\ref{Lemma:dual-bound-infinity} is analogous to~\cite[Lemma~3.2]{Lin-2019-Efficient} for the OT problem. However, the dual entropic-regularized FS-WBP is more complex and requires a novel constructive iterate, $(\lambda^\star, \tau^\star) = \sum_{j=1}^m \omega_j (\lambda^j, \tau^j)$. Moreover, the techniques in~\citet{Kroshnin-2019-Complexity} are not applicable for the analysis of the \textsc{FastIBP} algorithm, and, accordingly, Lemma~\ref{Lemma:dual-bound-infinity} is crucial for the analysis.
\end{remark}
The upper bound for the $\ell_\infty$-norm of the optimal solution of dual entropic-regularized FS-WBP in Lemma~\ref{Lemma:dual-bound-infinity} leads to the following straightforward consequence. 
\begin{corollary}\label{Corollary:dual-bound-l2}
For the dual entropic regularized FS-WBP, there exists an optimal solution $(\lambda^\star, \tau^\star)$ such that for all $k \in [m]$,
\begin{equation*}
\|\lambda_k^\star\| \leq \sqrt{n}R_\lambda, \quad \|\tau_k^\star\| \leq \sqrt{n}R_\tau, \quad \textnormal{for all } k \in [m],  
\end{equation*} 
where $R_\lambda, R_\tau > 0$ are defined in Lemma~\ref{Lemma:dual-bound-infinity}. 
\end{corollary}
Finally, we observe that $\varphi$ can be decomposed into the weighted sum of $m$ functions and prove that each component function $\varphi_k$ has Lipschitz continuous gradient with the constant $4$ in the following lemma. 
\begin{lemma}\label{Lemma:liptshitz-continuity}
The following statement holds true, $\varphi(\lambda, \tau) = \sum_{k=1}^m \varphi_k(\lambda_k, \tau_k)$ where $\varphi_k(\lambda_k, \tau_k) = \log(\one_n^\top B_k(\lambda_k, \tau_k)\one_n) - \lambda_k^\top u^k$ for all $k \in [m]$. Moreover, each $\varphi_k$ has Lipschitz continuous gradient in $\ell_2$-norm and the Lipschitz constant is upper bounded by $4$. Formally, 
\begin{equation*}
\|\nabla\varphi_k(\lambda, \tau) - \nabla\varphi_k(\lambda', \tau')\| \leq 4\left\|\begin{pmatrix} \lambda \\ \tau \end{pmatrix} - \begin{pmatrix} \lambda' \\ \tau' \end{pmatrix}\right\| \quad \textnormal{for all } k \in [m]. 
\end{equation*}
which is equivalent to
\begin{equation*}
\varphi(\lambda', \tau') - \varphi(\lambda, \tau) \leq \begin{pmatrix} \lambda' - \lambda \\ \tau' - \tau \end{pmatrix}^\top \nabla\varphi(\lambda, \tau) + 2\left(\sum_{k=1}^m \omega_k\left\|\begin{pmatrix} \lambda'_k - \lambda_k \\ \tau'_k - \tau_k \end{pmatrix} \right\|^2\right). 
\end{equation*}
\end{lemma}
\begin{proof}
The first statement directly follows from the definition of $\varphi$ in Eq.~\eqref{prob:barycenter_regularized_dual}. For the second statement, we provide the explicit form of the gradient of $\varphi_k$ as follows, 
\begin{equation*}
\nabla\varphi_k(\lambda, \tau) = \begin{pmatrix} \frac{B_k(\lambda, \tau)\one_n}{\one_n^\top B_k(\lambda, \tau)\one_n} - u^k \\ \frac{B_k(\lambda, \tau)^\top\one_n}{\one_n^\top B_k(\lambda, \tau)\one_n} \end{pmatrix}. 
\end{equation*}
Now we construct the following entropic regularized OT problem, 
\begin{equation*}
\min_{X:\|X\|_1=1} \langle C_k, X\rangle - \eta H(X), \quad \st X\one_n = (1/n)\one_n, \ X^\top\one_n = (1/n)\one_n. 
\end{equation*}
Since the function $-H(X)$ is strongly convex with respect to the $\ell_1$-norm on the probability simplex $Q \subseteq \br^{n^2}$, the above entropic regularized OT problem is a special case of the following linearly constrained convex optimization problem:
\begin{equation*}
\min_{x \in Q} \ f(x), \quad \st \ Ax = b, 
\end{equation*}
where $f$ is strongly convex with respect to the $\ell_1$-norm on the set $Q$. We use the $\ell_2$-norm for the dual space of the Lagrange multipliers. By~\citet[Theorem~1]{Nesterov-2005-Smooth} and the fact that $\|A\|_{1 \rightarrow 2} = 2$, the dual objective function $\tilde{\varphi}_k$ satisfies the following inequality:  
\begin{equation*}
\|\nabla\tilde{\varphi}_k(\alpha, \beta) - \nabla\tilde{\varphi}_k(\alpha', \beta')\| \leq \frac{4}{\eta}\left\|\begin{pmatrix} \alpha \\ \beta \end{pmatrix} - \begin{pmatrix} \alpha' \\ \beta' \end{pmatrix}\right\|. 
\end{equation*}
Recall that the function $\tilde{\varphi}_k$ is given by 
\begin{equation*}
\tilde{\varphi}_k(\alpha, \beta) = \eta\log\left(\sum_{i,j=1}^n e^{- \frac{(C_k)_{ij} - \alpha_i - \beta_j}{\eta} - 1}\right) - \langle \alpha, u\rangle - \langle\beta, v\rangle + \eta. 
\end{equation*}
This together with the definition of $B_k(\cdot, \cdot)$ implies that
\begin{equation*}
\nabla\tilde{\varphi}_k(\alpha, \beta) = \begin{pmatrix} \frac{B_k\left(\eta^{-1}\alpha - (1/2)\one_n, \eta^{-1}\beta - (1/2)\one_n\right)\one_n}{\one_n^\top B_k\left(\eta^{-1}\alpha - (1/2)\one_n, \eta^{-1}\beta - (1/2)\one_n\right)\one_n} - u \\ \frac{B_k\left(\eta^{-1}\alpha - (1/2)\one_n, \eta^{-1}\beta - (1/2)\one_n\right)^\top\one_n}{\one_n^\top B_k\left(\eta^{-1}\alpha - (1/2)\one_n, \eta^{-1}\beta - (1/2)\one_n\right)\one_n} - v \end{pmatrix}
\end{equation*}
Performing the change of variable $\lambda = \eta^{-1}\alpha - (1/2)\one_n$ and $\tau = \eta^{-1}\beta - (1/2)\one_n$ (resp. $\lambda'$ and $\tau'$), we have 
\begin{eqnarray*}
& & \|\nabla\varphi_k(\lambda, \tau) - \nabla\varphi_k(\lambda', \tau')\| \\ 
& = & \|\nabla\tilde{\varphi}_k(\eta(\lambda + (1/2)\one_n), \eta(\tau + (1/2)\one_n)) - \nabla\tilde{\varphi}_k(\eta(\lambda' + (1/2)\one_n), \eta(\tau' + (1/2)\one_n))\| \\
& \leq & 4\left\|\begin{pmatrix} \lambda' - \lambda \\ \tau' - \tau\end{pmatrix}\right\|. 
\end{eqnarray*}
This completes the proof. 
\end{proof}
\begin{remark}
It is worthy noting that Lemma~\ref{Lemma:liptshitz-continuity} exploits the decomposable structure of the dual function $\varphi$, and gives the a \textbf{weighted} smoothness inequality. This inequality is necessary for deriving the complexity bound which depends linearly on the number of probability measures.  
\end{remark}

\section{Computational Hardness}\label{sec:hardness}
In this section, we analyze the computational hardness of the FS-WBP in Eq.~\eqref{prob:barycenter}. After introducing some characterization theorems in combinatorial optimization, we show that the FS-WBP in Eq.~\eqref{prob:barycenter} is a minimum-cost flow (MCF) problem when $m = 2$ and $n \geq 3$ but not when $m \geq 3$ and $n \geq 3$.

\subsection{Combinatorial techniques}
We present some classical results in combinatorial optimization and graph theory, including Ghouila-Houri's celebrated characterization theorem~\citep{Ghouila-1962-Caracterisation}.
\begin{definition}\label{Def:TUM}
A totally unimodular (TU) matrix is one for which every square submatrix has determinant $-1$, $0$ or $1$.
\end{definition}
\begin{proposition}[Ghouila-Houri]\label{Prop:GHC}
A $\{-1, 0, 1\}$-valued matrix $A \in \br^{m \times n}$ is TU if and only if for each $I \subseteq [m]$ there is a partition $I = I_1 \cup I_2$ so that $\sum_{i \in I_1} a_{ij} - \sum_{i \in I_2} a_{ij} \in \{-1, 0, 1\}$ for $j \in [n]$.
\end{proposition}
The second result~\citep[Theorem~1, Chapter 15]{Berge-2001-Theory} shows that the incidence matrices of directed graphs and 2-colorable undirected graphs are TU.
\begin{proposition}\label{Prop:TUM}
Let $A$ be a $\{-1, 0, 1\}$-valued matrix. Then $A$ is TU if each column contains at most two nonzero entries and all rows are partitioned into two sets $I_1$ and $I_2$ such that: If two nonzero entries of a column have the same sign, they are in different sets. If these two entries have different signs, they are in the same set.
\end{proposition}
Finally, we characterize the constraint matrix arising in a MCF problem. 
\begin{definition}\label{Def:MCF}
The MCF problem finds the cheapest possible way of sending a certain amount of flow through a flow network. Formally, 
\begin{equation*}
\begin{array}{ll}
\min & \sum_{(u,v) \in E} f(u,v) \cdot a(u,v) \\
\st & f(u,v) \geq 0, \; \textnormal{ for all} \; (u,v) \in E, \\
& f(u,v) \leq c(u,v) \; \textnormal{ for all} \; (u,v) \in E, \\
& f(u,v) = - f(v,u) \;  \textnormal{ for all} \;   (u,v) \in E, \\
& \sum_{(u,w)\in E \;  \textnormal{or} \; (w, u) \in E } f(u,w) = 0, \\
& \sum_{w \in V} f(s,w) = d \;  \textnormal{ and } \;  \sum_{w \in V} f(w,t) = d.
\end{array}
\end{equation*}
The flow network $G = (V, E)$ is a directed graph $G = (V, E)$ with a source vertex $s \in V$ and a sink vertex $t \in V$, where each edge $(u, v) \in E$ has capacity $c(u,v) > 0$, flow $f(u,v) \geq 0$ and cost $a(u, v)$, with most MCF algorithms supporting edges with negative costs. The cost of sending this flow along an edge $(u, v)$ is $f(u,v) \cdot a(u,v)$. The problem requires an amount of flow $d$ to be sent from source $s$ to sink $t$. The definition of the problem is to minimize the total cost of the flow over all edges. 
\end{definition} 
\begin{proposition}\label{Prop:MCF}
The constraint matrix arising in a MCF problem is TU and its rows are categorized into a single set using Proposition~\ref{Prop:TUM}. 
\end{proposition}
\begin{proof}
The standard LP representation of the MCF problem is 
\begin{equation*}
\min_{x \in \br^{|E|}} \ c^\top x, \quad \st \ Ax = b, \ l \leq x \leq u. 
\end{equation*}
where $x \in \br^{|E|}$ with $x_j$ being the flow through arc $j$, $b \in \br^{|V|}$ with $b_i$ being external supply at node $i$ and $\one^\top b = 0$, $c_j$ is unit cost of flow through arc $j$, $l_j$ and $u_j$ are lower and upper bounds on flow through arc $j$ and $A \in \br^{|V| \times |E|}$ is the arc-node incidence matrix with entries 
\begin{equation*}
A_{ij} = \left\{\begin{array}{rl} -1 & \text{if arc $j$ starts at node $i$} \\ 1 & \text{if arc $j$ ends at node $i$} \\ 0 & \text{otherwise} \end{array}\right..
\end{equation*}
Since each arc has two endpoints, the constraint matrix $A$ is a $\{-1, 0, 1\}$-valued matrix in which each column contains two nonzero entries $1$ and $-1$. Using Proposition~\ref{Prop:TUM}, we obtain that $A$ is TU and the rows of $A$ are categorized into a single set. 
\end{proof}
\begin{figure}[!t]\small
\centering
\begin{tikzpicture}
\begin{scope}[every node/.style={circle, thick, draw, minimum size=1cm}]
\node (s1) at (0,3) {$u_{11}$};
\node (s2) at (0,1) {$u_{12}$};
\node (s3) at (0,-1) {$u_{13}$};
\node (s4) at (0,-3) {$u_{14}$};
\node (c1) at (3,3) {$0$};
\node (c2) at (3,1) {$0$};
\node (c3) at (3,-1) {$0$};
\node (c4) at (3,-3) {$0$};
\node (t1) at (6,3) {$u_{21}$};
\node (t2) at (6,1) {$u_{22}$};
\node (t3) at (6,-1) {$u_{23}$};
\node (t4) at (6,-3) {$u_{24}$};
\end{scope}
\begin{scope}[>={Stealth[black]},
every node/.style={fill=white,circle},
every edge/.style={draw=red,very thick}]
\path [->] (s1) edge (c1);
\path [->] (s1) edge (c2);
\path [->] (s1) edge (c3);
\path [->] (s1) edge (c4);
\path [->] (s2) edge (c1);
\path [->] (s2) edge (c2);
\path [->] (s2) edge (c3);
\path [->] (s2) edge (c4);
\path [->] (s3) edge (c1);
\path [->] (s3) edge (c2);
\path [->] (s3) edge (c3);
\path [->] (s3) edge (c4);
\path [->] (s4) edge (c1);
\path [->] (s4) edge (c2);
\path [->] (s4) edge (c3);
\path [->] (s4) edge (c4);
\end{scope}
\begin{scope}[>={Stealth[black]},
every node/.style={fill=white,circle},
every edge/.style={draw=blue,very thick}]
\path [->] (c1) edge (t1);
\path [->] (c1) edge (t2);
\path [->] (c1) edge (t3);
\path [->] (c1) edge (t4);
\path [->] (c2) edge (t1);
\path [->] (c2) edge (t2);
\path [->] (c2) edge (t3);
\path [->] (c2) edge (t4);
\path [->] (c3) edge (t1);
\path [->] (c3) edge (t2);
\path [->] (c3) edge (t3);
\path [->] (c3) edge (t4);
\path [->] (c4) edge (t1);
\path [->] (c4) edge (t2);
\path [->] (c4) edge (t3);
\path [->] (c4) edge (t4);
\end{scope}
\end{tikzpicture}
\caption{Represent the FS-WBP in Eq.~\eqref{prob:barycenter} as a MCF problem when $(m, n) = (2, 4)$.}\label{fig:example}\vspace{-1em}
\end{figure}
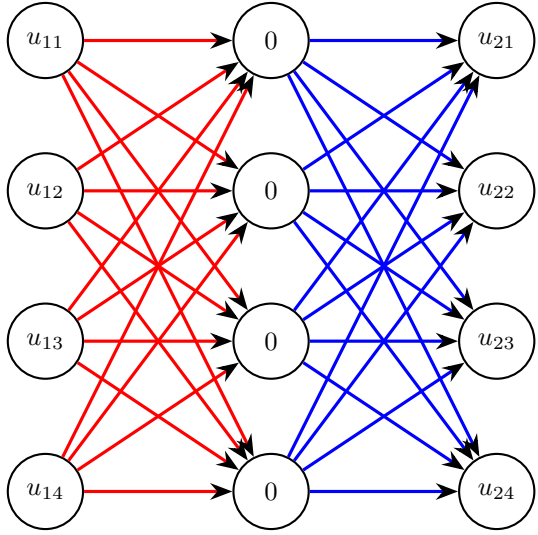

\subsection{Main result}
We present our main results on the computational hardness of the FS-WBP in Eq.~\eqref{prob:barycenter}. First, the FS-WBP in this LP form is an MCF problem when $m = 2$ and $n \geq 2$; see Figure~\ref{fig:example} for the graph when $(m, n) = (2, 4)$. Indeed, it is a transportation problem with $n$ warehouses, $n$ transshipment centers and $n$ retail outlets. Each $u_{1i}$ is the amount of supply provided by the $i$th warehouse and each $u_{2j}$ is the amount of demand requested by the $j$th retail outlet. $(X_1)_{ij}$ is the flow sent from $i$th warehouse to the $j$th transshipment center and $(X_2)_{ij}$ is the flow sent from the $i$th transshipment center to the $j$th retail outlet. $(C_1)_{ij}$ and $(C_2)_{ij}$ refer to the unit cost of the corresponding flow. See~\citep[Page 400]{Anderes-2016-Discrete}. 

Proceed to the setting $m \geq 3$, it is unclear whether the graph representation of the FS-WBP carries over. Instead, we turn to algebraic techniques and provide an explicit form as follows:
\begin{equation}\label{prob:barycenter_explicit}\small
\min \ \sum_{k=1}^m \langle C_k, X_k\rangle, \quad \st \begin{pmatrix}
-E & \cdots & \cdots & \cdots & \cdots \\
\vdots & E & \ddots & \ddots & \vdots \\
\vdots & \ddots & \ddots & \ddots & \vdots \\
\vdots & \ddots & \ddots & (-1)^{m-1} E & \vdots \\
\vdots & \ddots & \ddots & \ddots & (-1)^m E \\ \hdashline
G & -G & \ddots & \ddots & \vdots \\
\vdots & -G & G & \ddots & \vdots \\
\vdots & \ddots & \ddots & \ddots & \vdots \\
\cdots & \cdots & \cdots & (-1)^m G & (-1)^{m+1} G \\
\end{pmatrix}\begin{pmatrix}
\vct(X_1) \\ \vct(X_2) \\ \vdots \\ \vdots \\ \vdots \\ \vdots \\ \vct(X_m) 
\end{pmatrix} = \begin{pmatrix}
-u^1 \\ u^2 \\ \vdots \\ (-1)^{m-1} u^{m-1} \\ (-1)^m u^m \\ \zero_n \\ \vdots \\ \zero_n
\end{pmatrix},
\end{equation}
where $E = I_n \otimes \one_n^\top \in \br^{n \times n^2}$ and $G = \one_n^\top \otimes I_n \in \br^{n \times n^2}$. Each column of the constraint matrix arising in Eq.~\eqref{prob:barycenter_explicit} has either 2 or 3 nonzero entries in $\{-1, 0, 1\}$. In the following theorem, we study the structure of the constraint matrix when $m \geq 3$ and $n=2$.
\begin{theorem}\label{theorem:hardness-first}
The constraint matrix arising in Eq.~\eqref{prob:barycenter_explicit} is TU when $m \geq 3$ and $n=2$. 
\end{theorem}
\begin{proof}
When $n = 2$, the constraint matrix $A$ has $E = I_2 \otimes \one_2^\top$ and $G = \one_2^\top \otimes I_2$. The matrix $A \in \br^{(4m-2) \times 4m}$ is a $\{-1, 0, 1\}$-valued matrix with several redundant rows and each column has at most three nonzero entries in $\{-1, 0, 1\}$. Now we simplify the matrix $A$ by removing a specific set of redundant rows. In particular, we observe that 
\begin{equation*}
\sum_{i \in \{1, 2, 3, 4, 2m+1, 2m+2\}} a_{ij} = 0, \quad \forall j \in [4m], 
\end{equation*}
which implies that the $(2m+2)$th row is redundant. Similarly, we have
\begin{equation*}
\sum_{i \in \{3, 4, 5, 6, 2m+3, 2m+4\}} a_{ij} = 0, \quad \forall j \in [4m], 
\end{equation*}
which implies that the $(2m+3)$th row is redundant. Using this argument, we remove $m-1$ rows from the last $2m-2$ rows. The resulting matrix $\bar{A} \in \br^{(3m-1) \times 4m}$ has very nice structure such that each column has only two nonzero entries $1$ and $-1$; see the following matrix when $m$ is odd: 
\begin{equation*}
\bar{A} = \begin{pmatrix}
-E & \cdots & \cdots & \cdots & \cdots \\
\vdots & E & \ddots & \ddots & \vdots \\
\vdots & \ddots & \ddots & \ddots & \vdots \\
\vdots & \ddots & \ddots & (-1)^{m-1} E & \vdots \\
\vdots & \ddots & \ddots & \ddots & (-1)^m E \\ \hdashline
\one_2^\top \otimes e_1 & - \one_2^\top \otimes e_1 & \ddots & \ddots & \vdots \\
\vdots & - \one_2^\top \otimes e_2 &  \one_2^\top \otimes e_2 & \ddots & \vdots \\
\vdots & \ddots & \ddots & \ddots & \vdots \\
\cdots & \cdots & \cdots & (-1)^m \one_2^\top \otimes e_2 & (-1)^{m+1} \one_2^\top \otimes e_2 \\
\end{pmatrix}. 
\end{equation*}
where $e_1$ and $e_2$ are respectively the first and second standard basis (row) vectors in $\br^2$. Furthermore, the rows of $\bar{A}$ are categorized into a single set so that the criterion in Proposition~\ref{Prop:TUM} holds true (the dashed line in the formulation of $\bar{A}$ serves as a partition of this single set into two sets). Using Proposition~\ref{Prop:TUM}, we conclude that $\bar{A}$ is TU.
\end{proof}
To facilitate the reader, we provide an illustrative counterexample for showing that the FS-WBP in Eq.~\eqref{prob:barycenter_explicit} is not an MCF problem when $m = 3$ and $n = 3$.
\begin{example}\label{example:hardness}
When $m = 3$ and $n = 3$, the constraint matrix is 
\begin{equation*}
A = \begin{pmatrix}
-I_3 \otimes \one_3^\top & \zero_{3 \times 9} & \zero_{3 \times 9} \\
\zero_{3 \times 9} & I_3 \otimes \one_3^\top & \zero_{3 \times 9} \\
\zero_{3 \times 9} & \zero_{3 \times 9} & -I_3 \otimes \one_3^\top \\ \hdashline
\one_3^\top \otimes I_3 & -\one_3^\top \otimes I_3 & \zero_{3 \times 9} \\
\zero_{3 \times 9} & -\one_3^\top \otimes I_3 & \one_3^\top \otimes I_3 \\
\end{pmatrix}.
\end{equation*}
Setting the set $I = \{1, 4, 7, 10, 11, 13, 15\}$ and letting $e_1$, $e_2$ and $e_3$ be the first, second and third standard basis row vectors in $\br^n$, the resulting matrix with the rows in $I$ is 
\begin{equation*}
R = \begin{pmatrix}
-e_1 \otimes \one_3^\top & \zero_{1 \times 9} & \zero_{1 \times 9} \\
\zero_{1 \times 9} & e_1 \otimes \one_3^\top & \zero_{1 \times 9} \\
\zero_{1 \times 9} & \zero_{1 \times 9} & -e_1 \otimes \one_3^\top \\ \hdashline
\one_3^\top \otimes e_1 & -\one_3^\top \otimes e_1 & \zero_{1 \times 9} \\
\one_3^\top \otimes e_2  & -\one_3^\top \otimes e_2 & \zero_{1 \times 9} \\
\zero_{1 \times 9} & -\one_3^\top \otimes e_1 & \one_3^\top \otimes e_1 \\
\zero_{1 \times 9} & -\one_3^\top \otimes e_3 & \one_3^\top \otimes e_3 \\
\end{pmatrix}.
\end{equation*}
Instead of considering all columns of $R$, it suffices to show that no partition of $I$ guarantees for any $j \in \{1, 2, 11, 12, 13, 19, 21\}$ that 
\begin{equation*}
\sum_{i \in I_1} R_{ij} - \sum_{i \in I_2} R_{ij} \in \{-1, 0, 1\}. 
\end{equation*}
We write the submatrix of $R$ with these columns as 
\begin{equation*}
\bar{R} = \begin{pmatrix}
-1 & -1 & 0 & 0 & 0 & 0 & 0 \\ 
0 & 0 & 1 & 1 & 0 & 0 & 0 \\
0 & 0 & 0 & 0 & 0 & -1 & -1 \\
1 & 0 & 0 & 0 & -1 & 0 & 0 \\
0 & 1 & -1 & 0 & 0 & 0 & 0 \\
0 & 0 & 0 & 0 & -1 & 1 & 0 \\
0 & 0 & 0 & -1 & 0 & 0 & 1 \\
\end{pmatrix}
\end{equation*}
First, we claim that rows $1$, $2$, $4$, $5$ and $7$ are in the same set $I_1$. Indeed, columns $1$ and $2$ imply that rows $1$, $4$ and $5$ are in the same set. Column $3$ and $4$ imply that rows $2$, $5$ and $7$ are in the same set. Putting these pieces together yields the desired claim. Then we consider the set that the row $6$ belongs to and claim a contradiction. Indeed, row $6$ can not be in $I_1$ since column $5$ implies that rows $4$ and $6$ are not in the same set. However, row $6$ must be in $I_1$ since columns $6$ and $7$ imply that rows $3$, $6$ and $7$ are in the same set. Putting these pieces together yields the desired contradiction. Finally, by using Propositions~\ref{Prop:GHC} and~\ref{Prop:MCF}, we conclude that $A$ is not TU and problem~\eqref{prob:barycenter_explicit} is not a MCF problem when $m = 3$ and $n = 3$. 
\end{example}
Finally, we prove that the FS-WBP in Eq.~\eqref{prob:barycenter_explicit} is not a MCF when $m \geq 3$ and $n \geq 3$. The basic idea is to extend the construction in Example~\ref{example:hardness} to the general case.  
\begin{theorem}\label{theorem:hardness-second}
The FS-WBP in Eq.~\eqref{prob:barycenter_explicit} is not a MCF problem when $m \geq 3$ and $n \geq 3$. 
\end{theorem}
\begin{proof}
We use the proof by contradiction. In particular, assume that problem~\eqref{prob:barycenter_explicit} is a MCF problem when $m \geq 3$ and $n \geq 3$, Proposition~\ref{Prop:MCF} implies that the constraint matrix $A$ is TU. Since $A$ is a $\{-1, 0, 1\}$-valued matrix, Proposition~\ref{Prop:GHC} further implies that for each set $I \subseteq [2mn-n]$ there is a partition $I_1$, $I_2$ of $I$ such that 
\begin{equation}\label{eq:GU-criterion}
\sum_{i \in I_1} a_{ij} - \sum_{i \in I_2} a_{ij} \in \{-1, 0, 1\}, \quad \forall j \in [mn^2]. 
\end{equation}
In what follows, for any given $m \geq 3$ and $n \geq 3$, we construct a set of rows $I$ such that no partition of $I$ guarantees that Eq.~\eqref{eq:GU-criterion} holds true. For the ease of presentation, we rewrite the matrix $A \in \br^{(2mn-n) \times mn^2}$ as follows, 
\begin{equation*}
A = \begin{pmatrix}
-I_n \otimes \one_n^\top & \cdots & \cdots & \cdots & \cdots \\
\vdots & I_n \otimes \one_n^\top & \ddots & \ddots & \vdots \\
\vdots & \ddots & \ddots & \ddots & \vdots \\
\vdots & \ddots & \ddots & (-1)^{m-1}I_n \otimes \one_n^\top & \vdots \\
\vdots & \ddots & \ddots & \ddots & (-1)^m I_n \otimes \one_n^\top \\ \hdashline
\one_n^\top \otimes I_n & -\one_n^\top \otimes I_n & \ddots & \ddots & \vdots \\
\vdots & -\one_n^\top \otimes I_n & \one_n^\top \otimes I_n & \ddots & \vdots \\
\vdots & \ddots & \ddots & \ddots & \vdots \\
\cdots & \cdots & \cdots & (-1)^m \one_n^\top \otimes I_n & (-1)^{m+1}\one_n^\top \otimes I_n \\
\end{pmatrix}. 
\end{equation*}
Setting the set $I = \{1, n+1, 2n+1, 3n+1, 3n+2, 4n+1, 4n+3\}$ and letting $e_1$, $e_2$ and $e_3$ be the first, second and third standard basis row vectors in $\br^n$, the resulting matrix with the rows in $I$ is 
\begin{equation*}
R = \begin{pmatrix}
-e_1 \otimes \one_n^\top & \zero_{1 \times n^2} & \zero_{1 \times n^2} & \zero_{1 \times n^2} & \cdots & \zero_{1 \times n^2} \\
\zero_{1 \times n^2} & e_1 \otimes \one_n^\top & \zero_{1 \times n^2} & \zero_{1 \times n^2} & \cdots & \zero_{1 \times n^2} \\
\zero_{1 \times n^2} & \zero_{1 \times n^2} & -e_1 \otimes \one_n^\top & \zero_{1 \times n^2} & \cdots & \zero_{1 \times n^2} \\ \hdashline
\one_n^\top \otimes e_1 & -\one_n^\top \otimes e_1 & \zero_{1 \times n^2} & \zero_{1 \times n^2} & \cdots & \zero_{1 \times n^2} \\
\one_n^\top \otimes e_2 & -\one_n^\top \otimes e_2 & \zero_{1 \times n^2} & \zero_{1 \times n^2} & \cdots & \zero_{1 \times n^2} \\
\zero_{1 \times n^2} & -\one_n^\top \otimes e_1 & \one_n^\top \otimes e_1 & \zero_{1 \times n^2} & \cdots & \zero_{1 \times n^2} \\
\zero_{1 \times n^2} & -\one_n^\top \otimes e_3 & \one_n^\top \otimes e_3 & \zero_{1 \times n^2} & \cdots & \zero_{1 \times n^2} \\
\end{pmatrix}.
\end{equation*}
Instead of considering all columns of $R$, it suffices to show that no partition of $I$ guarantees
\begin{equation*}
\sum_{i \in I_1} R_{ij} - \sum_{i \in I_2} R_{ij} \in \{-1, 0, 1\},
\end{equation*}
for all $j \in \{1, 2, n^2+2, n^2+3, n^2+n+1, 2n^2+1, 2n^2+3\}$. We write the submatrix of $R$ with these columns as 
\begin{equation*}
\bar{R} = \begin{pmatrix}
-1 & -1 & 0 & 0 & 0 & 0 & 0 \\ 
0 & 0 & 1 & 1 & 0 & 0 & 0 \\
0 & 0 & 0 & 0 & 0 & -1 & -1 \\
1 & 0 & 0 & 0 & -1 & 0 & 0 \\
0 & 1 & -1 & 0 & 0 & 0 & 0 \\
0 & 0 & 0 & 0 & -1 & 1 & 0 \\
0 & 0 & 0 & -1 & 0 & 0 & 1 \\
\end{pmatrix}. 
\end{equation*}
Applying the same argument used in Example~\ref{example:hardness}, we obtain from Propositions~\ref{Prop:GHC} and~\ref{Prop:MCF} that $A$ is not TU when $m \geq 3$ and $n \geq 3$, which is a contradiction. As a consequence, the conclusion of the theorem follows.
\end{proof}
\begin{remark}
Theorem~\ref{theorem:hardness-second} resolves an open question and partially explains why the direct application of network flow algorithms to the FS-WBP in Eq.~\eqref{prob:barycenter_explicit} is inefficient. However, this does \textit{not} eliminate the possibility that the FS-WBP is equivalent to some other LP with good complexity. For example,~\citet{Ge-2019-Interior} have recently successfully identified an equivalent LP formulation of the FS-WBP which is suitable for the interior-point algorithm. Furthermore, our results support the problem reformulation of the FS-WBP which forms the basis for various algorithms; e.g.,~\citet{Benamou-2015-Iterative, Cuturi-2016-Smoothed, Kroshnin-2019-Complexity, Ge-2019-Interior, Guminov-2019-Accelerated}.  
\end{remark}

\section{Fast Iterative Bregman Projection}\label{sec:fast-IBP}
In this section, we present a fast \textit{deterministic} variant of the iterative Bregman projection (IBP) algorithm, named \textsc{FastIBP} algorithm, and prove that it achieves the complexity bound of $\bigOtil(mn^{7/3}\varepsilon^{-4/3})$. This improves over $\bigOtil(mn^2\varepsilon^{-2})$ from iterative Bregman projection algorithm~\citep{Benamou-2015-Iterative} in terms of $\varepsilon$ and $\bigOtil(mn^{5/2}\varepsilon^{-1})$ from the APDAGD and accelerated Sinkhorn algorithms~\citep{Kroshnin-2019-Complexity, Guminov-2019-Accelerated} in terms of $n$.
\begin{algorithm}[!t]\small
\caption{$\textsc{FastIBP}(\{C_k, u^k\}_{k \in [m]}, \varepsilon)$} \label{Algorithm:FastIBP}
\begin{algorithmic}
\STATE \textbf{Initialization:} $t = 0$, $\theta_0 = 1$ and $\check{\lambda}^0 = \tilde{\lambda}^0 = \check{\tau}^0 = \tilde{\tau}^0 = \zero_{mn}$.  
\WHILE{$E_t > \varepsilon$}
\STATE \textbf{Step 1:} Compute $\begin{pmatrix} \bar{\lambda}^t \\ \bar{\tau}^t \end{pmatrix} = (1 - \theta_t)\begin{pmatrix} \check{\lambda}^t \\ \check{\tau}^t \end{pmatrix} + \theta_t\begin{pmatrix} \tilde{\lambda}^t \\ \tilde{\tau}^t \end{pmatrix}$. 
\STATE \textbf{Step 2:} Compute $r_k = r(B_k(\bar{\lambda}_k^t, \bar{\tau}_k^t))$ and $c_k = c(B_k(\bar{\lambda}_k^t, \bar{\tau}_k^t))$ for all $k \in [m]$ and perform 
\begin{eqnarray*}
\tilde{\lambda}_k^{t+1} & = & \tilde{\lambda}_k^t - \frac{1}{4\theta_t}\left(\frac{r_k}{\one_n^\top r_k} - u^k\right), \quad \textnormal{for all } k \in [m], \\
\tilde{\tau}^{t+1} & = & \argmin\limits_{\sum_{k=1}^m \omega_k\tau_k = \zero_n} \sum_{k=1}^m \omega_k\left[(\tau_k - \bar{\tau}_k^t)^\top\frac{c_k}{\one_n^\top c_k} + 2\theta_t\|\tau_k - \tilde{\tau}_k^t\|^2\right]. 
\end{eqnarray*}
\STATE \textbf{Step 3:} Compute $\begin{pmatrix} \widehat{\lambda}^t \\ \widehat{\tau}^t \end{pmatrix} = \begin{pmatrix} \bar{\lambda}^t \\ \bar{\tau}^t \end{pmatrix} + \theta_t\begin{pmatrix} \tilde{\lambda}^{t+1} \\ \tilde{\tau}^{t+1} \end{pmatrix} - \theta_t\begin{pmatrix} \tilde{\lambda}^t \\ \tilde{\tau}^t \end{pmatrix}$. 
\STATE \textbf{Step 4:} Compute $\begin{pmatrix} \acute{\lambda}^t \\ \acute{\tau}^t \end{pmatrix} = \argmin\left\{\varphi(\lambda, \tau) \mid \begin{pmatrix} \lambda \\ \tau \end{pmatrix} \in \left\{\begin{pmatrix} \check{\lambda}^t \\ \check{\tau}^t \end{pmatrix}, \begin{pmatrix} \widehat{\lambda}^t \\ \widehat{\tau}^t \end{pmatrix} \right\}\right\}$.
\STATE \textbf{Step 5a:} Compute $c_k = c(B_k(\acute{\lambda}_k^t, \acute{\tau}_k^t))$ for all $k \in [m]$.  
\STATE \textbf{Step 5b:} Compute $\grave{\tau}_k^t = \acute{\tau}_k^t + \sum_{k=1}^m \omega_k \log(c_k) - \log(c_k)$ for all $k \in [m]$ and $\grave{\lambda}^{t+1} = \acute{\lambda}^t$.  
\STATE \textbf{Step 6a:} Compute $r_k = r(B_k(\grave{\lambda}_k^t, \grave{\tau}_k^t))$ for all $k \in [m]$. 
\STATE \textbf{Step 6b:} Compute $\lambda_k^t = \grave{\lambda}_k^t + \log(u^k) - \log(r_k)$ for all $k \in [m]$ and $\tau^t = \grave{\tau}^t$.
\STATE \textbf{Step 7a:} Compute $c_k = c(B_k(\lambda_k^t, \tau_k^t))$ for all $k \in [m]$.  
\STATE \textbf{Step 7b:} Compute $\check{\tau}_k^{t+1} = \tau_k^t + \sum_{k=1}^m \omega_k \log(c_k) - \log(c_k)$ for all $k \in [m]$ and $\check{\lambda}^{t+1} = \lambda^t$. 
\STATE \textbf{Step 8:} Compute $\theta_{t+1} = \theta_t(\sqrt{\theta_t^2 + 4} - \theta_t)/2$. 
\STATE \textbf{Step 9:} Increment by $t = t + 1$. 
\ENDWHILE
\STATE \textbf{Output:} $(B_1(\lambda_1^t, \tau_1^t), B_2(\lambda_2^t, \tau_2^t), \ldots, B_m(\lambda_m^t, \tau_m^t))$.  
\end{algorithmic}
\end{algorithm}
\subsection{Algorithmic scheme}
To facilitate the later discussion, we present the \textsc{FastIBP} algorithm in pseudocode form in Algorithm~\ref{Algorithm:FastIBP} and its application to entropic regularized FS-WBP in Algorithm~\ref{Algorithm:WBP_FastIBP}. Note that $(B_1(\lambda_1^t, \tau_1^t), \ldots, B_m(\lambda_m^t, \tau_m^t))$ stand for the primal variables while $(\lambda^t, \tau^t)$ are the dual variables for the entropic regularized FS-WBP.   

The \textsc{FastIBP} algorithm is a \textit{deterministic} variant of the iterative Bregman projection (IBP) algorithm~\citep{Benamou-2015-Iterative}. While the IBP algorithm can be interpreted as a dual coordinate descent, the acceleration achieved by the \textsc{FastIBP} algorithm mostly depends on the refined characterization of per-iteration progress using the scheme with momentum; see \textbf{Step 1-3} and \textbf{Step 8}. To the best of our knowledge, this scheme has been well studied by~\citep{Nesterov-2012-Efficiency, Nesterov-2013-Gradient, Fercoq-2015-Accelerated, Nesterov-2017-Efficiency} yet \textit{first} introduced to accelerate the optimal transport algorithms. 

Furthermore, \textbf{Step 4} guarantees that $\{\varphi(\check{\lambda}^t, \check{\tau}^t)\}_{t \geq 0}$ is monotonically decreasing and \textbf{Step 7} ensures the sufficient large progress from $(\lambda_k^t, \tau_k^t)$ to $(\check{\lambda}^{t+1}, \check{\tau}^{t+1})$. \textbf{Step 5} are performed such that $\tau_k^t = \grave{\tau}_k^t$ satisfies the bounded difference property: $\max_{1 \leq i \leq n} (\tau_k^t)_i - \min_{1 \leq i \leq n} (\tau_k^t)_i \leq R_\tau/2$ while \textbf{Step 6} guarantees that the marginal conditions hold true: $r(B_k(\lambda_k^t, \tau_k^t)) = u^k$ for all $k \in [m]$. We see from~\citet[Lemma~9]{Guminov-2019-Accelerated} that \textbf{Step 5-7} refer to the alternating minimization steps for the dual objective function $\varphi$ with respect to two-block variable $(\lambda, \tau)$. More specifically, these steps can be rewritten as follows, 
\begin{equation*}
\begin{array}{rll}
(\textbf{Step 5}) & \grave{\lambda}^t = \acute{\lambda}^t, & \grave{\tau}^t = \argmin \varphi(\acute{\lambda}^t, \tau), \quad \st \ \sum_{k=1}^m \tau_k = \zero_n, \\
(\textbf{Step 6}) & \lambda^t = \argmin \varphi(\lambda, \grave{\tau}^t), & \tau^t = \grave{\tau}^t, \\
(\textbf{Step 7}) & \check{\lambda}^{t+1} = \lambda^t, & \check{\tau}^{t+1} = \argmin \varphi(\lambda^t, \tau), \quad \st \ \sum_{k=1}^m \tau_k = \zero_n.
\end{array}
\end{equation*}  
We also remark that \textbf{Step 4-7} are \textit{specialized} to the FS-WBP in Eq.~\eqref{prob:barycenter} and have \textit{not} been appeared in the coordinate descent literature before. 

Finally, the optimality conditions of primal entropic regularized WBP in Eq.~\eqref{prob:barycenter_regularized} and dual entropic regularized WBP in Eq.~\eqref{prob:barycenter_regularized_dual} are
\begin{equation*}
\begin{array}{rcll}
\zero_n & = & \frac{r(B_k(\lambda_k, \tau_k))}{\|B_k(\lambda_k, \tau_k)\|_1} - u^k, & \textnormal{for all } k \in [m], \\
\zero_n & = & \frac{c(B_k(\lambda_k, \tau_k))}{\|B_k(\lambda_k, \tau_k)\|_1} - \sum_{i=1}^m \omega_i \frac{c(B_i(\lambda_i, \tau_i))}{\|B_i(\lambda_i, \tau_i)\|_1}, & \textnormal{for all } k \in [m], \\
\zero_n & = & \sum_{k=1}^m \omega_k\tau_k.  
\end{array}
\end{equation*}
Since the \textsc{FastIBP} algorithm guarantees that $\sum_{k=1}^m \omega_k\tau_k^t = \zero_n$ and $r(B_k(\lambda_k^t, \tau_k^t)) = u^k \in \Delta^n$ for all $k \in [m]$, we can solve simultaneously primal and dual entropic regularized FS-WBP with an adaptive stopping criterion which does not require to calculate any objective value. The criterion depends on the following quantity to measure the residue at each iteration:
\begin{equation}\label{Def:residue-FastIBP}
E_t \mydefn \sum_{k = 1}^m \omega_k\|c(B_k(\lambda_k^t, \tau_k^t)) - \sum_{i=1}^m \omega_i c(B_i(\lambda_i^t, \tau_i^t))\|_1. 
\end{equation}
For the existing algorithms, e.g., accelerated IBP and APDAGD, they are developed based on the primal-dual optimization framework which allows for achieving better dependence on $1/\varepsilon$ than \textsc{FastIBP} by directly optimizing $E_t$. In contrast, the \textsc{FastIBP} algorithm indirectly optimizes $E_t$ through the dual objective gap and the scheme with momentum (cf. \textbf{Step 1-3} and \textbf{Step 8}), which can lead to better dependence on $n$.
\begin{remark}
We provide some comments on the \textsc{FastIBP} algorithm. First, each iteration updates $\bigO(mn^2)$ entries which is similar to the IBP algorithm. Updating $\tilde{\lambda}$ and $\check{\lambda}$ can be efficiently implemented in distributed manner and each of $m$ machine updates $\bigO(n^2)$ entries at each iteration. Second, the computation of $4m$ marginals can be performed using implementation tricks. Indeed, this can be done effectively by using $r(e^{-\eta^{-1}C_k})$ and $c(e^{-\eta^{-1}C_k})$ for all $k \in [m]$, which are computed and stored at the beginning of the algorithm.
\end{remark}
\begin{remark}
First, we notice that $(\widehat{X}_1, \widehat{X}_2, \ldots, \widehat{X}_m)$ are one set of approximate optimal transportation plans between $m$ measures $\{u^k\}_{k \in [m]}$ and an $\varepsilon$-approximate barycenter $\widehat{u}$. These matrices are equivalent to those constructed by using~\cite[Algorithm~2]{Altschuler-2017-Near}. We also remark that the approximate barycenter $\widehat{u}$ can be constructed by only using $(\widetilde{X}_1, \widetilde{X}_2, \ldots, \widetilde{X}_m)$; see~\cite[Section~2.2]{Kroshnin-2019-Complexity} for the details. 
\end{remark}
\begin{algorithm}[!t]\small
\caption{Finding a Wasserstein barycenter by the \textsc{FastIBP} algorithm} \label{Algorithm:WBP_FastIBP}
\begin{algorithmic}
\STATE \textbf{Input:} $\eta =\varepsilon/(4\log(n))$ and $\bar{\varepsilon} = \varepsilon/(4\max_{1 \leq k \leq m} \|C_k\|_\infty)$. 
\STATE \textbf{Step 1:} Compute $(\tilde{u}^1, \ldots, \tilde{u}^m) = (1 - \bar{\varepsilon}/4)(u^1, \ldots, u^m) + (\bar{\varepsilon}/4n)(\one_n, \ldots, \one_n)$.    
\STATE \textbf{Step 2:} Compute $(\widetilde{X}_1, \widetilde{X}_2, \ldots, \widetilde{X}_m) = \textsc{FastIBP}(\{C_k, \tilde{u}^k\}_{k \in [m]}, \bar{\varepsilon}/2)$. 
\STATE \textbf{Step 3:} Round $(\widetilde{X}_1, \widetilde{X}_2, \ldots, \widetilde{X}_m)$ to $(\widehat{X}_1, \widehat{X}_2, \ldots, \widehat{X}_m)$ using~\citet[Algorithm~4]{Kroshnin-2019-Complexity} such that $(\widehat{X}_1, \widehat{X}_2, \ldots, \widehat{X}_m)$ is feasible to the FS-WBP in Eq.~\eqref{prob:barycenter}.  
\STATE \textbf{Step 4:} Compute $\widehat{u} = \sum_{k=1}^m \omega_k\widehat{X}_k^\top\one_n$
\STATE \textbf{Output:} $\widehat{u}$. 
\end{algorithmic}
\end{algorithm}

\subsection{Convergence analysis}
We present several technical lemmas which are important to analyzing the \textsc{FastIBP} algorithm. The first lemma provides the inductive formula and the upper bound for $\theta_t$. 
\begin{lemma}\label{Lemma:theta-inductive}
Let $\{\theta_t\}_{t \geq 0}$ be the iterates generated by the \textsc{FastIBP} algorithm. Then we have $0 < \theta_t \leq 2/(t+2)$ and $\theta_t^{-2} = (1 - \theta_{t+1})\theta_{t+1}^{-2}$ for all $t \geq 0$. 
\end{lemma}
\begin{proof}
By the definition of $\theta_t$, we have
\begin{equation*}
\left(\frac{\theta_{t+1}}{\theta_t}\right)^2 = \frac{1}{4}\left(\sqrt{\theta_t^2 + 4} - \theta_t\right)^2 = 1 + \frac{\theta_t}{2}\left(\theta_t - \sqrt{\theta_t^2 + 4}\right) = 1 - \theta_{t+1}, 
\end{equation*}
which implies the desired inductive formula and $\theta_t > 0$ for all $t \geq 0$. Then we proceed to prove that $0 < \theta_t \leq 2/(t+2)$ for all $t \geq 0$ using the induction. Indeed, the claim trivially holds when $t=0$ as $\theta_0 = 1$. Assume that the hypothesis holds for $t \leq t_0$, i.e., $\theta_{t_0} \leq 2/(t_0+2)$, we have
\begin{equation*}
\theta_{t_0+1} = \frac{2}{1 + \sqrt{1 + \frac{4}{\theta_{t_0}^2}}} \leq \frac{2}{t_0+3}. 
\end{equation*}
This completes the proof of the lemma. 
\end{proof}
The second lemma shows that all the iterates generated by the \textsc{FastIBP} algorithm are feasible to the dual entropic regularized FS-WBP for all $t \geq 1$.  
\begin{lemma}\label{Lemma:iterate-feasibility}
Let $\{(\check{\lambda}^t, \check{\tau}^t)\}_{t \geq 0}$, $\{(\tilde{\lambda}^t, \tilde{\tau}^t)\}_{t \geq 0}$, $\{(\bar{\lambda}^t, \bar{\tau}^t)\}_{t \geq 0}$, $\{(\widehat{\lambda}^t, \widehat{\tau}^t)\}_{t \geq 0}$, $\{(\acute{\lambda}^t, \acute{\tau}^t)\}_{t \geq 0}$, and $\{(\lambda^t, \tau^t)\}_{t \geq 0}$ be the iterates generated by the \textsc{FastIBP} algorithm. Then, we have 
\begin{equation*}
\sum_{k=1}^m \omega_k \check{\tau}_k^t = \sum_{k=1}^m \omega_k \tilde{\tau}_k^t = \sum_{k=1}^m \omega_k \bar{\tau}_k^t = \sum_{k=1}^m \omega_k \widehat{\tau}_k^t = \sum_{k=1}^m \omega_k \acute{\tau}_k^t  = \sum_{k=1}^m \omega_k \grave{\tau}_k^t = \sum_{k=1}^m \omega_k \tau_k^t  = \textnormal{\zero}_n \quad \textnormal{for all } t \geq 0. 
\end{equation*}
\end{lemma}
\begin{proof}
We first verify Lemma~\ref{Lemma:iterate-feasibility} when $t=0$. Indeed, 
\begin{equation*}
\sum_{k=1}^m \omega_k \check{\tau}_k^0 = \sum_{k=1}^m \omega_k \tilde{\tau}_k^0 = \zero_n. 
\end{equation*}
By the definition, $\bar{\tau}^0$ is a convex combination of $\check{\tau}^0$ and $\tilde{\tau}^0$ and $\widehat{\tau}^0$ is a linear combination of $\bar{\tau}^0$, $\tilde{\tau}^1$ and $\tilde{\tau}^0$. Thus, we have 
\begin{equation*}
\sum_{k=1}^m \omega_k \bar{\tau}_k^0 = \sum_{k=1}^m \omega_k \widehat{\tau}_k^0 = \zero_n. 
\end{equation*}
This also implies that $\sum_{k=1}^m \omega_k \acute{\tau}_k^0 = \zero_n$. Using the update formula for $\grave{\tau}^0$, $\tau^0$ and $\check{\tau}^1$, we have 
\begin{equation*}
\sum_{k=1}^m \omega_k \grave{\tau}_k^0 = \sum_{k=1}^m \omega_k \tau_k^0 = \sum_{k=1}^m \omega_k \check{\tau}_k^1 = \zero_n. 
\end{equation*}
Besides that, the update formula for $\tilde{\tau}^1$ implies $\sum_{k=1}^m \omega_k \tilde{\tau}_k^1 = \zero_n$. Repeating this argument, we obtain the desired equality in the conclusion of Lemma~\ref{Lemma:iterate-feasibility} for all $t \geq 0$. 
\end{proof}
The third lemma shows that the iterates $\{\tau^t\}_{t \geq 0}$ generated by the \textsc{FastIBP} algorithm satisfies the bounded difference property: $\max_{1 \leq i \leq n} (\tau_k^t)_i - \min_{1 \leq i \leq n} (\tau_k^t)_i \leq R_\tau/2$. 
\begin{lemma}\label{Lemma:iterate-boundedness}
Let $\{(\lambda^t, \tau^t)\}_{t \geq 0}$ be the iterates generated by the \textsc{FastIBP} algorithm. Then the following statement holds true: 
\begin{equation*}
\max_{1 \leq i \leq n} (\tau_k^t)_i - \min_{1 \leq i \leq n} (\tau_k^t)_i \leq R_\tau/2, 
\end{equation*}
where $R_\tau > 0$ is defined in Lemma~\ref{Lemma:dual-bound-infinity}.
\end{lemma}
\begin{proof}
We observe that $\tau_k^t = \grave{\tau}_k^t$ for all $k \in [m]$. By the update formula for $\grave{\tau}_k^t$, we have
\begin{equation*}
\grave{\tau}_k^t = \acute{\tau}_k^t + \sum_{i=1}^m \omega_i \log(c_i) - \log(c_k) = \sum_{i=1}^m \omega_i\log(e^{-\eta^{-1}C_i}\diag(e^{\acute{\lambda}^t_i})\one_n)- \log(e^{-\eta^{-1}C_k}\diag(e^{\acute{\lambda}^t_k})\one_n). 
\end{equation*}
After the simple calculation, we have
\begin{equation*}
-\eta^{-1}\|C_k\|_\infty + \one_n^\top e^{\acute{\lambda}^t_k} \leq \log(e^{-\eta^{-1}C_k}\diag(e^{\acute{\lambda}^t_k})\one_n)]_j \leq \one_n^\top e^{\acute{\lambda}^t_k}. 
\end{equation*}
Therefore, the following inequality holds true for all $k \in [m]$, 
\begin{equation*}
\max_{1 \leq i \leq n} (\tau_k^t)_i - \min_{1 \leq i \leq n} (\tau_k^t)_i \leq \eta^{-1}\|C_k\|_\infty + \eta^{-1}\left(\sum_{i=1}^m \omega_i\|C_i\|_\infty\right) = 2\eta^{-1}(\max_{1 \leq k \leq m} \|C_k\|_\infty).  
\end{equation*}
This together with the definition of $R_\tau$ yields the desired inequality. 
\end{proof}
The final lemma presents a key descent inequality for the \textsc{FastIBP} algorithm. 
\begin{lemma}\label{Lemma:FastIBP-descent}
Let $\{(\check{\lambda}^t, \check{\tau}^t)\}_{t \geq 0}$ be the iterates generated by the \textsc{FastIBP} algorithm and let $(\lambda^\star, \tau^\star)$ be an optimal solution in Lemma~\ref{Lemma:dual-bound-infinity}. Then the following statement holds true: 
\begin{equation*}
\varphi(\check{\lambda}^{t+1}, \check{\tau}^{t+1}) - (1-\theta_t)\varphi(\check{\lambda}^t, \check{\tau}^t) - \theta_t\varphi(\lambda^\star, \tau^\star) \leq 2\theta_t^2\left(\sum_{k=1}^m \omega_k\left(\left\|\begin{pmatrix} \lambda^\star_k - \tilde{\lambda}_k^t \\ \tau^\star_k - \tilde{\tau}_k^t \end{pmatrix} \right\|^2 - \left\|\begin{pmatrix} \lambda^\star_k - \tilde{\lambda}_k^{t+1} \\ \tau^\star_k - \tilde{\tau}_k^{t+1} \end{pmatrix} \right\|^2\right)\right). 
\end{equation*}
\end{lemma}
\begin{proof}
Using Lemma~\ref{Lemma:liptshitz-continuity} with $(\lambda', \tau') = (\widehat{\lambda}^{t+1}, \widehat{\tau}^{t+1})$ and $(\lambda, \tau) = (\bar{\lambda}^t, \bar{\tau}^t)$, we have
\begin{equation*}
\varphi(\widehat{\lambda}^{t+1}, \widehat{\tau}^{t+1}) \leq \varphi(\bar{\lambda}^t, \bar{\tau}^t) + \theta_t\begin{pmatrix} \tilde{\lambda}^{t+1} - \tilde{\lambda}^t \\ \tilde{\tau}^{t+1} - \tilde{\tau}^t \end{pmatrix}^\top\nabla\varphi(\bar{\lambda}^t, \bar{\tau}^t) + 2\theta_t^2\left(\sum_{k=1}^m \omega_k\left\|\begin{pmatrix} \tilde{\lambda}_k^{t+1} - \tilde{\lambda}_k^t \\ \tilde{\tau}_k^{t+1} - \tilde{\tau}_k^t \end{pmatrix} \right\|^2\right). 
\end{equation*}
After some simple calculations, we find that
\begin{eqnarray*}
\varphi(\bar{\lambda}^t, \bar{\tau}^t) & = & (1-\theta_t)\varphi(\bar{\lambda}^t, \bar{\tau}^t) + \theta_t\varphi(\bar{\lambda}^t, \bar{\tau}^t), \\
\begin{pmatrix} \tilde{\lambda}^{t+1} - \tilde{\lambda}^t \\ \tilde{\tau}^{t+1} - \tilde{\tau}^t \end{pmatrix}^\top\nabla\varphi(\bar{\lambda}^t, \bar{\tau}^t) & = & -\begin{pmatrix} \tilde{\lambda}^t - \bar{\lambda}^t \\ \tilde{\tau}^t - \bar{\tau}^t\end{pmatrix}^\top\nabla\varphi(\bar{\lambda}^t, \bar{\tau}^t) + \begin{pmatrix} \tilde{\lambda}^{t+1} - \bar{\lambda}^t \\ \tilde{\tau}^{t+1} - \bar{\tau}^t \end{pmatrix}^\top\nabla\varphi(\bar{\lambda}^t, \bar{\tau}^t). 
\end{eqnarray*}
Putting these pieces together yields that 
\begin{eqnarray}\label{claim-AIBP-descent-main}
\varphi(\widehat{\lambda}^{t+1}, \widehat{\tau}^{t+1}) & \leq & \underbrace{(1-\theta_t)\varphi(\bar{\lambda}^t, \bar{\tau}^t) - \theta_t\begin{pmatrix} \tilde{\lambda}^t - \bar{\lambda}^t \\ \tilde{\tau}^t - \bar{\tau}^t\end{pmatrix}^\top\nabla\varphi(\bar{\lambda}^t, \bar{\tau}^t)}_{\textnormal{I}} \\
& & \hspace*{-6em} + \theta_t\left(\underbrace{\varphi(\bar{\lambda}^t, \bar{\tau}^t) + \begin{pmatrix} \tilde{\lambda}^{t+1} - \bar{\lambda}^t \\ \tilde{\tau}^{t+1} - \bar{\tau}^t \end{pmatrix}^\top\nabla\varphi(\bar{\lambda}^t, \bar{\tau}^t) + 2\theta_t\left(\sum_{k=1}^m \omega_k\left\|\begin{pmatrix} \tilde{\lambda}^{t+1} - \tilde{\lambda}^t \\ \tilde{\tau}^{t+1} - \tilde{\tau}^t \end{pmatrix} \right\|^2\right)}_{\textnormal{II}}\right). \nonumber
\end{eqnarray}
For the term $\textnormal{I}$ in equation~\eqref{claim-AIBP-descent-main}, we derive from the definition of $(\bar{\lambda}^t, \bar{\tau}^t)$ that
\begin{equation*}
-\theta_t\begin{pmatrix} \tilde{\lambda}^t - \bar{\lambda}^t \\ \tilde{\tau}^t  - \bar{\tau}^t \end{pmatrix} = \theta_t\begin{pmatrix} \bar{\lambda}^t \\ \bar{\tau}^t \end{pmatrix} + (1 - \theta_t)\begin{pmatrix} \check{\lambda}^t \\ \check{\tau}^t \end{pmatrix} - \begin{pmatrix} \bar{\lambda}^t \\ \bar{\tau}^t \end{pmatrix} = (1-\theta_t)\begin{pmatrix} \check{\lambda}^t - \bar{\lambda}^t \\ \check{\tau}^t - \bar{\tau}^t \end{pmatrix}. 
\end{equation*}
Using this equality and the convexity of $\varphi$, we have
\begin{equation}\label{claim-AIBP-descent-first}
\textnormal{I} = (1-\theta_t)\left(\varphi(\bar{\lambda}^t, \bar{\tau}^t) + \begin{pmatrix} \check{\lambda}^t - \bar{\lambda}^t \\ \check{\tau}^t - \bar{\tau}^t \end{pmatrix}^\top\nabla\varphi(\bar{\lambda}^t, \bar{\tau}^t)\right) \leq (1-\theta_t)\varphi(\check{\lambda}^t, \check{\tau}^t). 
\end{equation}
For the term $\textnormal{II}$ in equation~\eqref{claim-AIBP-descent-main}, the definition of $(\tilde{\lambda}^{t+1}, \tilde{\tau}^{t+1})$ implies that 
\begin{equation*}
\begin{pmatrix} \lambda - \tilde{\lambda}^{t+1} \\ \tau - \tilde{\tau}^{t+1} \end{pmatrix}^\top\left(\nabla\varphi(\bar{\lambda}^t, \bar{\tau}^t) + 4\theta_t\begin{pmatrix} \omega_1(\tilde{\lambda}_1^{t+1} - \tilde{\lambda}_1^t) \\ \vdots \\ \omega_m(\tilde{\lambda}_m^{t+1} - \tilde{\lambda}_m^t) \\ \omega_1(\tilde{\tau}_1^{t+1} - \tilde{\tau}_1^t) \\ \vdots \\ \omega_m(\tilde{\tau}_m^{t+1} - \tilde{\tau}_m^t) \end{pmatrix}\right) \geq 0, \quad \textnormal{for all } (\lambda, \tau) \in \br^{mn} \times \PCal. 
\end{equation*}
Letting $(\lambda, \tau) = (\lambda^\star, \tau^\star)$ and rearranging the resulting inequality yields that 
\begin{eqnarray*}
& & \begin{pmatrix} \tilde{\lambda}^{t+1} - \bar{\lambda}^t \\ \tilde{\tau}^{t+1} - \bar{\tau}^t \end{pmatrix}^\top\nabla\varphi(\bar{\lambda}^t, \bar{\tau}^t) + 2\theta_t\left(\sum_{k=1}^m \omega_k\left\|\begin{pmatrix} \tilde{\lambda}_k^{t+1} - \tilde{\lambda}_k^t \\ \tilde{\tau}_k^{t+1} - \tilde{\tau}_k^t \end{pmatrix} \right\|^2\right)  \\
& \leq & \begin{pmatrix} \lambda^\star - \bar{\lambda}^t \\ \tau^\star - \bar{\tau}^t \end{pmatrix}^\top\nabla\varphi(\bar{\lambda}^t, \bar{\tau}^t) + 2\theta_t\left(\sum_{k=1}^m \omega_k\left(\left\|\begin{pmatrix} \lambda^\star_k - \tilde{\lambda}_k^t \\ \tau^\star_k - \tilde{\tau}_k^t \end{pmatrix} \right\|^2 - \left\|\begin{pmatrix} \lambda^\star_k - \tilde{\lambda}_k^{t+1} \\ \tau^\star_k - \tilde{\tau}_k^{t+1}\end{pmatrix} \right\|^2\right)\right).  
\end{eqnarray*}
Using the convexity of $\varphi$ again, we have 
\begin{equation*}
\begin{pmatrix} \lambda^\star - \bar{\lambda}^t \\ \tau^\star - \bar{\tau}^t \end{pmatrix}^\top\nabla\varphi(\bar{\lambda}^t, \bar{\tau}^t) \leq \varphi(\lambda^\star, \tau^\star) - \varphi(\bar{\lambda}^t, \bar{\tau}^t). 
\end{equation*}
Putting these pieces together yields that 
\begin{equation}\label{claim-AIBP-descent-second}
\textnormal{I} \leq \varphi(\lambda^\star, \tau^\star) + 2\theta_t\left(\sum_{k=1}^m \omega_k\left(\left\|\begin{pmatrix} \lambda^\star_k - \tilde{\lambda}_k^t \\ \tau^\star_k - \tilde{\tau}_k^t \end{pmatrix} \right\|^2 - \left\|\begin{pmatrix} \lambda^\star_k - \tilde{\lambda}_k^{t+1} \\ \tau^\star_k - \tilde{\tau}_k^{t+1} \end{pmatrix} \right\|^2\right)\right). 
\end{equation}
Plugging Eq.~\eqref{claim-AIBP-descent-first} and Eq.~\eqref{claim-AIBP-descent-second} into Eq.~\eqref{claim-AIBP-descent-main} yields that 
\begin{equation*}
\varphi(\widehat{\lambda}^{t+1}, \widehat{\tau}^{t+1}) \leq (1-\theta_t)\varphi(\check{\lambda}^t, \check{\tau}^t) + \theta_t \varphi(\lambda^\star, \tau^\star) + 2\theta_t^2\left(\sum_{k=1}^m \omega_k\left(\left\|\begin{pmatrix} \lambda^\star_k - \tilde{\lambda}_k^t \\ \tau^\star_k - \tilde{\tau}_k^t \end{pmatrix} \right\|^2 - \left\|\begin{pmatrix} \lambda^\star_k - \tilde{\lambda}_k^{t+1} \\ \tau^\star_k - \tilde{\tau}_k^{t+1} \end{pmatrix} \right\|^2\right)\right). 
\end{equation*}
Since $(\check{\lambda}^{t+1}, \check{\tau}^{t+1})$ is obtained by an exact coordinate update from $(\lambda^t, \tau^t)$, we have $\varphi(\lambda^t, \tau^t) \geq \varphi(\check{\lambda}^{t+1}, \check{\tau}^{t+1})$. Using the similar argument, we have $\varphi(\acute{\lambda}^t, \acute{\tau}^t) \geq \varphi(\grave{\lambda}^t, \grave{\tau}^t) \geq \varphi(\lambda^t, \tau^t)$. By the definition of $(\acute{\lambda}^t, \acute{\tau}^t)$, we have $\varphi(\widehat{\lambda}^t, \widehat{\tau}^t) \geq \varphi(\acute{\lambda}^t, \acute{\tau}^t)$. Putting these pieces together yields the desired inequality. 
\end{proof}

\subsection{Main result}
We present an upper bound for the iteration numbers required by the \textsc{FastIBP} algorithm. 
\begin{theorem}\label{Theorem:FastIBP-iteration-complexity}
Let $\{(\lambda^t, \tau^t)\}_{t \geq 0}$ be the iterates generated by the \textsc{FastIBP} algorithm. Then the number of iterations required to reach the stopping criterion $E_t \leq \varepsilon$ satisfies
\begin{equation*}
t \leq 1 + 10\left(\frac{n(R_\lambda^2+R_\tau^2)}{\varepsilon^2}\right)^{1/3},
\end{equation*}
where $R_\lambda, R_\tau > 0$ are defined in Lemma~\ref{Lemma:dual-bound-infinity}.
\end{theorem}
\begin{proof}
First, let $\delta_t = \varphi(\check{\lambda}^t, \check{\tau}^t) - \varphi(\lambda^\star, \tau^\star)$, we show that 
\begin{equation}\label{inequality:FastIBP-iteration-first}
\delta_t \leq \frac{8n(R_\lambda^2 + R_\tau^2)}{(t+1)^2}.
\end{equation}
Indeed, by Lemma~\ref{Lemma:theta-inductive} and~\ref{Lemma:FastIBP-descent}, we have
\begin{equation*}
\left(\frac{1-\theta_{t+1}}{\theta_{t+1}^2}\right)\delta_{t+1} - \left(\frac{1-\theta_t}{\theta_t^2}\right)\delta_t \leq 2\left(\sum_{k=1}^m \omega_k\left(\left\|\begin{pmatrix} \lambda_k^\star - \tilde{\lambda}_k^t \\ \tau_k^\star - \tilde{\tau}_k^t \end{pmatrix} \right\|^2 - \left\|\begin{pmatrix} \lambda_k^\star - \tilde{\lambda}_k^{t+1} \\ \tau_k^\star - \tilde{\tau}_k^{t+1} \end{pmatrix} \right\|^2\right)\right).
\end{equation*}
By unrolling the recurrence and using $\theta_0 = 1$ and $\tilde{\lambda}_0 = \tilde{\tau}_0 = \zero_{mn}$, we have
\begin{eqnarray*}
\left(\frac{1-\theta_t}{\theta_t^2}\right)\delta_t + 2\left(\sum_{k=1}^m \omega_k\left\|\begin{pmatrix} \lambda_k^\star - \tilde{\lambda}_k^t \\ \tau_k^\star - \tilde{\tau}_k^t \end{pmatrix} \right\|^2\right) & \leq & \left(\frac{1-\theta_0}{\theta_0^2}\right)\delta_0 + 2\left(\sum_{k=1}^m \omega_k\left\|\begin{pmatrix} \lambda_k^\star - \tilde{\lambda}_k^0 \\ \tau_k^\star - \tilde{\tau}_k^0 \end{pmatrix} \right\|^2\right) \\
& & \hspace*{-12em} \leq 2\left(\sum_{k=1}^m \omega_k\left\|\begin{pmatrix} \lambda_k^\star \\ \tau_k^\star \end{pmatrix} \right\|^2\right) \ \overset{\text{Corollary}~\ref{Corollary:dual-bound-l2}}{\leq} \ 2n(R_\lambda^2 + R_\tau^2). 
\end{eqnarray*}
For $t \geq 1$, Lemma~\ref{Lemma:theta-inductive} implies that $\theta_{t-1}^{-2} = (1 - \theta_t)\theta_t^{-2}$. Therefore, we conclude that 
\begin{equation*}
\delta_t \leq 2\theta_{t-1}^2n(R_\lambda^2 + R_\tau^2). 
\end{equation*}
This together with the fact that $0 < \theta_{t-1} \leq 2/(t+1)$ yields the desired inequality. 

Furthermore, we show that 
\begin{equation}\label{inequality:FastIBP-iteration-second}
\delta_t - \delta_{t+1} \geq \frac{E_t^2}{11}.
\end{equation}
Indeed, by the definition of $\Delta_t$, we have
\begin{equation*}
\delta_t - \delta_{t+1} = \varphi(\check{\lambda}^t, \check{\tau}^t) - \varphi(\check{\lambda}^{t+1}, \check{\tau}^{t+1}) \geq \varphi(\lambda^t, \tau^t) - \varphi(\check{\lambda}^{t+1}, \check{\tau}^{t+1}). 
\end{equation*}
By the definition of $\varphi$, we have
\begin{equation*}
\varphi(\lambda^t, \tau^t) - \varphi(\check{\lambda}^{t+1}, \check{\tau}^{t+1}) = \sum_{k=1}^m \omega_k(\log(\|B_k(\lambda_k^t, \tau_k^t)\|_1) - \log(\|B_k(\check{\lambda}_k^{t+1}, \check{\tau}_k^{t+1})\|_1)).  
\end{equation*}
Since $r(B_k(\lambda_k^t, \tau_k^t)) = u^k \in \Delta_n$ for all $k \in [m]$, we have $\|B_k(\lambda_k^t, \tau_k^t)\|_1 = 1$. This together with the update formula of $(\check{\lambda}^{t+1}, \check{\tau}^{t+1})$ yields that 
\begin{equation*}
\varphi(\lambda^t, \tau^t) - \varphi(\check{\lambda}^{t+1}, \check{\tau}^{t+1}) = - \log\left(\one_n^\top e^{\sum_{k=1}^m \omega_k\log(c(B_k(\lambda_k^t, \tau_k^t)))}\right). 
\end{equation*}
Recall that $\log(1+x) \leq x$ for all $x \in \br$, we have 
\begin{equation*}
\varphi(\lambda^t, \tau^t) - \varphi(\check{\lambda}^{t+1}, \check{\tau}^{t+1}) \geq 1 - \one_n^\top e^{\sum_{k=1}^m \omega_k\log(c(B_k(\lambda_k^t, \tau_k^t)))}. 
\end{equation*}
Since $r(B_k(\lambda_k^t, \tau_k^t)) = u^k \in \Delta_n$ for all $k \in [m]$, we have $\one_n^\top c(B_k(\lambda_k^t, \tau_k^t)) = 1$. In addition, $(\omega_1, \omega_2, \ldots, \omega_m) \in \Delta^m$. Thus, we have
\begin{equation*}
\varphi(\lambda^t, \tau^t) - \varphi(\check{\lambda}^{t+1}, \check{\tau}^{t+1}) \geq \one_n^\top\left(\sum_{k=1}^m \omega_k c(B_k(\lambda_k^t, \tau_k^t)) - e^{\sum_{k=1}^m \omega_k\log(c(B_k(\lambda_k^t, \tau_k^t)))}\right).  
\end{equation*}
Combining $c(B_k(\lambda_k^t, \tau_k^t)) \in \Delta^n$ with the arguments in~\citet[Lemma~6]{Kroshnin-2019-Complexity} yields
\begin{equation*}
\varphi(\lambda^t, \tau^t) - \varphi(\check{\lambda}^{t+1}, \check{\tau}^{t+1}) \geq \frac{1}{11}\sum_{k=1}^m \omega_k\|c(B_k(\lambda_k^t, \tau_k^t)) - \sum_{i=1}^m \omega_i c(B_i(\lambda_i^t, \tau_i^t))\|_1^2. 
\end{equation*}
Using the Cauchy-Schwarz inequality together with $\sum_{k=1}^m \omega_k = 1$, we have 
\begin{equation*}
E_t^2 \leq \sum_{k=1}^m \omega_k\|c(B_k(\lambda_k^t, \tau_k^t)) - \sum_{i=1}^m \omega_i c(B_i(\lambda_i, \tau_i))\|_1^2.  
\end{equation*}
Putting these pieces together yields the desired inequality. 

Finally, we derive from Eq.~\eqref{inequality:FastIBP-iteration-first} and~\eqref{inequality:FastIBP-iteration-second} and the non-negativeness of $\delta_t$ that 
\begin{equation*}
\sum_{i=t}^{+\infty} E_i^2 \leq 11\left(\sum_{i=t}^{+\infty} (\delta_i - \delta_{i+1})\right) \leq 11\delta_t \leq \frac{88n(R_\lambda^2 + R_\tau^2)}{(t+1)^2} 
\end{equation*}
Let $T > 0$ satisfy $E_T \leq \varepsilon$, we have $E_t > \varepsilon$ for all $t \in [T]$. Without loss of generality, we assume $T$ is even. Then the following statement holds true:
\begin{equation*}
\varepsilon^2 \leq \frac{704n(R_\lambda^2+R_\tau^2)}{T^3}. 
\end{equation*}
Rearranging the above inequality yields the desired inequality.
\end{proof}
Equipped with the result of Theorem~\ref{Theorem:FastIBP-iteration-complexity}, we are ready to present the complexity bound of Algorithm~\ref{Algorithm:WBP_FastIBP} for approximating the FS-WBP in Eq.~\eqref{prob:barycenter}.
\begin{theorem}\label{Theorem:FastIBP-barycenter}
The \textsc{FastIBP} algorithm for approximately solving the FS-WBP in Eq.~\eqref{prob:barycenter} (Algorithm~\ref{Algorithm:WBP_FastIBP}) returns an $\varepsilon$-approximate barycenter $\widehat{u} \in \br^n$ within 
\begin{equation*}
\bigO\left(mn^{7/3}\left(\frac{(\max_{1 \leq k \leq m}\|C_k\|_\infty)\sqrt{\log(n)}}{\varepsilon}\right)^{4/3}\right) 
\end{equation*}
arithmetic operations.
\end{theorem}
\begin{proof}
Consider the iterate $(\widetilde{X}_1, \widetilde{X}_2, \ldots, \widetilde{X}_m)$ be generated by the \textsc{FastIBP} algorithm (cf. Algorithm~\ref{Algorithm:FastIBP}), the rounding scheme (cf.~\citet[Algorithm~4]{Kroshnin-2019-Complexity}) returns the feasible solution $(\widehat{X}_1, \widehat{X}_2, \ldots, \widehat{X}_m)$ to the FS-WBP in Eq.~\eqref{prob:barycenter} and $c(\widehat{X}_k)$ are the same for all $k \in [m]$. 

To show that $\widehat{u} = \sum_{k=1}^m \omega_k c(\widehat{X}_k)$ is an $\varepsilon$-approximate barycenter (cf. Definition~\ref{def:approx_barycenter}), it suffices to show that 
\begin{equation}\label{inequality:FastIBP-complexity-main}
\sum_{k=1}^m \omega_k \langle C_k, \widehat{X}_k\rangle \leq \sum_{k=1}^m \omega_k \langle C_k, X_k^\star\rangle + \varepsilon, 
\end{equation}
where $(X_1^\star, X_2^\star, \ldots, X_m^\star)$ is a set of optimal transportation plan between $m$ measures $\{u^k\}_{k \in [m]}$ and the barycenter of the FS-WBP. 

First, we derive from the scheme of~\citet[Algorithm~4]{Kroshnin-2019-Complexity} that the following inequality holds for all $k \in [m]$, 
\begin{equation*}
\|\widehat{X}_k - \widetilde{X}_k\|_1 \leq \|c(\widetilde{X}_k) - \sum_{i=1}^m \omega_i c(\widetilde{X}_i)\|_1.  
\end{equation*}
This together with the H\"{o}lder's inequality implies that 
\begin{equation}\label{inequality:FastIBP-complexity-first}
\sum_{k=1}^m \omega_k \langle C_k, \widehat{X}_k - \widetilde{X}_k\rangle \leq \left(\max_{1 \leq k \leq m} \|C_k\|_\infty\right)\left(\sum_{k=1}^m \omega_k\|c(\widetilde{X}_k) - \sum_{i=1}^m \omega_i c(\widetilde{X}_i)\|_1\right). 
\end{equation}
Furthermore, we have
\begin{eqnarray*}
\sum_{k=1}^m \omega_k \langle C_k, \widetilde{X}_k - X_k^\star\rangle & = & \sum_{k=1}^m \omega_k (\langle C_k, \widetilde{X}_k\rangle - \eta H(\widetilde{X}_k)) - \sum_{k=1}^m \omega_k (\langle C_k, X_k^\star\rangle - \eta H(X_k^\star)) \\
& & + \sum_{k=1}^m \omega_k \eta H(\widetilde{X}_k) - \sum_{k=1}^m \omega_k \eta H(X_k^\star). 
\end{eqnarray*} 
Since $0 \leq H(X) \leq 2\log(n)$ for any $X \in \br_+^{n \times n}$ satisfying that $\|X\|_1=1$~\citep{Cover-2012-Elements} and $\sum_{k=1}^m \omega_k = 1$, we have
\begin{equation*}
\sum_{k=1}^m \omega_k \langle C_k, \widetilde{X}_k - X_k^\star\rangle \leq 2\eta\log(n) + \sum_{k=1}^m \omega_k (\langle C_k, \widetilde{X}_k\rangle - \eta H(\widetilde{X}_k)) - \sum_{k=1}^m \omega_k (\langle C_k, X_k^\star\rangle - \eta H(X_k^\star)). 
\end{equation*}
Let $(X_1^\eta, X_2^\eta, \ldots, X_m^\eta)$ be a set of optimal transportation plans to the entropic regularized FS-WBP in Eq.~\eqref{prob:barycenter_regularized}, we have
\begin{equation*}
\sum_{k=1}^m \omega_k (\langle C_k, X_k^\eta\rangle - \eta H(X_k^\eta)) \leq \sum_{k=1}^m \omega_k (\langle C_k, X_k^\star\rangle - \eta H(X_k^\star)). 
\end{equation*}
By the optimality of $(X_1^\eta, X_2^\eta, \ldots, X_m^\eta)$, we have
\begin{equation*}
\sum_{k=1}^m \omega_k (\langle C_k, X_k^\eta\rangle - \eta H(X_k^\eta)) = -\eta\left(\min_{\lambda \in \br^{mn}, \tau \in \PCal} \varphi(\lambda, \tau)\right) \geq - \eta\varphi(\lambda^t, \tau^t). 
\end{equation*}
Since $(\widetilde{X}_1, \widetilde{X}_2, \ldots, \widetilde{X}_m)$ is generated by the \textsc{FastIBP} algorithm, we have $\widetilde{X}_k = B_k(\lambda_k^t, \tau_k^t)$ for all $k \in [m]$ where $(\lambda^t, \tau^t)$ are the dual iterates. Then 
\begin{eqnarray*}
\sum_{k=1}^m \omega_k (\langle C_k, \widetilde{X}_k\rangle - \eta H(\widetilde{X}_k)) & = & \sum_{k=1}^m \omega_k (\langle C_k, B_k(\lambda_k^t, \tau_k^t)\rangle - \eta H(B_k(\lambda_k^t, \tau_k^t))) \\
& & \hspace*{-10em} = \ -\eta\left(\sum_{k=1}^m \omega_k (\one_n^\top B_k(\lambda_k^t, \tau_k^t)\one_n - (\lambda_k^t)^\top u^k)\right) + \eta\sum_{k=1}^m \omega_k (\tau_k^t)^\top c(B_k(\lambda_k^t, \tau_k^t)) \\
& & \hspace*{-10em} = \ - \eta\varphi(\lambda^t, \tau^t) + \eta\left(\sum_{k=1}^m \omega_k (\tau_k^t)^\top \left(c(B_k(\lambda_k^t, \tau_k^t)) - \sum_{i=1}^m \omega_i c(B_i(\lambda_i^t, \tau_i^t))\right)\right). 
\end{eqnarray*}
Putting these pieces together yields that 
\begin{equation*}
\sum_{k=1}^m \omega_k \langle C_k, \widetilde{X}_k - X_k^\star\rangle \leq 2\eta\log(n) + \eta\left(\sum_{k=1}^m \omega_k (\tau_k^t)^\top \left(c(B_k(\lambda_k^t, \tau_k^t)) - \sum_{i=1}^m \omega_i c(B_i(\lambda_i^t, \tau_i^t))\right)\right). 
\end{equation*}
Since $\one_n^\top c(B_k(\lambda_k^t, \tau_k^t)) = 1$ for all $k \in [m]$, we have
\begin{eqnarray*}
& & \left(\sum_{k=1}^m \omega_k (\tau_k^t)^\top \left(c(B_k(\lambda_k^t, \tau_k^t)) - \sum_{i=1}^m \omega_i c(B_i(\lambda_i^t, \tau_i^t))\right)\right) \\
& = & \left(\sum_{k=1}^m \omega_k \left(\tau_k^t - \frac{\max_{1 \leq i \leq n} (\tau_k^t)_i + \min_{1 \leq i \leq n} (\tau_k^t)_i}{2}\one_n\right)^\top \left(c(B_k(\lambda_k^t, \tau_k^t)) - \sum_{i=1}^m \omega_i c(B_i(\lambda_i^t, \tau_i^t))\right)\right) \\ 
& \leq & \left\|\tau_k^t - \frac{\max_{1 \leq i \leq n} (\tau_k^t)_i + \min_{1 \leq i \leq n} (\tau_k^t)_i}{2}\one_n\right\|_\infty\left(\sum_{k=1}^m \omega_k\|c(\widetilde{X}_k) - \sum_{i=1}^m \omega_i c(\widetilde{X}_i)\|_1\right). 
\end{eqnarray*}
Using Lemma~\ref{Lemma:iterate-boundedness}, we have
\begin{equation*}
\left\|\tau_k^t - \frac{\max_{1 \leq i \leq n} (\tau_k^t)_i + \min_{1 \leq i \leq n} (\tau_k^t)_i}{2}\one_n\right\|_\infty \leq \frac{R_\tau}{2}. 
\end{equation*} 
Putting these pieces together yields that 
\begin{equation}\label{inequality:FastIBP-complexity-second}
\sum_{k=1}^m \omega_k \langle C_k, \widetilde{X}_k - X_k^\star\rangle \leq 2\eta\log(n) + \frac{\eta R_\tau}{2}\left(\sum_{k=1}^m \omega_k\|c(\widetilde{X}_k) - \sum_{i=1}^m \omega_i c(\widetilde{X}_i)\|_1\right). 
\end{equation}
Recall that $E_t = \sum_{k=1}^m \omega_k\|c(\widetilde{X}_k) - \sum_{i=1}^m \omega_i c(\widetilde{X}_i)\|_1$ and $R_\tau = 4\eta^{-1}(\max_{1 \leq k \leq m} \|C_k\|_\infty)$. Then Eq.~\eqref{inequality:FastIBP-complexity-first} and Eq.~\eqref{inequality:FastIBP-complexity-second} together imply that
\begin{equation*}
\sum_{k=1}^m \omega_k \langle C_k, \widehat{X}_k - X_k^\star\rangle \leq 2\eta\log(n) + 3\left(\max_{1 \leq k \leq m} \|C_k\|_\infty\right)E_t.  
\end{equation*}
This together with $E_t \leq \bar{\varepsilon}/2$ and the choice of $\eta$ and $\bar{\varepsilon}$ implies Eq.~\eqref{inequality:FastIBP-complexity-main} as desired.  
\paragraph{Complexity bound estimation.} We first bound the number of iterations required by the \textsc{FastIBP} algorithm (cf. Algorithm~\ref{Algorithm:FastIBP}) to reach $E_t \leq \bar{\varepsilon}/2$. Indeed, Theorem~\ref{Theorem:FastIBP-iteration-complexity} implies that 
\begin{equation*}
t \leq 1 + 20\left(\frac{n(R_\lambda^2+R_\tau^2)}{\bar{\varepsilon}^2}\right)^{1/3} \leq 20\sqrt[3]{n}\left(\frac{R_\lambda + R_\tau}{\bar{\varepsilon}}\right)^{2/3}. 
\end{equation*}
For the simplicity, we let $\bar{C} = \max_{1 \leq k \leq m}\left\|C_k\right\|_\infty$. Using the definition of $R_\lambda$ and $R_\tau$ in Lemma~\ref{Lemma:dual-bound-infinity}, the construction of $\{\tilde{u}^k\}_{k \in [m]}$ and the choice of $\eta$ and $\bar{\varepsilon}$, we have
\begin{equation*}
t \ \leq \ 1 + 20\sqrt[3]{n}\left(\frac{4\bar{C}}{\varepsilon}\left(\frac{52\log(n)\bar{C}}{\varepsilon} + 2\log(n) - \log\left(\frac{16n\bar{C}}{\varepsilon}\right)\right)\right)^{2/3} \ = \ \bigO\left(\sqrt[3]{n}\left(\frac{\bar{C}\sqrt{\log(n)}}{\varepsilon}\right)^{4/3}\right).
\end{equation*}
Recall that each iteration of the \textsc{FastIBP} algorithm requires $\OCal(mn^2)$ arithmetic operations, the total arithmetic operations required by the \textsc{FastIBP} algorithm as the subroutine in Algorithm~\ref{Algorithm:WBP_FastIBP} is bounded by 
\begin{equation*}
\bigO\left(mn^{7/3}\left(\frac{\bar{C}\sqrt{\log(n)}}{\varepsilon}\right)^{4/3}\right). 
\end{equation*}
Computing a collection of vectors $\{\tilde{u}^k\}_{k \in [m]}$ needs $\OCal(mn)$ arithmetic operations while the rounding scheme in~\citet[Algorithm~4]{Kroshnin-2019-Complexity} requires $\OCal(mn^2)$ arithmetic operations. Putting these pieces together yields that the desired complexity bound of Algorithm~\ref{Algorithm:WBP_FastIBP}. 
\end{proof}
\begin{remark}
For the simplicity, we assume that all measures have the same support size. This assumption is not necessary and our analysis is still valid when each measure has fixed support of different size. However, our results can not be generalized to the free-support Wasserstein barycenter problem in general since the computation of free-support barycenters requires solving a multimarginal OT problem where the complexity bounds of algorithms become much worse; see~\citet{Lin-2019-Complexity} for the details. 
\end{remark}

\section{Experiments}\label{sec:experiments}
In this section, we report the results of extensive numerical experiments to evaluate the \textsc{FastIBP} algorithm for computing fixed-support Wasserstein barycenters. In all our experiments, we consider the Wasserstein distance with $\ell_2$-norm, i.e., 2-Wasserstein distance, and compare the \textsc{FastIBP} algorithm with Gurobi, iterative Bregman projection (IBP) algorithm~\citep{Benamou-2015-Iterative} and Bregman ADMM (BADMM)~\citep{Ye-2017-Fast}\footnote{We implement ADMM~\citep{Yang-2018-ADMM}, APDAGD~\citep{Kroshnin-2019-Complexity} and accelerated IBP~\citep{Guminov-2019-Accelerated} and find that they perform worse than our algorithm. However, we believe it is largely due to our own implementation issue since these algorithms require fine hyper-parameter tuning. We are also unaware of any public codes available online. Thus, we exclude them for a fair comparison.}. All the experiments are conducted in MATLAB R2020a on a workstation with an Intel Core i5-9400F (6 cores and 6 threads) and 32GB memory, equipped with Ubuntu 18.04.

\subsection{Implementation details}
For the \textsc{FastIBP} algorithm, the regularization parameter $\eta$ is chosen from $\{0.01, 0.001\}$ in our experiments. We follow~\citet[Remark~3]{Benamou-2015-Iterative} to implement the algorithm and terminate it when
\begin{eqnarray*}
\frac{\sum_{k = 1}^m \omega_k\|c(B_k(\lambda_k^t, \tau_k^t)) - \sum_{i=1}^m \omega_i c(B_i(\lambda_i^t, \tau_i^t))\|}{1 + \sum_{k=1}^m \omega_k\|c(B_k(\lambda_k^t, \tau_k^t))\| + \|\sum_{i=1}^m \omega_i c(B_i(\lambda_i^t, \tau_i^t))\|} & \leq & \textnormal{Tol}_{\textsf{fibp}}, \\
\frac{\sum_{k = 1}^m \omega_k\| r(B_k(\lambda_k^t, \tau_k^t)) - u^k\|}{1 + \sum_{k=1}^m \omega_k\|r(B_k(\lambda_k^t, \tau_k^t))\| + \sum_{k=1}^m \omega_k\|u^k\|} & \leq & \textnormal{Tol}_{\textsf{fibp}}, \\
\frac{\|\sum_{i=1}^m \omega_i c(B_i(\lambda_i^t, \tau_i^t)) - \sum_{i=1}^m \omega_i c(B_i(\lambda_i^{t-1}, \tau_i^{t-1}))\|}{1 + \|\sum_{i=1}^m \omega_i c(B_i(\lambda_i^t, \tau_i^t))\| + \|\sum_{i=1}^m \omega_i c(B_i(\lambda_i^{t-1}, \tau_i^{t-1}))\|} & \leq & \textnormal{Tol}_{\textsf{fibp}}, \\
\frac{\sum_{k = 1}^m \omega_k\|B_k(\lambda_k^t, \tau_k^t) - B_k(\lambda_k^{t-1}, \tau_k^{t-1})\|_F}{1 + \sum_{k=1}^m \omega_k\|B_k(\lambda_k^t, \tau_k^t)\|_F + \sum_{k=1}^m \omega_k\|B_k(\lambda_k^{t-1}, \tau_k^{t-1})\|_F} & \leq & \textnormal{Tol}_{\textsf{fibp}}, \\
\frac{\sum_{k=1}^m \omega_k\|\lambda_k^t - \lambda_k^{t-1}\|}{1 + \sum_{k=1}^m \omega_k\|\lambda_k^t\| +  \sum_{k=1}^m \omega_k\|\lambda_k^{t-1}\|} & \leq & \textnormal{Tol}_{\textsf{fibp}}, \\
\frac{\sum_{k=1}^m \omega_k\|\tau_k^t - \tau_k^{t-1}\|}{1 + \sum_{k=1}^m \omega_k\|\tau_k^t\| +  \sum_{k=1}^m \omega_k\|\tau_k^{t-1}\|} & \leq & \textnormal{Tol}_{\textsf{fibp}}. \\
\end{eqnarray*}
These inequalities guarantee that (i) the infeasibility violations for marginal constraints, (ii) the iterative gap between approximate barycenters, and (iii) the iterative gap between dual variables are relatively small. Computing all the above residuals is expensive. Thus, in our implementations, we only compute them and check the termination criteria at every 20 iterations when $\eta = 0.01$ and every 200 iteration when $\eta = 0.001$. We set $\textnormal{Tol}_{\textsf{fibp}} = 10^{-6}$ and $\textnormal{MaxIter}_{\textsf{fibp}} = 10000$ on synthetic data and $\textnormal{Tol}_{\textsf{fibp}} = 10^{-10}$ on MNIST images. 

For IBP and BADMM, we use the Matlab code\footnote{Available in https://github.com/bobye/WBC{\_}Matlab} implemented by~\citet{Ye-2017-Fast} and terminate them with the refined stopping criterion provided by~\cite{Yang-2018-ADMM}. The regularization parameter $\eta$ for the IBP algorithm is still chosen from $\{0.01, 0.001\}$. For synthetic data, we set $\textnormal{Tol}_{\textsf{badmm}} = 10^{-5}$ and $\textnormal{Tol}_{\textsf{ibp}} = 10^{-6}$ with $\textnormal{MaxIter}_{\textsf{badmm}} = 5000$ and $\textnormal{MaxIter}_{\textsf{ibp}} = 10000$. For MNIST images, we set $\textnormal{Tol}_{\textsf{ibp}} = 10^{-10}$.

For the linear programming algorithm, we apply Gurobi 9.0.2 (Gurobi Optimization, 2019) (with an academic license) to solve the FS-WBP in Eq.~\eqref{prob:barycenter}. Since Gurobi can provide high quality solutions when the problem of medium size, we use the solution obtained by Gurobi as a benchmark to evaluate the qualities of solution obtained by different algorithms on synthetic data. In our experiments, we force Gurobi to \textit{only run the dual simplex algorithm} and use other parameters in the default settings. 

For the evaluation metrics, ``\textbf{normalized obj}" stands for the normalized objective value which is defined by
\begin{equation*}
\textbf{normalized obj} \mydefn \frac{|\sum_{k=1}^m \omega_k\langle C_k, \widehat{X}_k\rangle - \sum_{k=1}^m \omega_k\langle C_k, X_k^{\textsf{g}}\rangle|}{|\sum_{k=1}^m \omega_k\langle C_k, X_k^{\textsf{g}}\rangle|}, 
\end{equation*}
where $(\widehat{X}_1, \ldots, \widehat{X}_m)$ is the solution obtained by each algorithm and $(X_1^{\textsf{g}}, \ldots, X_m^{\textsf{g}})$ denotes the solution obtained by Gurobi. ``\textbf{feasibility}" denotes the the deviation of the terminating solution from the feasible set\footnote{Since we do not put the iterative gap between dual variables in ``\textbf{feasibility}" and the FS-WBP is relatively easier than general WBP, our results for BADMM and IBP are consistently smaller than that presented by~\cite{Ye-2017-Fast, Yang-2018-ADMM, Ge-2019-Interior}.}; see~\citet[Section~5.1]{Yang-2018-ADMM}. ``\textbf{iteration}" denotes the number of iterations. ``\textbf{time (in seconds)}" denotes the computational time.

In what follows, we present our experimental results. In Section~\ref{subsec:synthetic}, we evaluate all the candidate algorithms on synthetic data and compare their computational performance in terms of accuracy and speed. In Section~\ref{subsec:mnist}, we compare our algorithm with IBP on the MNIST dataset to visualize the quality of approximate barycenters obtained by each algorithm. For the simplicity of the presentation, in our figures ``g" stands for Gurobi; ``b" stands for BADMM; ``i1" and ``i2" stand for the IBP algorithm with $\eta=0.01$ and $\eta=0.001$; ``f1" and ``f2" stand for the \textsc{FastIBP} algorithm with $\eta=0.01$ and $\eta=0.001$.
\begin{figure}[!t]
\centering
\includegraphics[scale=0.45]{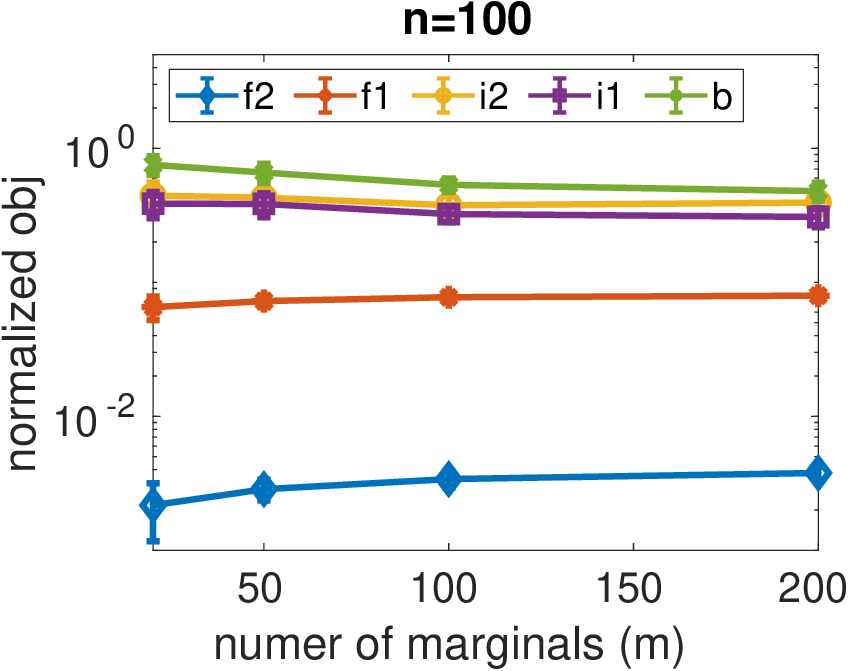} \qquad
\includegraphics[scale=0.45]{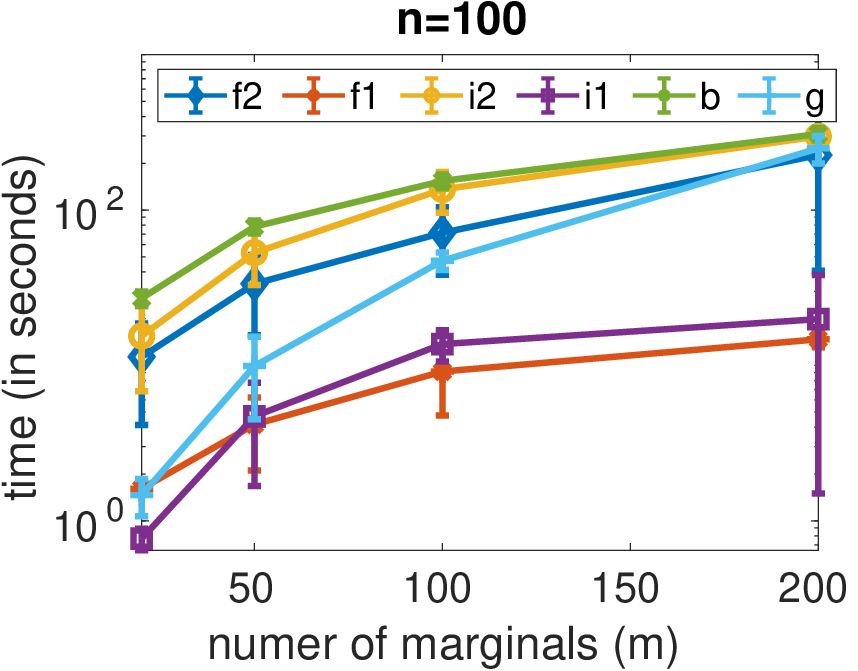}
\caption{The average normalized objective value and computational time (in seconds) of \textsc{FastIBP}, \textsc{IBP}, BADMM, and Gurobi from 10 independent trials.}\label{fig:random}
\end{figure}
\begin{figure}[!t]
\centering
\includegraphics[scale=0.45]{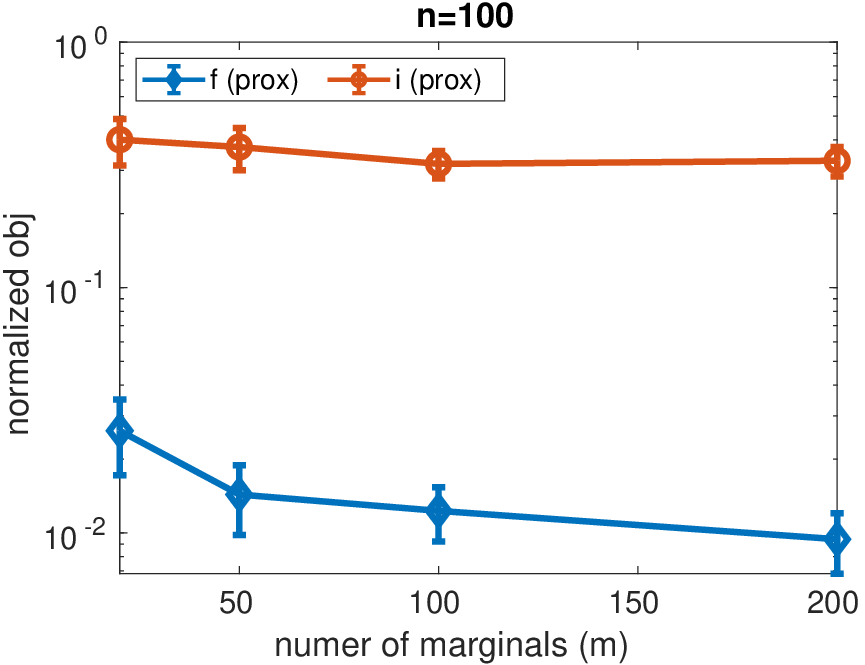} \qquad 
\includegraphics[scale=0.45]{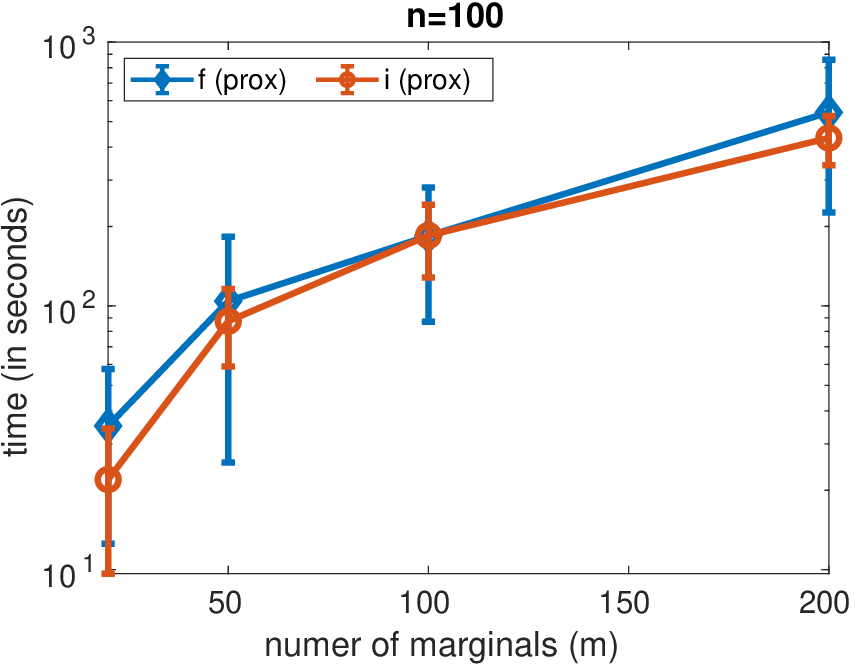}
\caption{The average normalized objective value and computational time (in seconds) of the proximal variants of \textsc{FastIBP} and \textsc{IBP} from 10 independent trials.}\label{fig:prox}
\end{figure}  

\subsection{Experiments on synthetic data}\label{subsec:synthetic}
In this section, we generate a set of discrete probability distributions $\{\mu_k\}_{k=1}^m$ with $\mu_k = \{(u_i^k, \x_i) \in \br_+ \times \br^d \mid i \in [n]\}$ and $\sum_{i=1}^n u_i^k = 1$. The fixed-support Wasserstein barycenter $\widehat{\mu} = \{(\widehat{u}_i, \x_i) \in \br_+ \times \br^d \mid i \in [n]\}$ where $(\x_1, \x_2, \ldots, \x_n)$ are known. In our experiment, we set $d = 3$ and choose different values of $(m, n)$. Then, given each tuple $(m, n)$, we randomly generate a trial as follows. 

First, we generate the support points $(\x_1^k, \x_2^k, \ldots, \x_n^k)$ whose entries are drawn from a Gaussian mixture distribution via the Matlab commands provided by~\citet{Yang-2018-ADMM}: 
\begin{flushleft}
\hspace*{7em} \textsf{gm{\_}num = 5; gm{\_}mean = [-20; -10; 0; 10; 20];} \\
\hspace*{7em} \textsf{sigma = zeros(1, 1, gm{\_}num); sigma(1, 1, :) = 5*ones(gm{\_}num, 1);} \\
\hspace*{7em} \textsf{gm{\_}weights  = rand(gm{\_}num, 1); gm{\_}weights  = gm{\_}weights/sum(gm{\_}weights);} \\
\hspace*{7em} \textsf{distrib = gmdistribution(gm{\_}mean, sigma, gm{\_}weights);}
\end{flushleft}
For each $k \in [m]$, we generate the weight vector $(u_1^k, u_2^k, \ldots, u_n^k)$ whose entries are drawn from the uniform distribution on the interval (0, 1), and normalize it such that $\sum_{i=1}^n u_i^k = 1$. After generating all $\{\mu_k\}_{k=1}^m$, we use the k-means\footnote{In our experiments, we call the Matlab function \textsf{kmeans}, which is built in machine learning toolbox.} method to choose $n$ points from $\{\x_i^k \mid i \in [n], k \in [m]\}$ to be the support points of the barycenter. Finally, we generate the weight vector $(\omega_1, \omega_2, \ldots, \omega_m)$ whose entries are drawn from the uniform distribution on the interval $(0, 1)$, and normalize it such that $\sum_{k=1}^m \omega_k = 1$.

\begin{wrapfigure}{r}{0.4\textwidth}
\centering
\vspace*{-1.5em}\includegraphics[width=2.2in]{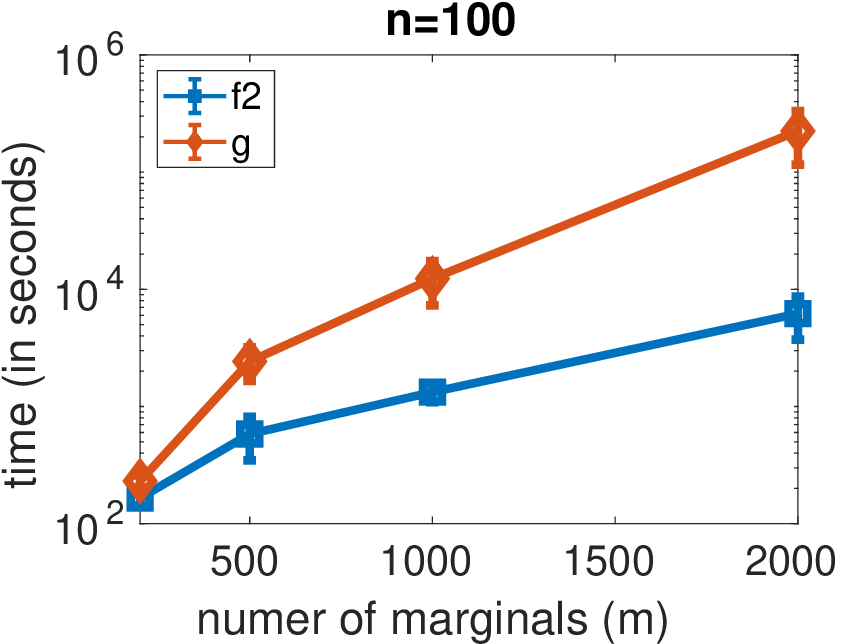}
\hspace*{.5em}\scriptsize\begin{tabular}{|c|cc|}\hline
m & g & f2 \\ \hline
& \multicolumn{2}{|c|}{\textbf{normalized obj}} \\ \hline 
200  & - & 3.6e-03$\pm$3.1e-04 \\
500  & - & 4.4e-03$\pm$6.2e-04 \\
1000 & - & 4.8e-03$\pm$5.4e-04 \\
2000 & - & 5.0e-03$\pm$3.8e-04 \\ \hline
& \multicolumn{2}{|c|}{\textbf{feasibility}} \\ \hline 
200  & 3.2e-07$\pm$1.8e-07 & 7.4e-07$\pm$1.8e-07 \\
500  & 2.8e-07$\pm$5.0e-08 & 7.0e-07$\pm$2.8e-07 \\
1000 & 2.1e-07$\pm$1.0e-07 & 7.1e-07$\pm$2.0e-07 \\
2000 & 2.0e-07$\pm$1.3e-07 & 8.7e-07$\pm$2.0e-07 \\ \hline
& \multicolumn{2}{|c|}{\textbf{iteration}} \\ \hline 
200  & 1.5e+05$\pm$2.4e+04 & 2.4e+03$\pm$3.2e+02 \\
500  & 5.0e+05$\pm$8.8e+04 & 3.3e+03$\pm$1.4e+03 \\
1000 & 1.3e+06$\pm$1.5e+05 & 1.9e+03$\pm$3.1e+02 \\
2000 & 4.9e+06$\pm$1.6e+06 & 4.5e+03$\pm$1.7e+03 \\ \hline
\end{tabular}
\vspace*{-1em}\caption{\footnotesize{Preliminary results with Gurobi and the \textsc{FastIBP} algorithm ($\eta=0.001$).}}\label{fig:gurobi}\vspace*{-1em}
\end{wrapfigure}
We present some preliminary numerical results in Figure~\ref{fig:random} and~\ref{fig:prox}. Given $n=100$, we evaluate the performance of \textsc{FastIBP}, \textsc{IBP}, BADMM algorithms, and Gurobi by varying $m \in \{20, 50, 100, 200\}$ and use the same setup to compare the proximal variants of \textsc{FastIBP} and \textsc{IBP}. We use the proximal framework~\citep{Kroshnin-2019-Complexity} with the same parameter setting as provided by their paper. As indicated in Figure~\ref{fig:random}, the \textsc{FastIBP} algorithm performs better than BADMM and IBP algorithms in the sense that it consistently returns an objective value closer to that of Gurobi in less computational time. More specifically, IBP converges very fast when $\eta = 0.01$, but suffers from a crude solution with poor objective value; BADMM takes much more time with unsatisfactory objective value, and is not provably convergent in theory; Gurobi is highly optimized and can solve the problem of relatively small size very efficiently. However, when the problem size becomes larger, Gurobi would take much more time. As an example, for the case where $(m, n) = (200, 100)$, we see that Gurobi is about 10 times slower than the \textsc{FastIBP} algorithm with $\eta = 0.001$ while keeping relatively small normalized objective value. As indicated in Figure~\ref{fig:prox}, the proximal variant of \textsc{FastIBP} algorithm also outperforms that of  \textsc{IBP} algorithm in terms of objective value while not sacrificing the time efficiency. To facilitate the readers, we present the averaged results from 10 independent trials with \textsc{FastIBP}, \textsc{IBP}, BADMM algorithms, and Gurobi in Table~\ref{tab:random}. Note that we implement \textit{the rounding scheme} after each algorithm (except Gurobi) so the terms in \textbf{``feasibility"} are zero \textit{up to numerical errors} for most of medium-size problems. 
\begin{table}[!ht]\footnotesize
\centering\caption{Numerical results on synthetic data where each distribution has different dense weights but same support size. The support points of the barycenter is fixed. }\label{tab:random}
\hspace*{-5.5em}\begin{tabular}{|cc|cccccc|} \hline
m & n & g & b & i1 & i2 & f1 & f2 \\ \hline
& & \multicolumn{6}{|c|}{\textbf{normalized obj}} \\ \hline 
20 & 50 & - & 5.9e-01$\pm$1.2e-01 & 2.0e-01$\pm$8.0e-02 & 2.1e-01$\pm$1.1e-01 & 5.7e-02$\pm$1.1e-02 & 1.7e-03$\pm$9.7e-04 \\
20 & 100 & - & 6.7e-01$\pm$8.2e-02 & 3.2e-01$\pm$5.5e-02 & 3.6e-01$\pm$9.5e-02 & 6.7e-02$\pm$8.0e-03 & 2.1e-03$\pm$8.2e-04 \\
20 & 200 & - & 7.8e-01$\pm$7.4e-02 & 4.8e-01$\pm$5.9e-02 & 6.0e-01$\pm$7.6e-02 & 6.3e-02$\pm$4.7e-03 & 2.9e-03$\pm$3.8e-04 \\
50 & 50 & - & 4.5e-01$\pm$4.3e-02 & 1.7e-01$\pm$3.7e-02 & 1.6e-01$\pm$4.9e-02 & 6.8e-02$\pm$1.0e-02 & 2.2e-03$\pm$8.3e-04 \\
50 & 100 & - & 6.4e-01$\pm$1.0e-01 & 3.7e-01$\pm$6.8e-02 & 4.3e-01$\pm$6.4e-02 & 7.6e-02$\pm$8.3e-03 & 3.0e-03$\pm$6.1e-04 \\
50 & 200 & - & 8.2e-01$\pm$7.8e-02 & 5.9e-01$\pm$5.7e-02 & 6.7e-01$\pm$8.5e-02 & 6.2e-02$\pm$6.9e-03 & 4.0e-03$\pm$6.4e-04 \\
100 & 50 & - & 3.1e-01$\pm$3.0e-02 & 1.1e-01$\pm$2.9e-02 & 7.2e-02$\pm$2.8e-02 & 6.7e-02$\pm$1.4e-02 & 3.9e-03$\pm$2.3e-03 \\
100 & 100 & - & 6.1e-01$\pm$9.3e-02 & 3.8e-01$\pm$6.0e-02 & 4.6e-01$\pm$7.2e-02 & 7.7e-02$\pm$6.0e-03 & 3.6e-03$\pm$7.0e-04 \\
100 & 200 & - & 8.3e-01$\pm$5.0e-02 & 6.1e-01$\pm$4.0e-02 & 7.5e-01$\pm$4.2e-02 & 5.6e-02$\pm$4.7e-03 & 4.3e-03$\pm$6.9e-04 \\
200 & 50 & - & 2.8e-01$\pm$4.2e-02 & 1.1e-01$\pm$3.9e-02 & 6.0e-02$\pm$3.8e-02 & 6.9e-02$\pm$1.4e-02 & 3.2e-03$\pm$1.8e-03 \\
200 & 100 & - & 4.4e-01$\pm$4.6e-02 & 2.8e-01$\pm$3.0e-02 & 3.7e-01$\pm$5.8e-02 & 7.9e-02$\pm$3.1e-03 & 3.7e-03$\pm$3.3e-04 \\
200 & 200 & - & 8.0e-01$\pm$8.7e-02 & 6.0e-01$\pm$6.8e-02 & 7.2e-01$\pm$4.6e-02 & 5.7e-02$\pm$4.5e-03 & 5.2e-03$\pm$4.7e-04 \\ \hline

& & \multicolumn{6}{|c|}{\textbf{feasibility}} \\ \hline 
20 & 50 & 4.9e-07$\pm$0.0e+00 & 1.9e-07$\pm$0.0e+00 & 9.0e-07$\pm$0.0e+00 & 9.6e-07$\pm$0.0e+00 & 1.8e-14$\pm$0.0e+00 & 3.6e-09$\pm$0.0e+00 \\ 
20 & 100 & 0.0e+00$\pm$0.0e+00 & 0.0e+00$\pm$0.0e+00 & 0.0e+00$\pm$0.0e+00 & 0.0e+00$\pm$0.0e+00 & 0.0e+00$\pm$0.0e+00 & 0.0e+00$\pm$0.0e+00 \\ 
20 & 200 & 0.0e+00$\pm$0.0e+00 & 0.0e+00$\pm$0.0e+00 & 0.0e+00$\pm$0.0e+00 & 0.0e+00$\pm$0.0e+00 & 0.0e+00$\pm$0.0e+00 & 0.0e+00$\pm$0.0e+00 \\  
50 & 50 & 0.0e+00$\pm$0.0e+00 & 0.0e+00$\pm$0.0e+00 & 0.0e+00$\pm$0.0e+00 & 0.0e+00$\pm$0.0e+00 & 0.0e+00$\pm$0.0e+00 & 0.0e+00$\pm$0.0e+00 \\ 
50 & 100 & 0.0e+00$\pm$0.0e+00 & 0.0e+00$\pm$0.0e+00 & 0.0e+00$\pm$0.0e+00 & 0.0e+00$\pm$0.0e+00 & 0.0e+00$\pm$0.0e+00 & 0.0e+00$\pm$0.0e+00 \\  
50 & 200 & 0.0e+00$\pm$0.0e+00 & 0.0e+00$\pm$0.0e+00 & 0.0e+00$\pm$0.0e+00 & 0.0e+00$\pm$0.0e+00 & 0.0e+00$\pm$0.0e+00 & 0.0e+00$\pm$0.0e+00 \\ 
100 & 50 & 0.0e+00$\pm$0.0e+00 & 0.0e+00$\pm$0.0e+00 & 0.0e+00$\pm$0.0e+00 & 0.0e+00$\pm$0.0e+00 & 0.0e+00$\pm$0.0e+00 & 0.0e+00$\pm$0.0e+00 \\
100 & 100 & 0.0e+00$\pm$0.0e+00 & 0.0e+00$\pm$0.0e+00 & 0.0e+00$\pm$0.0e+00 & 0.0e+00$\pm$0.0e+00 & 0.0e+00$\pm$0.0e+00 & 0.0e+00$\pm$0.0e+00 \\ 
100 & 200 & 0.0e+00$\pm$0.0e+00 & 0.0e+00$\pm$0.0e+00 & 0.0e+00$\pm$0.0e+00 & 0.0e+00$\pm$0.0e+00 & 0.0e+00$\pm$0.0e+00 & 0.0e+00$\pm$0.0e+00 \\ 
200 & 50 & 0.0e+00$\pm$0.0e+00 & 0.0e+00$\pm$0.0e+00 & 0.0e+00$\pm$0.0e+00 & 0.0e+00$\pm$0.0e+00 & 0.0e+00$\pm$0.0e+00 & 0.0e+00$\pm$0.0e+00 \\ 
200 & 100 & 0.0e+00$\pm$0.0e+00 & 0.0e+00$\pm$0.0e+00 & 0.0e+00$\pm$0.0e+00 & 0.0e+00$\pm$0.0e+00 & 0.0e+00$\pm$0.0e+00 & 0.0e+00$\pm$0.0e+00 \\ 
200 & 200 & 2.7e-07$\pm$1.9e-07 & 2.2e-07$\pm$2.3e-08 & 6.2e-07$\pm$1.5e-07 & 9.2e-07$\pm$3.9e-08 & 7.5e-14$\pm$1.9e-13 & 1.0e-07$\pm$1.6e-07 \\ \hline

& & \multicolumn{6}{|c|}{\textbf{iteration}} \\ \hline 
20 & 50 & 3115$\pm$532 & 5000$\pm$0 & 440$\pm$158 & 7140$\pm$3912 & 200$\pm$0 & 2760$\pm$1560 \\
20 & 100 & 6415$\pm$999 & 3400$\pm$94 & 400$\pm$0 & 8660$\pm$5464 & 200$\pm$0 & 2860$\pm$1678 \\
20 & 200 & 12430$\pm$2339 & 3580$\pm$175 & 400$\pm$0 & 5520$\pm$2368 & 200$\pm$0 & 1240$\pm$497 \\ 
50 & 50 & 10139$\pm$1192 & 5000$\pm$0 & 360$\pm$227 & 7320$\pm$6351 & 200$\pm$0 & 3560$\pm$1786 \\  
50 & 100 & 21697$\pm$4377 & 3480$\pm$103 & 580$\pm$63 & 9920$\pm$5351 & 200$\pm$0 & 1540$\pm$1116 \\  
50 & 200 & 39564$\pm$6916 & 3740$\pm$97 & 580$\pm$63 & 6800$\pm$2512 & 220$\pm$63 & 1240$\pm$833 \\ 
100 & 50 & 26729$\pm$2731 & 4580$\pm$887 & 380$\pm$274 & 5180$\pm$1901 & 240$\pm$84 & 5940$\pm$2406 \\
100 & 100 & 52799$\pm$9610 & 3560$\pm$84 & 700$\pm$287 & 10020$\pm$3400 & 200$\pm$0 & 1360$\pm$815 \\
100 & 200 & 97357$\pm$10615 & 3780$\pm$114 & 920$\pm$103 & 9720$\pm$3737 & 200$\pm$0 & 440$\pm$280 \\  
200 & 50 & 55841$\pm$6359 & 4280$\pm$1038 & 300$\pm$141 & 11140$\pm$4724 & 200$\pm$0 & 8000$\pm$3749 \\
200 & 100 & 149230$\pm$18051 & 3600$\pm$0 & 980$\pm$537 & 12840$\pm$3650 & 200$\pm$0 & 1880$\pm$634 \\ 
200 & 200 & 258059$\pm$62104 & 3800$\pm$94 & 1440$\pm$158 & 12160$\pm$2609 & 200$\pm$0 & 340$\pm$165 \\ \hline

& & \multicolumn{6}{|c|}{\textbf{time (in seconds)}} \\ \hline
20 & 50 & 3.6e-01$\pm$1.1e-01 & 9.4e+00$\pm$3.6e+00 & 2.5e-01$\pm$1.1e-01 & 4.1e+00$\pm$2.2e+00 & 3.1e-01$\pm$5.7e-02 & 5.2e+00$\pm$2.4e+00 \\ 
20 & 100 & 1.9e+00$\pm$1.2e+00 & 2.4e+01$\pm$3.9e+00 & 1.2e+00$\pm$6.2e-01 & 2.5e+01$\pm$1.5e+01 & 1.6e+00$\pm$9.0e-01 & 2.1e+01$\pm$1.3e+01 \\  
20 & 200 & 6.2e+00$\pm$1.7e+00 & 1.2e+02$\pm$5.1e+00 & 4.0e+00$\pm$5.6e-01 & 5.9e+01$\pm$2.7e+01 & 5.9e+00$\pm$7.5e-01 & 3.9e+01$\pm$1.5e+01 \\ 
50 & 50 & 3.1e+00$\pm$1.3e+00 & 2.3e+01$\pm$4.8e+00 & 5.4e-01$\pm$3.8e-01 & 1.2e+01$\pm$1.1e+01 & 1.1e+00$\pm$6.0e-01 & 1.8e+01$\pm$9.2e+00 \\ 
50 & 100 & 1.1e+01$\pm$2.1e+00 & 6.9e+01$\pm$5.0e+00 & 3.7e+00$\pm$7.1e-01 & 7.2e+01$\pm$4.1e+01 & 3.5e+00$\pm$5.3e-01 & 3.0e+01$\pm$2.2e+01 \\  
50 & 200 & 2.7e+01$\pm$5.8e+00 & 3.2e+02$\pm$1.4e+01 & 1.7e+01$\pm$5.2e+00 & 2.0e+02$\pm$7.5e+01 & 1.5e+01$\pm$4.5e+00 & 8.8e+01$\pm$5.5e+01 \\  

100 & 50 & 1.3e+01$\pm$4.3e+00 & 4.3e+01$\pm$7.8e+00 & 1.1e+00$\pm$9.0e-01 & 1.7e+01$\pm$7.0e+00 & 2.7e+00$\pm$1.2e+00 & 6.1e+01$\pm$2.7e+01 \\ 
100 & 100 & 3.6e+01$\pm$1.1e+01 & 1.4e+02$\pm$3.9e+00 & 7.9e+00$\pm$3.5e+00 & 1.4e+02$\pm$4.9e+01 & 7.2e+00$\pm$1.0e+00 & 5.2e+01$\pm$3.0e+01 \\
100 & 200 & 1.0e+02$\pm$2.1e+01 & 6.6e+02$\pm$2.7e+01 & 5.1e+01$\pm$6.0e+00 & 5.7e+02$\pm$2.3e+02 & 2.7e+01$\pm$3.8e+00 & 6.6e+01$\pm$4.1e+01 \\ 
200 & 50 & 5.4e+01$\pm$1.2e+01 & 9.3e+01$\pm$2.5e+01 & 2.0e+00$\pm$8.9e-01 & 9.0e+01$\pm$4.2e+01 & 4.8e+00$\pm$2.9e+00 & 1.8e+02$\pm$1.0e+02 \\ 
200 & 100 & 2.8e+02$\pm$6.7e+01 & 3.2e+02$\pm$2.7e+01 & 3.0e+01$\pm$1.9e+01 & 4.0e+02$\pm$1.3e+02 & 1.5e+01$\pm$4.5e+00 & 1.5e+02$\pm$5.1e+01 \\ 
200 & 200 & 4.9e+02$\pm$2.0e+02 & 1.9e+03$\pm$9.5e+01 & 2.8e+02$\pm$3.2e+01 & 2.5e+03$\pm$5.6e+02 & 1.1e+02$\pm$8.1e+00 & 1.9e+02$\pm$9.5e+01 \\ \hline
\end{tabular}
\end{table}

To further compare the performances of Gurobi and the \textsc{FastIBP} algorithm, we conduct the experiment with $n=100$ and the varying number of marginals $m \in \{200, 500, 1000, 2000\}$. We fix $\textnormal{Tol}_{\textsf{fibp}} = 10^{-6}$ but without setting the maximum iteration number. Figure~\ref{fig:gurobi} shows the average running time taken by two algorithms over 5 independent trials. We see that the \textsc{FastIBP} algorithm is competitive with Gurobi in terms of objective value and feasibility violation. In terms of computational time, the \textsc{FastIBP} algorithm increases linearly with respect to the number of marginals, while Gurobi increases much more rapidly. Compared to the similar results of Gurobi presented before~\citep{Yang-2018-ADMM, Ge-2019-Interior}, we find that the feasibility violation in our paper is better but the computational time grows much faster. This makes sense since we run the dual simplex algorithm, which iterates over the feasible solutions but is more computationally expensive than the interior-point algorithm. Figure~\ref{fig:gurobi} demonstrates that the structure of the FS-WBP is not favorable to the dual simplex algorithm, confirming our computational hardness results in Section~\ref{sec:hardness}.  
\begin{table}[!ht]
\centering
\begin{tabular}{cccccccccc}
& \multicolumn{9}{c}{\textsc{FastIBP} ($\eta = 0.001$)} \\
100s & \begin{minipage}{0.05\textwidth}
\includegraphics[scale = 0.18]{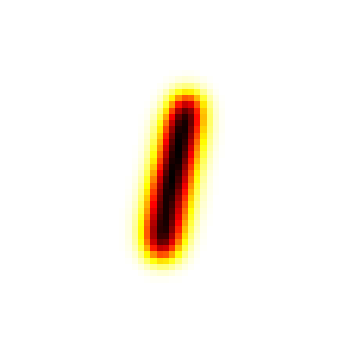}
\end{minipage}
& \begin{minipage}{0.05\textwidth}
\includegraphics[scale = 0.18]{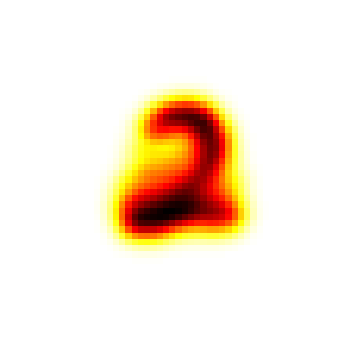}
\end{minipage}
& \begin{minipage}{0.05\textwidth}
\includegraphics[scale = 0.18]{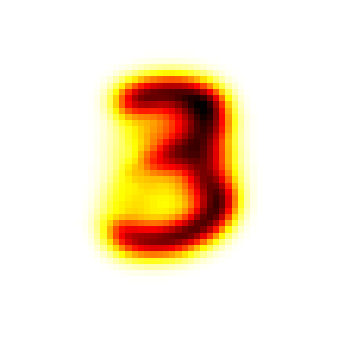}
\end{minipage}
& \begin{minipage}{0.05\textwidth}
\includegraphics[scale = 0.18]{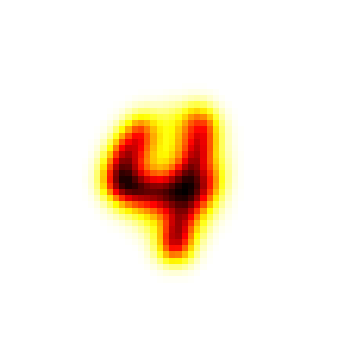}
\end{minipage}
& \begin{minipage}{0.05\textwidth}
\includegraphics[scale = 0.18]{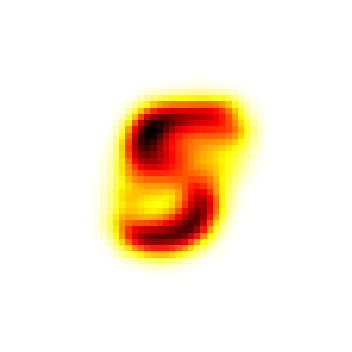}
\end{minipage}
& \begin{minipage}{0.05\textwidth}
\includegraphics[scale = 0.18]{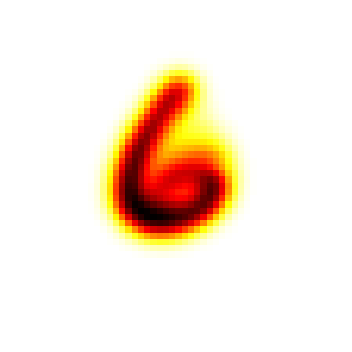}
\end{minipage}
& \begin{minipage}{0.05\textwidth}
\includegraphics[scale = 0.18]{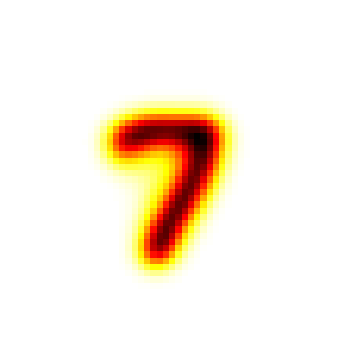}
\end{minipage}
& \begin{minipage}{0.05\textwidth}
\includegraphics[scale = 0.18]{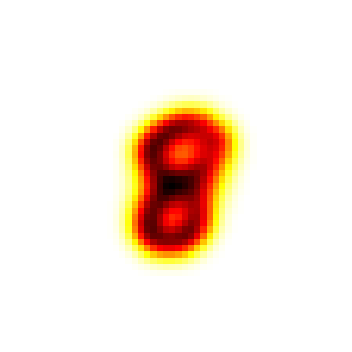}
\end{minipage} 
& \begin{minipage}{0.05\textwidth}
\includegraphics[scale = 0.18]{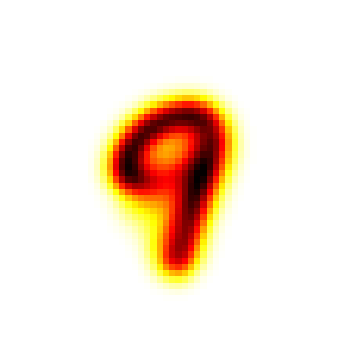}
\end{minipage} \\ 
200s & \begin{minipage}{0.05\textwidth}
\includegraphics[scale = 0.18]{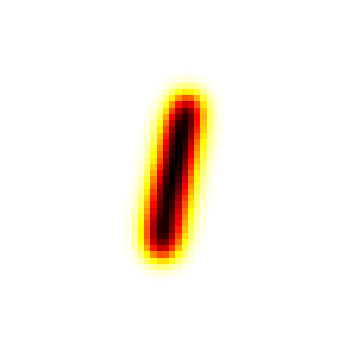}
\end{minipage}
& \begin{minipage}{0.05\textwidth}
\includegraphics[scale = 0.18]{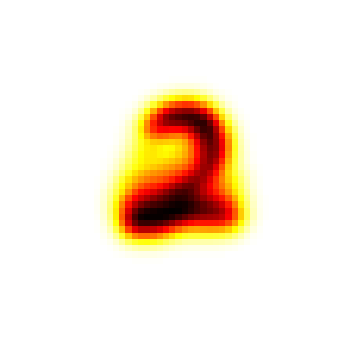}
\end{minipage}
& \begin{minipage}{0.05\textwidth}
\includegraphics[scale = 0.18]{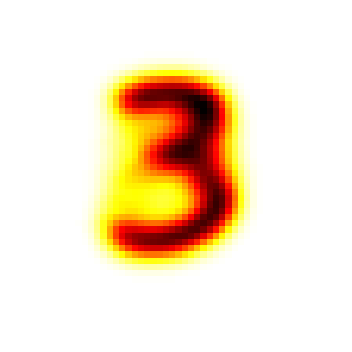}
\end{minipage}
& \begin{minipage}{0.05\textwidth}
\includegraphics[scale = 0.18]{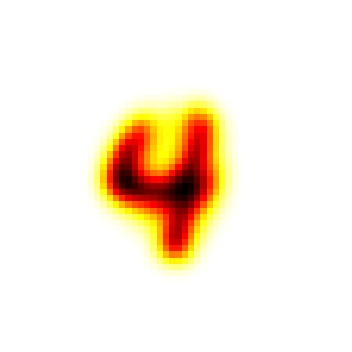}
\end{minipage}
& \begin{minipage}{0.05\textwidth}
\includegraphics[scale = 0.18]{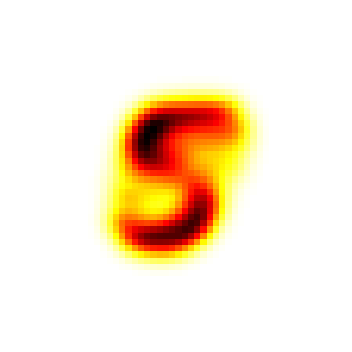}
\end{minipage}
& \begin{minipage}{0.05\textwidth}
\includegraphics[scale = 0.18]{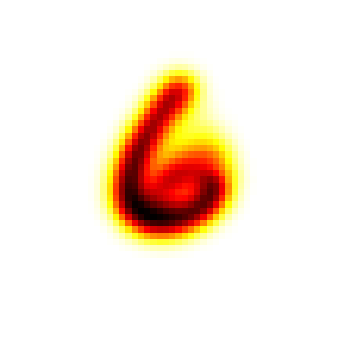}
\end{minipage}
& \begin{minipage}{0.05\textwidth}
\includegraphics[scale = 0.18]{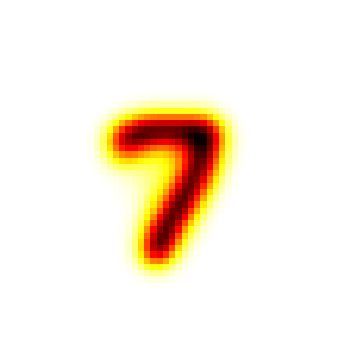}
\end{minipage}
& \begin{minipage}{0.05\textwidth}
\includegraphics[scale = 0.18]{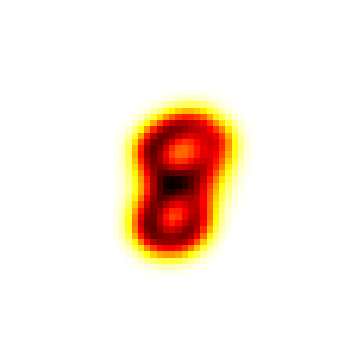}
\end{minipage} 
& \begin{minipage}{0.05\textwidth}
\includegraphics[scale = 0.18]{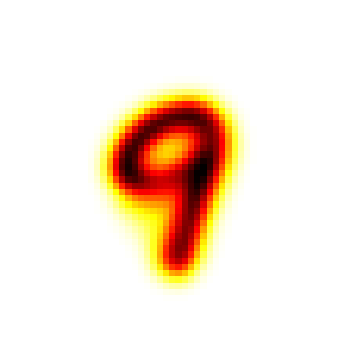}
\end{minipage} \\  
400s & \begin{minipage}{0.05\textwidth}
\includegraphics[scale = 0.18]{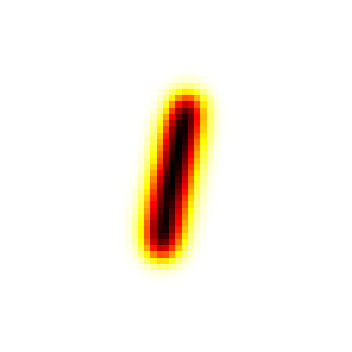}
\end{minipage}
& \begin{minipage}{0.05\textwidth}
\includegraphics[scale = 0.18]{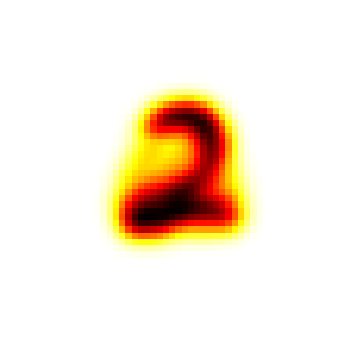}
\end{minipage}
& \begin{minipage}{0.05\textwidth}
\includegraphics[scale = 0.18]{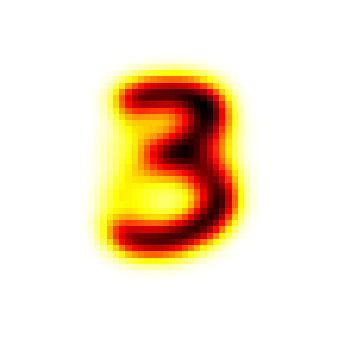}
\end{minipage}
& \begin{minipage}{0.05\textwidth}
\includegraphics[scale = 0.18]{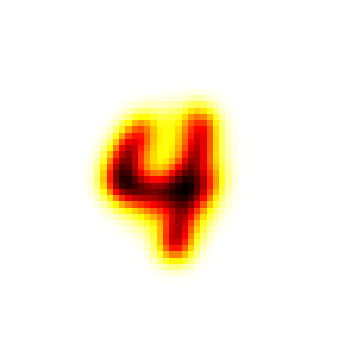}
\end{minipage}
& \begin{minipage}{0.05\textwidth}
\includegraphics[scale = 0.18]{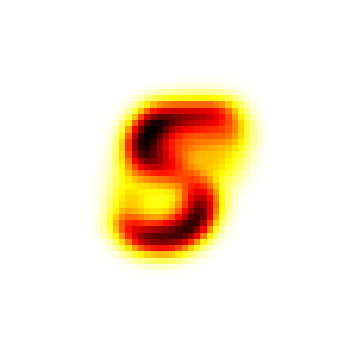}
\end{minipage}
& \begin{minipage}{0.05\textwidth}
\includegraphics[scale = 0.18]{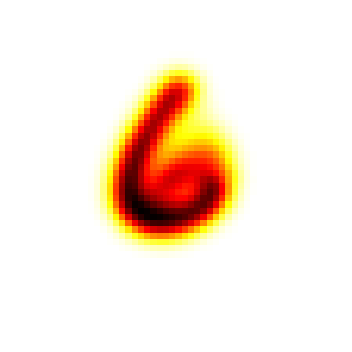}
\end{minipage}
& \begin{minipage}{0.05\textwidth}
\includegraphics[scale = 0.18]{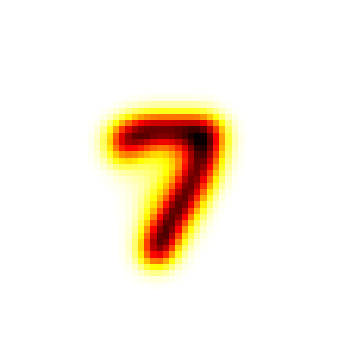}
\end{minipage}
& \begin{minipage}{0.05\textwidth}
\includegraphics[scale = 0.18]{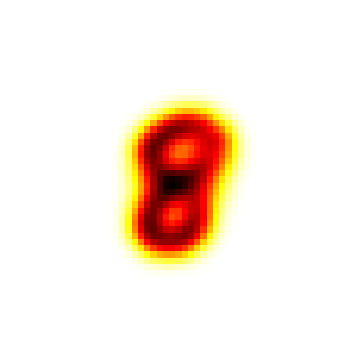}
\end{minipage} 
& \begin{minipage}{0.05\textwidth}
\includegraphics[scale = 0.18]{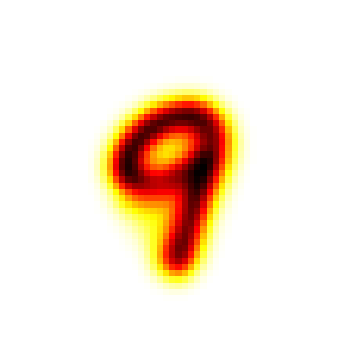}
\end{minipage} \\
800s & \begin{minipage}{0.05\textwidth}
\includegraphics[scale = 0.18]{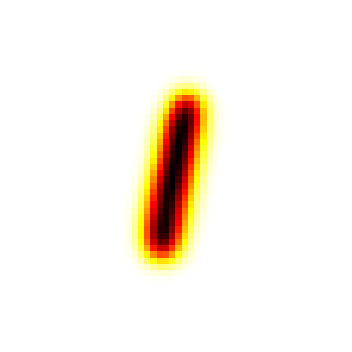}
\end{minipage}
& \begin{minipage}{0.05\textwidth}
\includegraphics[scale = 0.18]{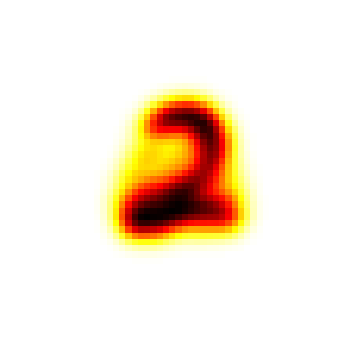}
\end{minipage}
& \begin{minipage}{0.05\textwidth}
\includegraphics[scale = 0.18]{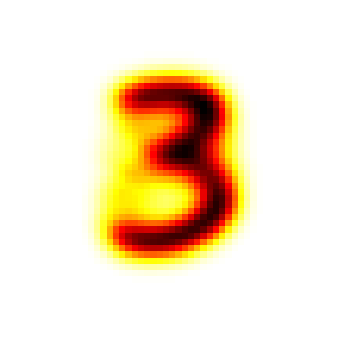}
\end{minipage}
& \begin{minipage}{0.05\textwidth}
\includegraphics[scale = 0.18]{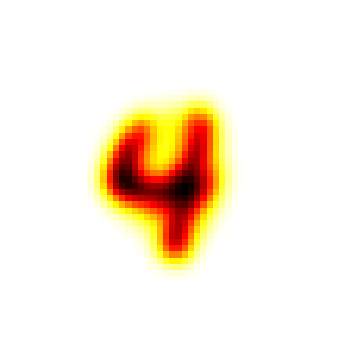}
\end{minipage}
& \begin{minipage}{0.05\textwidth}
\includegraphics[scale = 0.18]{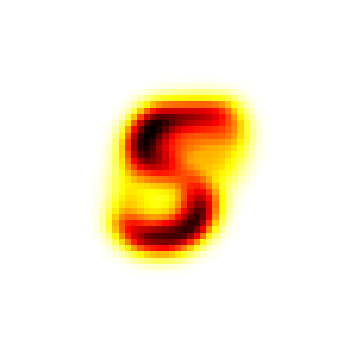}
\end{minipage}
& \begin{minipage}{0.05\textwidth}
\includegraphics[scale = 0.18]{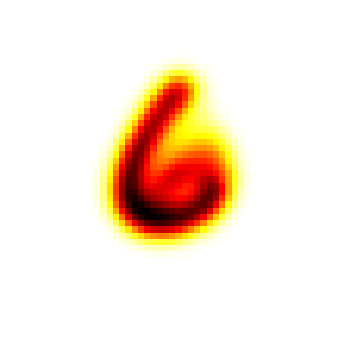}
\end{minipage}
& \begin{minipage}{0.05\textwidth}
\includegraphics[scale = 0.18]{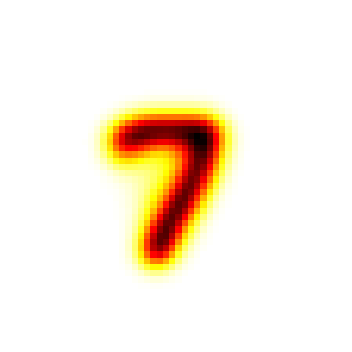}
\end{minipage}
& \begin{minipage}{0.05\textwidth}
\includegraphics[scale = 0.18]{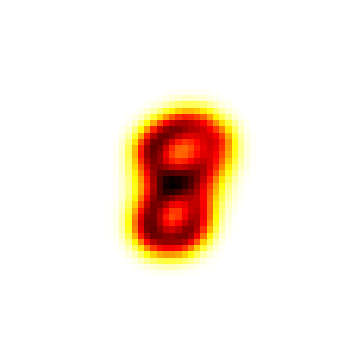}
\end{minipage} 
& \begin{minipage}{0.05\textwidth}
\includegraphics[scale = 0.18]{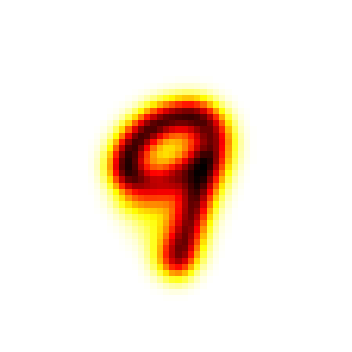}
\end{minipage} \\
& \multicolumn{9}{c}{IBP ($\eta = 0.001$)} \\
100s & \begin{minipage}{0.05\textwidth}
\includegraphics[scale = 0.18]{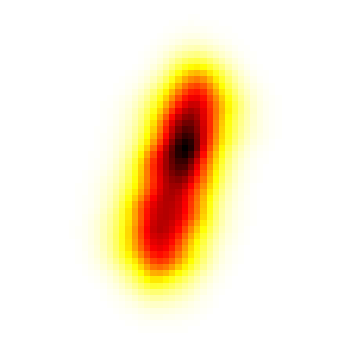}
\end{minipage}
& \begin{minipage}{0.05\textwidth}
\includegraphics[scale = 0.18]{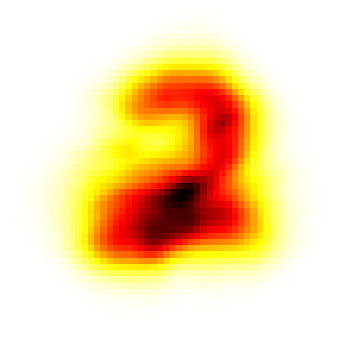}
\end{minipage}
& \begin{minipage}{0.05\textwidth}
\includegraphics[scale = 0.18]{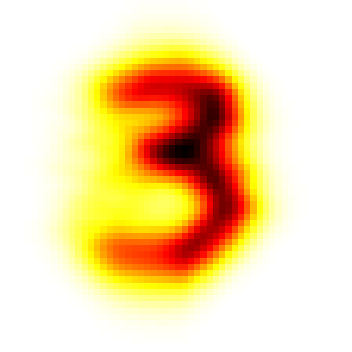}
\end{minipage}
& \begin{minipage}{0.05\textwidth}
\includegraphics[scale = 0.18]{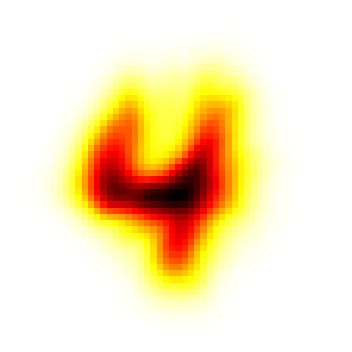}
\end{minipage}
& \begin{minipage}{0.05\textwidth}
\includegraphics[scale = 0.18]{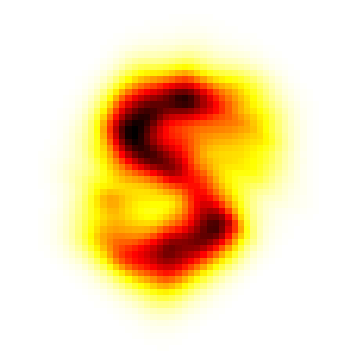}
\end{minipage}
& \begin{minipage}{0.05\textwidth}
\includegraphics[scale = 0.18]{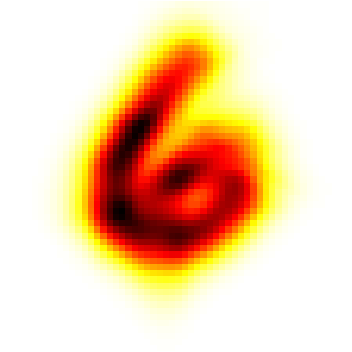}
\end{minipage}
& \begin{minipage}{0.05\textwidth}
\includegraphics[scale = 0.18]{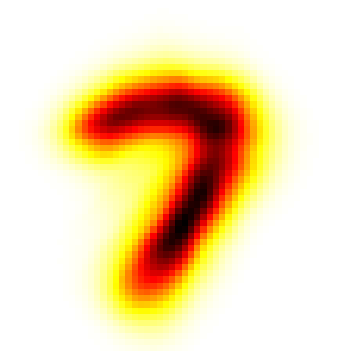}
\end{minipage}
& \begin{minipage}{0.05\textwidth}
\includegraphics[scale = 0.18]{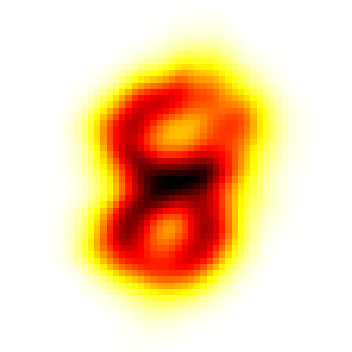}
\end{minipage} 
& \begin{minipage}{0.05\textwidth}
\includegraphics[scale = 0.18]{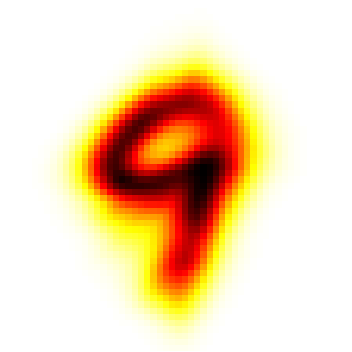}
\end{minipage} \\ 
200s & \begin{minipage}{0.05\textwidth}
\includegraphics[scale = 0.18]{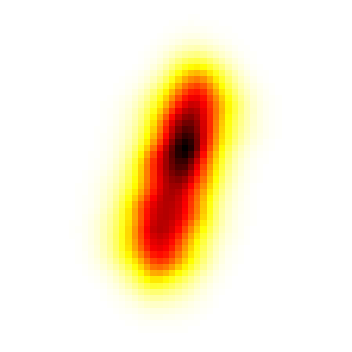}
\end{minipage}
& \begin{minipage}{0.05\textwidth}
\includegraphics[scale = 0.18]{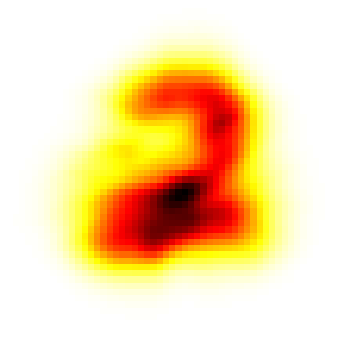}
\end{minipage}
& \begin{minipage}{0.05\textwidth}
\includegraphics[scale = 0.18]{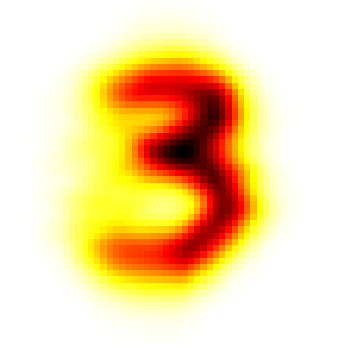}
\end{minipage}
& \begin{minipage}{0.05\textwidth}
\includegraphics[scale = 0.18]{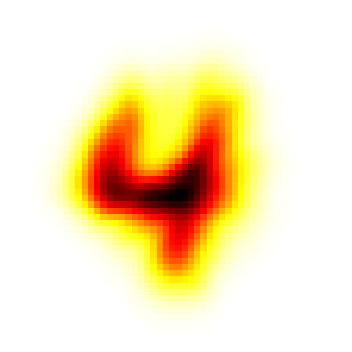}
\end{minipage}
& \begin{minipage}{0.05\textwidth}
\includegraphics[scale = 0.18]{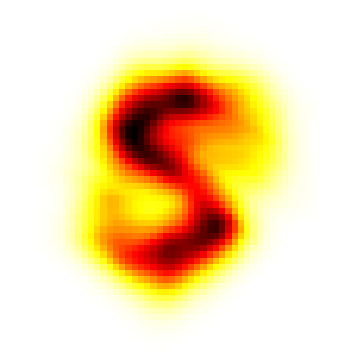}
\end{minipage}
& \begin{minipage}{0.05\textwidth}
\includegraphics[scale = 0.18]{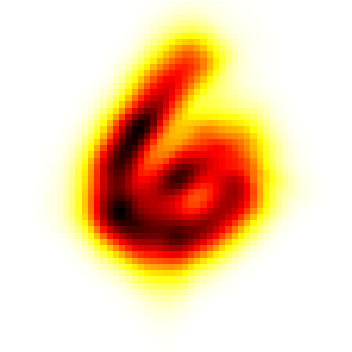}
\end{minipage}
& \begin{minipage}{0.05\textwidth}
\includegraphics[scale = 0.18]{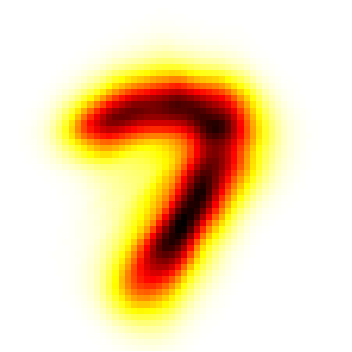}
\end{minipage}
& \begin{minipage}{0.05\textwidth}
\includegraphics[scale = 0.18]{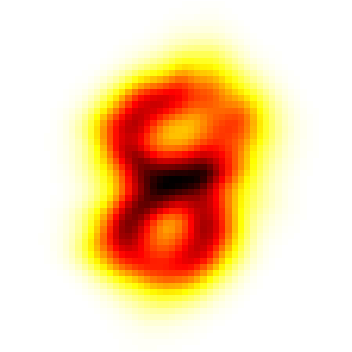}
\end{minipage} 
& \begin{minipage}{0.05\textwidth}
\includegraphics[scale = 0.18]{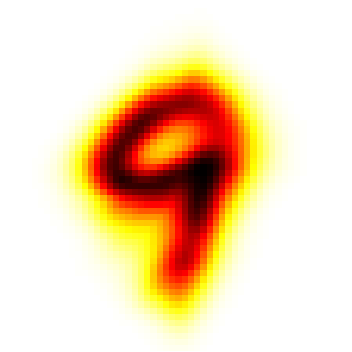}
\end{minipage} \\  
400s & \begin{minipage}{0.05\textwidth}
\includegraphics[scale = 0.18]{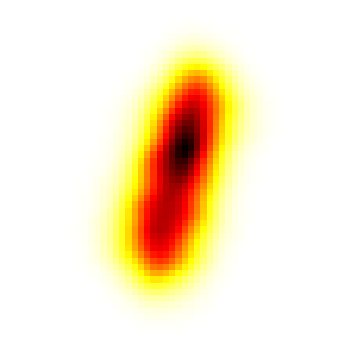}
\end{minipage}
& \begin{minipage}{0.05\textwidth}
\includegraphics[scale = 0.18]{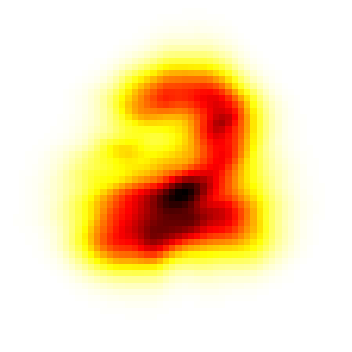}
\end{minipage}
& \begin{minipage}{0.05\textwidth}
\includegraphics[scale = 0.18]{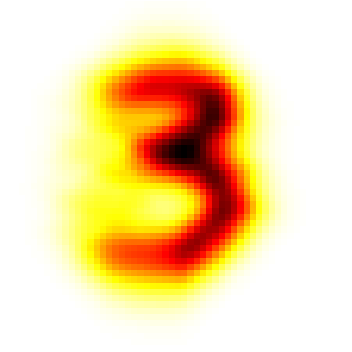}
\end{minipage}
& \begin{minipage}{0.05\textwidth}
\includegraphics[scale = 0.18]{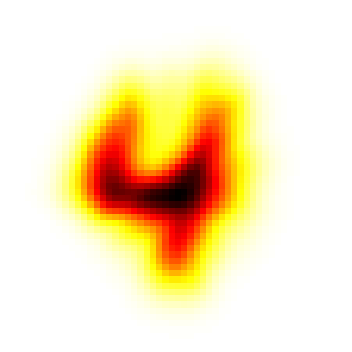}
\end{minipage}
& \begin{minipage}{0.05\textwidth}
\includegraphics[scale = 0.18]{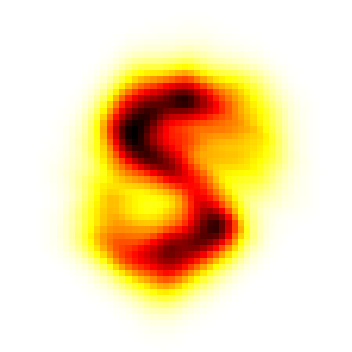}
\end{minipage}
& \begin{minipage}{0.05\textwidth}
\includegraphics[scale = 0.18]{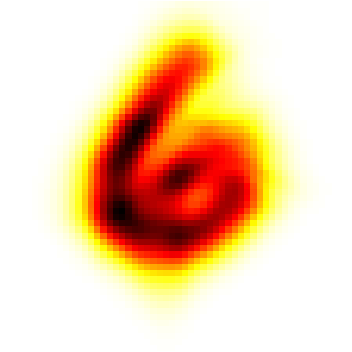}
\end{minipage}
& \begin{minipage}{0.05\textwidth}
\includegraphics[scale = 0.18]{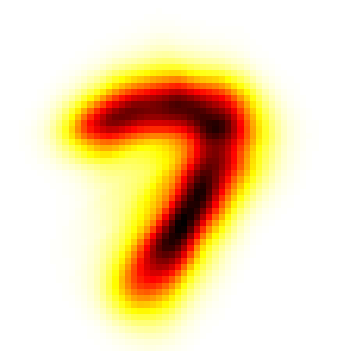}
\end{minipage}
& \begin{minipage}{0.05\textwidth}
\includegraphics[scale = 0.18]{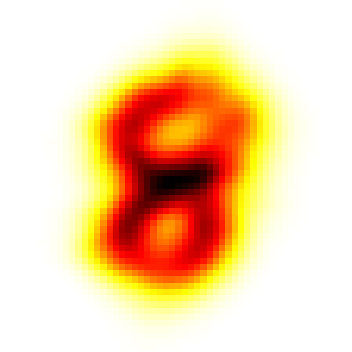}
\end{minipage} 
& \begin{minipage}{0.05\textwidth}
\includegraphics[scale = 0.18]{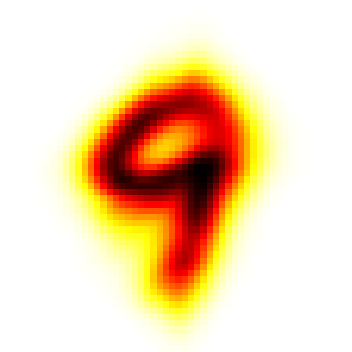}
\end{minipage} \\
800s & \begin{minipage}{0.05\textwidth}
\includegraphics[scale = 0.18]{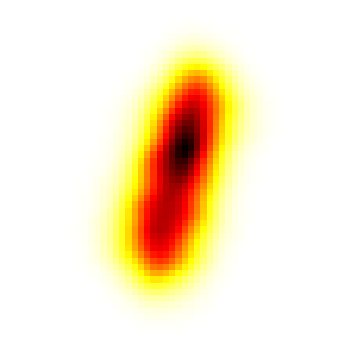}
\end{minipage}
& \begin{minipage}{0.05\textwidth}
\includegraphics[scale = 0.18]{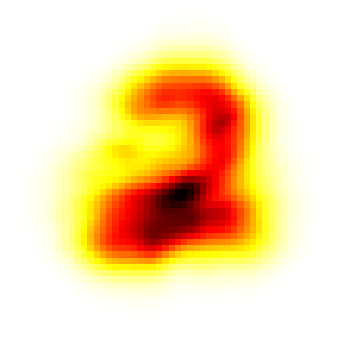}
\end{minipage}
& \begin{minipage}{0.05\textwidth}
\includegraphics[scale = 0.18]{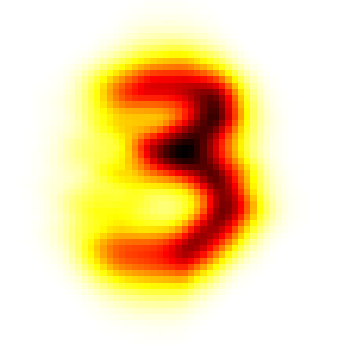}
\end{minipage}
& \begin{minipage}{0.05\textwidth}
\includegraphics[scale = 0.18]{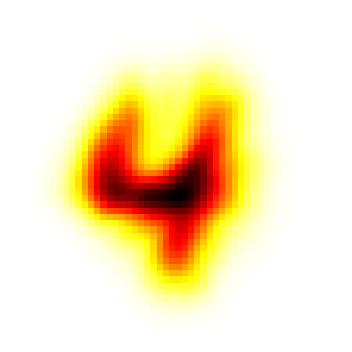}
\end{minipage}
& \begin{minipage}{0.05\textwidth}
\includegraphics[scale = 0.18]{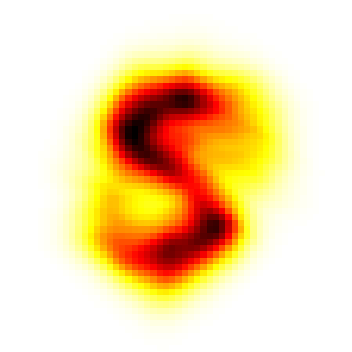}
\end{minipage}
& \begin{minipage}{0.05\textwidth}
\includegraphics[scale = 0.18]{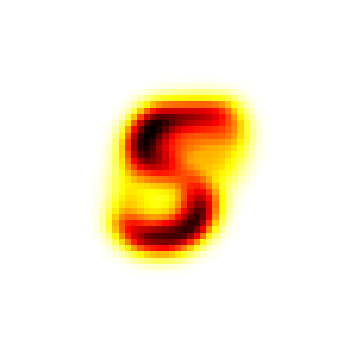}
\end{minipage}
& \begin{minipage}{0.05\textwidth}
\includegraphics[scale = 0.18]{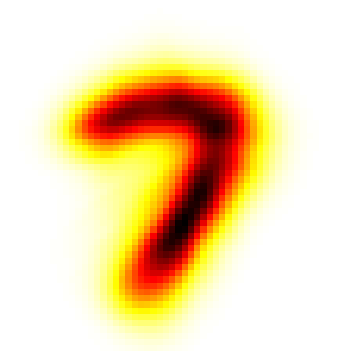}
\end{minipage}
& \begin{minipage}{0.05\textwidth}
\includegraphics[scale = 0.18]{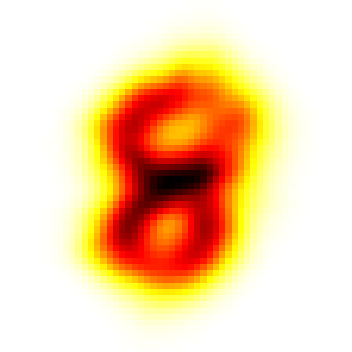}
\end{minipage} 
& \begin{minipage}{0.05\textwidth}
\includegraphics[scale = 0.18]{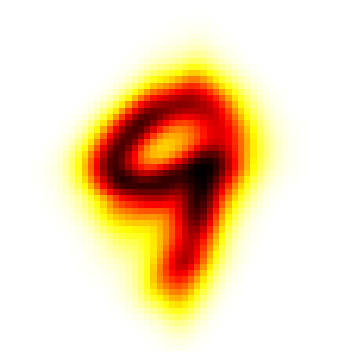}
\end{minipage}
\end{tabular}
\caption{\footnotesize{Approximate barycenters obtained by running \textsc{FastIBP} and IBP for 100s, 200s, 400s, 800s.}}\label{tab:MNIST}\vspace*{-1em}
\end{table}

\subsection{Experiments on MNIST}\label{subsec:mnist}
To better visualize the quality of approximate barycenters obtained by each algorithm, we follow~\citet{Cuturi-2014-Fast} on the MNIST\footnote{Available in http://yann.lecun.com/exdb/mnist/} dataset~\citep{Lecun-1998-Gradient}. We randomly select 50 images for each digit (1$\sim$9) and resize each image to $\zeta$ times of its original size of 28 $\times$ 28, where $\zeta$ is drawn uniformly at random from $[0.5, 2]$. We randomly put each resized image in a larger 56 $\times$ 56 blank image and normalize the resulting image so that all pixel values add up to 1. Each image can be viewed as a discrete distribution supported on grids. Additionally, we set the weight vector $(\omega_1, \omega_2, \ldots, \omega_m)$ such that $\omega_k = 1/m$ for all $k \in [m]$.

We apply the \textsc{FastIBP} algorithm ($\eta = 0.001$) to compute the Wasserstein barycenter of the resulting images for each digit on the MNIST dataset and compare it to IBP ($\eta = 0.001$). We exclude BADMM since~\citet[Figure~3]{Yang-2018-ADMM} and~\citet[Table~1]{Ge-2019-Interior} have shown that IBP outperforms BADMM on the MNIST dataset. The size of barycenter is set to 56 $\times$ 56. For a fair comparison, we do not implement convolutional technique~\citep{Solomon-2015-Convolutional} and its stabilized version~\citep[Section~4.1.2]{Schmitzer-2019-Stabilized}, which can be used to substantially improve IBP with small $\eta$. The approximate barycenters obtained by the \textsc{FastIBP} and IBP algorithms are presented in Table~\ref{tab:MNIST}. It can be seen that the \textsc{FastIBP} algorithm provides a ``sharper" approximate barycenter than IBP when $\eta = 0.001$ is set for both. This demonstrates the good quality of the solution obtained by our algorithm.

\section{Conclusions}\label{sec:discussions}
In this paper, we study the computational hardness for solving the fixed-support Wasserstein barycenter problem (FS-WBP) and proves that the FS-WBP in the standard linear programming form is not a minimum-cost flow (MCF) problem when $m \geq 3$ and $n \geq 3$. Our results suggest that the direct application of network flow algorithms to the FS-WBP in standard LP form is inefficient, shedding the light on the practical performance of various existing algorithms, which are developed based on problem reformulation of the FS-WBP. Moreover, we propose a \textit{deterministic} variant of iterative Bregman projection (IBP) algorithm, namely \textsc{FastIBP}, and prove that the complexity bound is $\bigOtil(mn^{7/3}\varepsilon^{-4/3})$. This bound is better than the complexity bound of $\bigOtil(mn^2\varepsilon^{-2})$ from the IBP algorithm in terms of $\varepsilon$, and that of $\bigOtil(mn^{5/2}\varepsilon^{-1})$ from other accelerated algorithms in terms of $n$. Experiments on synthetic and real datasets demonstrate the favorable performance of the \textsc{FastIBP} algorithm in practice. 

% Acknowledgements should only appear in the accepted version.
\section{Acknowledgments}
We would like to thank Lei Yang for very helpful discussion with the experiments on MNIST datasets and four anonymous referees for constructive suggestions that improve the quality of this paper. Xi Chen is supported by National Science Foundation via the Grant IIS-1845444. This work is supported in part by the Mathematical Data Science program of the Office of Naval Research under grant number N00014-18-1-2764.
%%%%%%%%%%%%%%%%%%%%%%%%%%%%%%%%%%%%%%%%%%%%%%%%%%%%%%%%%%%%%%%%%%%%%

\bibliographystyle{plainnat}
\bibliography{ref}

\end{document}